\documentclass[12pt]{article} 
\pdfoutput=1

\usepackage[english]{babel}           % style anglais
\usepackage{lmodern}
\usepackage[T1]{fontenc} % codage moderne des caractÃšÂres sous Latex
\usepackage[latin9]{inputenc}

\usepackage{cancel}
\usepackage[pdftex]{graphicx}
\usepackage{epsfig}
\usepackage{graphicx}
\usepackage{comment}
\usepackage{latexsym}
\usepackage[hypertexnames=false]{hyperref}
\usepackage{ulem}
\usepackage{amsmath}
\usepackage{mathrsfs}
\usepackage[usenames, dvipsnames]{color}
\usepackage{amsbsy}
\usepackage{amssymb}
\usepackage{amsthm}
\usepackage{amsfonts}
\usepackage{cite}
\usepackage{enumitem}
\usepackage{xcolor}
\usepackage{color}
\usepackage{mathtools}
\usepackage{caption} 
\usepackage{subcaption} %sous-figures 
\usepackage{euscript} % FOR HAND CALLIGRAPHIC 
\usepackage{float}
\usepackage{diagbox}
\usepackage[title]{appendix}
\usepackage{tabularx}
\usepackage{tikz}
\usepackage{pgfplots} %package to plot graphs
\pgfplotsset{compat=1.16} %same
\usepackage{tcolorbox}
\usetikzlibrary{decorations.pathmorphing,arrows.meta}%\usepackage{ulem}

\usepackage{amsmath}
\allowdisplaybreaks
\usepackage{amsfonts}
\usepackage{amssymb}
\usepackage{bbm}
\usepackage{verbatim}
\usepackage{dsfont}
\usepackage{booktabs}
\usepackage{slashed}

\newcommand{\Le}{{\rm L}}
\newcommand{\Ex}{{\rm E}}
\newcommand{\Ri}{{\rm R}}
\newcommand{\hA}{\widehat{\cal A}}
\newcommand{\tA}{\widetilde{\cal A}}

\renewcommand\d{{\rm d}}

\newcommand{\F}{{\cal F}}
\newcommand{\A}{{\cal A}}

\newcommand{\C}{{\cal C}}

\renewcommand{\S}{{\cal S}}

\newcommand{\be}{\begin{equation}}
\newcommand{\ee}{\end{equation}}
\newcommand{\dis}{\displaystyle}
\renewcommand{\thefootnote}{\fnsymbol{footnote}}

\newcommand{\Eq}[1]{Eq.~\eqref{#1}}
\newcommand{\Eqs}[1]{Eqs~\eqref{#1}}
\newcommand{\Refe}[1]{Ref.~\cite{#1}}
\newcommand{\Refs}[1]{Refs~\cite{#1}}
\newcommand{\Sect}[1]{Section~\ref{#1}}

\newcommand{\Appendix}[1]{Appendix~\ref{#1}}
\newcommand{\Fig}[1]{Figure~\ref{#1}}
\newcommand{\Figs}[1]{Figures~\ref{#1}}

\newcommand{\R}{\mathbb{R}}

\renewcommand{\O}{{\cal O}}

\DeclareMathOperator\sign{sign}

\DeclareMathOperator\tr{tr}

\renewcommand{\o}{\overset{\scriptscriptstyle *}}
\newcommand{\oo}{\overset{\scriptscriptstyle **}}

\newcommand{\ie}{{\it i.e.} }

\newcommand{\eg}{{\it e.g.} }
\newcommand{\via}{{\it via} }
\newcommand{\apriori}{{\it a priori}}
\newcommand{\where}{\mbox{where}}

\newcommand{\when}{\mbox{when}}
\renewcommand{\and}{\mbox{and}}

\newcommand{\esps}{\phantom{\!\!\!\overset{|}{a}}}
\newcommand{\esp}{\phantom{\!\!\overset{\displaystyle |}{|}}}

\newcommand{\bm}{\boldmath} 
\newcommand{\red}{\color{red}}

\newcommand{\blue}{\color{blue}}

\catcode`\@=11
\def\marginnote#1{}
\newcount\hour
\newcount\minute
\newtoks\amorpm
\hour=\time\divide\hour by60 \minute=\time{\multiply\hour by60
\global\advance\minute by-\hour}
\edef\standardtime{{\ifnum\hour<12 \global\amorpm={am}%
        \else\global\amorpm={pm}\advance\hour by-12 \fi
        \ifnum\hour=0 \hour=12 \fi
        \number\hour:\ifnum\minute<10 0\fi\number\minute\the\amorpm}}
\edef\militarytime{\number\hour:\ifnum\minute<10 0\fi\number\minute}
\def\draftlabel#1{{\@bsphack\if@filesw {\let\thepage\relax
   \xdef\@gtempa{\write\@auxout{\string
      \newlabel{#1}{{\@currentlabel}{\thepage}}}}}\@gtempa
   \if@nobreak \ifvmode\nobreak\fi\fi\fi\@esphack}
        \gdef\@eqnlabel{#1}}
\def\@eqnlabel{}
\def\@vacuum{}
\def\draftmarginnote#1{\marginpar{\raggedright\scriptsize\tt#1}}
\def\draft{\oddsidemargin -.2truein
        \def\@oddfoot{\sl preliminary draft \hfil
        \rm\thepage\hfil\sl\today\quad\militarytime}
        \let\@evenfoot\@oddfoot \overfullrule 3pt
        \let\label=\draftlabel
        \let\marginnote=\draftmarginnote
   \def\@eqnnum{(\theequation)\rlap{\kern\marginparsep\tt\@eqnlabel}%
\global\let\@eqnlabel\@vacuum}  }
\def\thebibliography#1{
\vskip 0.5cm \centerline{\bf \Large References}
\list{
[\arabic{enumi}]}{\settowidth\labelwidth{[#1]}
\leftmargin\labelwidth
\advance\leftmargin\labelsep
\usecounter{enumi}}
\def\newblock{\hskip .11em plus .33em minus .07em}
\sloppy\clubpenalty4000\widowpenalty4000
\sfcode`\.=1000\relax}

\renewcommand{\theequation}{\arabic{section}.\arabic{equation}}
\renewcommand{\section}{\setcounter{equation}{0}\@startsection
{section}{1}{0mm}{-\baselineskip}{0.5\baselineskip} {\normalfont\Large\bfseries}}
\renewcommand{\subsection}{\@startsection
{subsection}{2}{0mm}{-\baselineskip}{0.5\baselineskip} {\normalfont\large\bfseries}}
\renewcommand{\subsubsection}{\@startsection
{subsubsection}{3}{0mm}{-\baselineskip}{0.5\baselineskip}
{\normalfont\normalsize\slshape}}

\begin{document}

%%%%%

\begin{titlepage}
\begin{flushright}
CPHT-RR066.102023, October 2023
\vspace{0.0cm}
\end{flushright}
\begin{centering}
{\bm\bf \Large Closed FRW holography: A time-dependent \\ ER=EPR realization}

\vspace{6mm}

 {\bf Victor Franken,$^1$\footnote{victor.franken@polytechnique.edu} Herv\'e Partouche,$^1$\footnote{herve.partouche@polytechnique.edu} Fran\c cois Rondeau$^2$\footnote{rondeau.francois@ucy.ac.cy} and Nicolaos Toumbas$^2$\footnote{nick@ucy.ac.cy}}

 \vspace{3mm}

$^1$  {\it CPHT, CNRS, Ecole polytechnique, IP Paris, \\F-91128 Palaiseau, France}

$^2$ {\it Department of Physics, University of Cyprus, \\Nicosia 1678, Cyprus}

\end{centering}
\vspace{0.5cm}
$~$\\
\centerline{\bf\large Abstract}\vspace{0.2cm}
We extend a recent de Sitter holographic proposal and entanglement entropy prescription to generic closed FRW cosmologies in arbitrary dimensions, and propose that for large classes of bouncing and Big Bang/Big Crunch cosmologies, the full spacetime can be encoded holographically on two holographic screens, associated to two antipodal observers. In the expanding phase, the two screens lie at the apparent horizons. In the contracting phase, there is an infinite number of possible trajectories of the holographic screens, which can be grouped in equivalence classes. In each class the effective holographic theory can be derived from a pair of ``parent'' screens on the apparent horizons. A number of cases including moduli dominated cosmologies escape our discussion, and it is expected that two antipodal observers and their associated screens do not suffice to reconstruct these cosmologies.
The leading contributions to the entanglement entropy between the screens arise from a minimal extremal trapped or anti-trapped surface lying in the region between them. This picture entails a time-dependent realization of the ER=EPR conjecture, where an effective geometrical bridge connecting the screens via the minimal extremal surface emerges from entanglement. For the Big Crunch contracting cases, the screens disentangle and the geometrical bridge closes off when the minimal extremal trapped sphere hits the Big Crunch singularity at a finite time before the collapse of the Universe. Semiclassical, thermal corrections are incorporated in the cases of radiation dominated cosmologies.

\vspace{-0.6cm}

\begin{quote}

\end{quote}

\end{titlepage}
\newpage
\setcounter{footnote}{0}
\renewcommand{\thefootnote}{\arabic{footnote}}
 \setlength{\baselineskip}{.7cm} \setlength{\parskip}{.2cm}

\setcounter{section}{0}

%%%%%%%%%%%%%%%%%%%%%%%%%%%%%%%%%%%%%%%%
\newpage
{\small \tableofcontents}
\newpage
%%%%%%%%%%%%%%%%%%%%%%%%%%%%%%%%%%%%%%%%%%%%%%

\section{Introduction}
\label{intro}

The holographic principle \cite{tHooft:1993dmi, Susskind:1994vu} has given us powerful tools to study quantum gravitational systems. There are by now robust non-perturbative formulations of string theory and M-theory on certain backgrounds such as Matrix theory \cite{Banks:1996vh} and the Anti-de Sitter space/conformal field theory (AdS/CFT) correspondence \cite{Maldacena:1997re, Gubser:1998bc, Witten:1998qj}. Extending such holographic formulations to more generic cases, one would hope to gain insight into difficult problems of quantum gravity associated with black hole physics, the information loss paradox and the initial singularity in cosmology.

Indeed thus far, most concrete holographic examples involve asymptotically AdS or flat backgrounds characterized by conditions of ``asymptotic coldness'' \cite{Susskind:2007pv,Seiberg:2006wf}, a property that requires the fluctuations of the metric and other bulk fields to tend to zero as the boundary of space is approached. Thanks to this property, asymptotic, weakly coupled gravity regions can be established, where string theoretic tools and symmetries can be applied in order to formulate precise observables and construct the holographic dictionary. Still very little is known about the exact nature of the holographic dual theories describing cosmological backgrounds, where fluctuations persist across the entire region of space and a spatial boundary may be absent. Such cosmological backgrounds include accelerating Universes dominated by dark energy or a positive cosmological constant, closed FRW cosmologies with Big Bang and Big Crunch singularities, as well as our own Universe with early inflationary and high temperature phases. It is thus important to understand how to formulate holography for cosmological backgrounds and develop the holographic dictionary describing the bulk spacetime.

A significant step towards this direction has been the recent understanding of a deep connection between geometry and quantum entanglement \cite{Maldacena:2001kr, Ryu:2006bv,Hubeny:2007xt, VanRaamsdonk:2009ar, VanRaamsdonk:2010pw, Wall:2012uf, Lewkowycz:2013nqa, Maldacena:2013xja, Faulkner:2013ana, Engelhardt:2014gca, Freedman:2016zud, Dong:2016eik, Cotler:2017erl, Headrick:2022nbe, Almheiri:2020cfm}, culminating with the formulation of a well defined prescription for the calculation of the fine grained entropy of gravitational systems \cite{Ryu:2006bv, Hubeny:2007xt, Wall:2012uf, Lewkowycz:2013nqa, Faulkner:2013ana, Engelhardt:2014gca, Headrick:2022nbe, Almheiri:2020cfm}. The calculation involves extremizing a generalized entropy formula with both geometrical area and semiclassical entropy contributions. Starting with \cite{Ryu:2006bv, Hubeny:2007xt}, the prescription has been applied with great success to compute the entanglement entropy of strongly coupled CFT systems, using the AdS/CFT correspondence, and later on the entropy of Hawking radiation produced during the evaporation of a black hole \cite{Almheiri:2020cfm, Penington:2019npb, Almheiri:2019psf, Almheiri:2019hni, Penington:2019kki, Almheiri:2019qdq}. There have been also interesting attempts to apply it in the context of cosmology, see e.g. \cite{Hartman:2020khs, Chen:2020tes, VanRaamsdonk:2020tlr, Balasubramanian:2020coy, Balasubramanian:2020xqf, Manu:2020tty, Kames-King:2021etp, Aguilar-Gutierrez:2021bns, Azarnia:2021uch, Goswami:2021ksw, Bousso:2022gth, Ben-Dayan:2022nmb}. Recently, it has also been applied in the context of static patch holography for de Sitter space \cite{Dyson:2002nt, Dyson:2002pf, Goheer:2002vf, Susskind:2021omt} in order to compute the fine grained entropy of subsystems of the dual holographic theory \cite{Susskind:2021esx, Shaghoulian:2021cef, Shaghoulian:2022fop, Franken:2023pni}.\footnote{See also \cite{Narayan:2015vda, Narayan:2017xca, Arias:2019pzy ,Geng:2021wcq, Sybesma:2020fxg, Aalsma:2021bit, Levine:2022wos, Banihashemi:2022htw,Arenas-Henriquez:2022pyh, Kawamoto:2023nki, Galante:2023uyf, A:2023psv, Susskind:2023rxm, Antonini:2023hdh} for related interesting work.} 

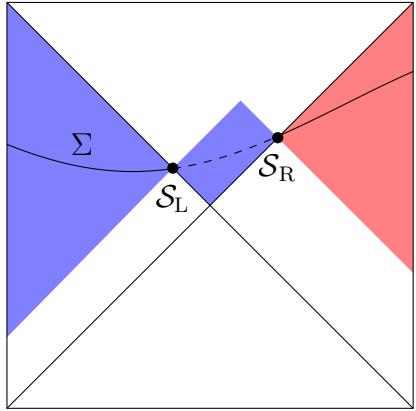
\begin{figure}[!h]

    \centering
\begin{tikzpicture}[scale=0.9]
\begin{scope}[transparency group]
\begin{scope}[blend mode=multiply]
\path
       +(3,3)  coordinate (IItopright)
       +(-3,3) coordinate (IItopleft)
       +(3,-3) coordinate (IIbotright)
       +(-3,-3) coordinate(IIbotleft)
      
       ;
\draw (IItopleft) -- (IItopright) -- (IIbotright) -- (IIbotleft) --(IItopleft) -- cycle;

\node at (-0.55,0.55) [circle, fill, inner sep=1.5 pt, label = below:$\S_\Le$
]{};
\node at (1,1) [circle, fill, inner sep=1.5 pt, label = below:$\S_\Ri$
]{};

\draw[domain=-3:-0.55, smooth, variable=\x] plot ({\x}, {sin(deg((\x/2-1)))+1.5});
\draw[domain=-0.55:1, smooth, variable=\x, black,dashed] plot ({\x}, {sin(deg((\x/2-1)))+1.5});
\draw[domain=1:3, smooth, variable=\x] plot ({\x}, {sin(deg((\x/2-1)))+1.5});

\node at (-1.9,1.4) [label=below:$\Sigma$]{};

%\fill[fill=red!40] (-0.55,0.55) -- (0.45,1.55) -- (1,1) -- (0,0) -- cycle;

\fill[fill=blue!50] (-0.55,0.55) -- (0.45,1.55) -- (1,1) -- (0,0) -- cycle;

\fill [
       fill=blue!50
       ] (-3,3) -- (-0.55,0.55) -- (-3,-1.95) --  cycle;

\fill [
       fill=red!50
       ] (1,1) -- (3,3) -- (3,-1) --  cycle;

\draw (IItopleft) -- (IIbotright)
              (IItopright) -- (IIbotleft) ;

\end{scope}
\end{scope}
\end{tikzpicture}
    \caption{\footnotesize Penrose diagram for the de Sitter spacetime. Spacelike slices have the topology of a sphere and the worldlines of two antipodal observers follow the left and right edges of the diagram. Holographic degrees of freedom lie on the two cosmological horizons depicted by the diagonal lines. The black dots depict the intersections of the cosmological horizons with a bulk Cauchy slice $\rm{\Sigma}$, on which the holographic screens $\S_\Le$ and $\S_\Ri$ are located. The blue rectangle in the exterior region emerges from the entanglement between $\S_\Le$ and $\S_\Ri$. The state on the blue region can be reconstructed from $\S_\Le$, while the state on the red region can be reconstructed from $\S_\Ri$. The union $\S_\Le\cup \S_\Ri$ encodes the state of the bulk domains in blue and red, spanned by complete Cauchy slices. The dashed line indicates an effective geometrical bridge between the two static patches.
    \label{fig:introdS}
    }
\end{figure}

Recall that the static patch holographic proposal for de Sitter space entails anchoring holographic screens on the two cosmological horizons associated with comoving observers at antipodal points of the spatial sphere. The Penrose diagram for de Sitter spacetime is shown in Figure~\ref{fig:introdS}. We can always choose coordinates so that the worldlines of these two observers coincide with the left and right edges of the Penrose diagram. The cosmological horizons, which are depicted by the diagonal lines of the Penrose diagram, bound the two static causal patches of the observers, respectively. More specifically, given a bulk Cauchy slice $\Sigma$, one can locate two holographic screens, denoted ${\cal{S}}_{\rm L}$ and ${\cal{S}}_{\rm R}$, at the intersections of $\rm{\Sigma}$ with the two cosmological horizons, respectively, as shown in Figure~\ref{fig:introdS}. A bulk foliation with Cauchy slices induces an evolution of the screens along the cosmological horizons. Until recently, only very general properties of the holographic dual theory could be established, based on symmetries \cite{Dyson:2002nt, Dyson:2002pf, Goheer:2002vf, Susskind:2021omt, Banks:2000fe, Witten:2001kn, Banks:2002wr, Banks:2003cg} and the holographic covariant entropy bound \cite{Bousso:1999xy, Bousso:2002ju}. 

Two prescriptions for holographic entropy computations have been put forward, the monolayer and the bilayer proposals \cite{Susskind:2021esx, Shaghoulian:2021cef, Shaghoulian:2022fop}, uncovering new interesting properties of the holographic dual system and its subsystems. In the monolayer proposal, the leading geometrical contributions and the quantum corrections to the generalized entropy arise solely from the region between the cosmological horizons, lying in the exterior of the two static patches. In the bilayer proposal, the generalized entropy receives geometrical and quantum contributions from the exterior as well as the two interior regions. In \cite{Franken:2023pni}, we argued that the bilayer prescription is the consistent one and leads to a stronger holographic conjecture, where the entire de Sitter spacetime, the two interior regions and the exterior region, can be reconstructed in terms of the holographic dual on the horizons. In a direct manifestation of the ER=EPR paradigm \cite{Maldacena:2013xja, Geng:2020kxh}, quantum entanglement connects the two static patches via effective bridges that span the exterior causal diamond region.
In particular, the entanglement wedge of the screen with the larger quantum area, which is the screen closer to the bifurcate horizon, extends and covers the exterior causal diamond region \cite{Franken:2023pni}, as shown in Figure \ref{fig:introdS}. 
On the other hand, the entanglement wedge of the screen with the smaller quantum area confines to its interior region. The entanglement wedge of the full two-screen system consists of complete bulk Cauchy slices, which account for the full set of bulk degrees of freedom. The parts of these slices lying in the exterior region have the topology of a barrel and end on the two screens. They behave as effective geometrical bridges between the two static patches manifesting the ER=EPR paradigm -- see Figure \ref{fig:introdS}. 

In parallel to the developments above, the importance of observers in quantum cosmology has been a subject of great interest, leading to important results on the gravity side that may be applicable in the context of holography \cite{Chandrasekaran:2022cip, Witten:2023qsv, Gomez:2023upk, Balasubramanian:2023xyd}. 

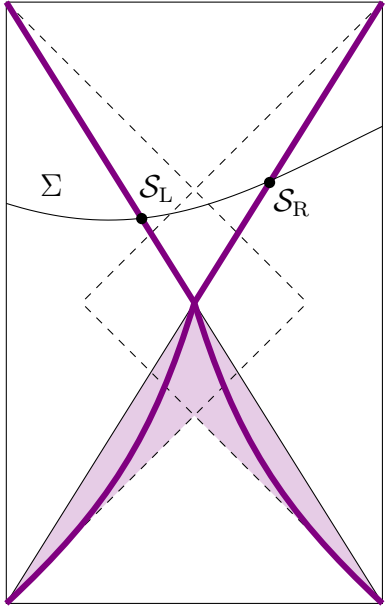
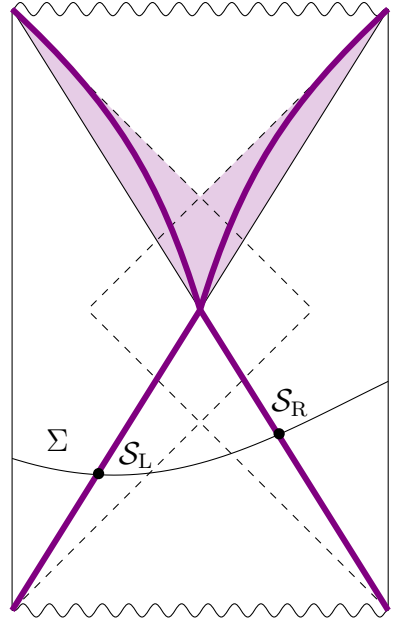
\begin{figure}[h!]
   % \centering
\begin{subfigure}[t]{0.48\linewidth}
\centering
\begin{tikzpicture}

\path
       +(2.5,4)  coordinate (IItopright)
       +(-2.5,4) coordinate (IItopleft)
       +(2.5,-4) coordinate (IIbotright)
       +(-2.5,-4) coordinate(IIbotleft)
      
       ;

\fill[fill=violet!20] (-2.5,-4) -- (0.577,-0.923) -- (0,0) -- cycle; 
\fill[fill=violet!20] (2.5,-4) -- (-0.577,-0.923) -- (0,0) -- cycle;

\draw (IItopleft) --
          node[midway, above, sloped]    {}
      (IItopright) --
          node[midway, above, sloped] {}
      (IIbotright) -- 
          node[midway, below, sloped] {}
      (IIbotleft) --
          node[midway, above , sloped] {}
      (IItopleft) -- cycle;
      
\draw (IItopleft) -- (IIbotright)
              (IItopright) -- (IIbotleft) ;

\draw[dashed] (IItopleft) -- (1.5,0)
                (IItopright) -- (-1.5,0)
                (IIbotright) -- (-1.5,0)
                (IIbotleft) -- (1.5,0);
\draw[violet,line width=0.8mm] (IItopleft) -- (0,0);
\draw[violet,line width=0.8mm] (0,0) to [bend left=15] (-2.5,-4) ;
\draw[violet,line width=0.8mm] (IItopright) -- (0,0);
\draw[violet,line width=0.8mm] (0,0) to [bend right=15] (2.5,-4) ;

\node at (-0.7,1.12) [circle, fill, inner sep=1.5 pt]{};
\node at (-0.5,1.05)[label=above:$\S_\Le$]{};
\node at (1,1.6) [circle, fill, inner sep=1.5 pt]{};
\node at (1.3,1.8)[label=below:$\S_\Ri$]{};

\draw[domain=-2.5:2.5, smooth, variable=\x] plot ({\x}, {sin(deg((\x/2-1)))+2.1});

\node at (-1.9,2) [label=below:$\Sigma$]{};

\end{tikzpicture} 
\caption{\footnotesize Bouncing cosmology.}
\end{subfigure}
\quad \,
\begin{subfigure}[t]{0.48\linewidth}
\centering
\begin{tikzpicture}

\path
       +(2.5,4)  coordinate (IItopright)
       +(-2.5,4) coordinate (IItopleft)
       +(2.5,-4) coordinate (IIbotright)
       +(-2.5,-4) coordinate(IIbotleft)
      
       ;

\fill[fill=violet!20] (-2.5,4) -- (0.577,0.923) -- (0,0) -- cycle; 
\fill[fill=violet!20] (2.5,4) -- (-0.577,0.923) -- (0,0) -- cycle; 
       
\draw[decorate,decoration=snake] (IItopleft) --
          node[midway, above, sloped]    {}
      (IItopright);
      
\draw (IItopright) --
          node[midway, above, sloped] {}
      (IIbotright);
      
\draw[decorate,decoration=snake]  (IIbotright) -- 
          node[midway, below, sloped] {}
      (IIbotleft);
      
\draw (IIbotleft) --
          node[midway, above , sloped] {}
      (IItopleft);
      
\draw (IItopleft) -- (IIbotright)
              (IItopright) -- (IIbotleft) ;

\draw[dashed] (IItopleft) -- (1.5,0)
                (IItopright) -- (-1.5,0)
                (IIbotright) -- (-1.5,0)
                (IIbotleft) -- (1.5,0);
                
\draw[violet,line width=0.8mm] (IIbotleft) -- (0,0);
\draw[violet,line width=0.8mm] (0,0) to [bend right=15] (-2.5,4) ;
\draw[violet,line width=0.8mm] (IIbotright) -- (0,0);
\draw[violet,line width=0.8mm] (0,0) to [bend left=15] (2.5,4) ;

\node at (-1.35,-2.18) [circle, fill, inner sep=1.5 pt]{};
\node at (-0.85,-2.4)[label=above:$\S_\Le$]{};
\node at (1.05,-1.65) [circle, fill, inner sep=1.5 pt]{};
\node at (1.2,-1.7)[label=above:$\S_\Ri$]{};

\draw[domain=-2.5:2.5, smooth, variable=\x] plot ({\x}, {sin(deg((\x/2-1)))-1.2});

\node at (-1.9,-1.3) [label=below:$\Sigma$]{};

\end{tikzpicture}     
\caption{\footnotesize Big Bang/Big Crunch cosmology.}
\end{subfigure}
    \caption{\footnotesize Holographic picture in closed FRW spacetimes. Figure (a) depicts an example of a non-singular bouncing cosmology and Figure (b) depicts an example of a Big Bang/Big Crunch cosmology, with the wavy lines depicting the initial and final singularities. A spacelike Cauchy slice $\rm{\Sigma}$ has the topology of a sphere. We choose coordinates such that the worldlines of two antipodal comoving observers follow the left and right edges of the diagrams. The region causally accessible to each observer, the causal patch, is delimited by a particle and an event horizon depicted by dashed lines. The apparent horizons are the diagonal lines. The holographic degrees of freedom lie on two holographic screens $\S_{\rm L}$ and $\S_{\rm R}$ following the timelike trajectories in purple. In the expanding phase, the trajectories coincide with the apparent horizons, while there is an infinite number of possible timelike trajectories in the contracting phase, spanning the purple regions. These regions are the overlaps of the causal patches of the observers with the region of trapped spheres. \label{fig:introFRW} }
\end{figure}

In this paper, we extend the causal patch holographic proposal and the bilayer holographic entropy prescription to generic closed FRW cosmologies in arbitrary dimensions, which are characterized by a more involved horizon structure.\footnote{See \cite{Fischler:1998st, Hellerman:2001yi, Bak:1999hd, Bousso:1999cb, Diaz:2007mh, Bousso:2015mqa, Bousso:2015qqa, Sanches:2016sxy, Caginalp:2019fyt} for other related attempts.} We consider both bouncing, non-singular cosmologies, for which the perfect fluid index satisfies $-1<w<2/n -1$, where $n$ is the number of spatial dimensions, as well as Big Bang/Big Crunch cosmologies with $2/n-1<w\leq 4/n-1$. Examples of representative Penrose diagrams for both cases are shown in Figure \ref{fig:introFRW}. Given a pair of comoving observers at two antipodal points of the spatial sphere, one can identify particle and event horizons delimiting the corresponding causal patches, and two apparent horizons bounding the regions of trapped and anti-trapped surfaces in between them. Unlike the de Sitter case ($w=-1$) and the Big Bang/Big Crunch case $w=4/n - 1$, the apparent horizons are timelike and lie inside the causal patches of the observers, as shown in Figure \ref{fig:introFRW}. They divide complete bulk Cauchy slices into three parts (except for those passing through the bifurcation point, which are divided into two parts), two interior parts with the topology of spherical caps and a third part in between the apparent horizons with the topology of a barrel. In the expanding phase of the cosmologies, the covariant entropy conjecture \cite{Bousso:1999xy, Bousso:2002ju} motivates us to place a holographic screen on each of the two apparent horizons, respectively -- see also, \cite{Bak:1999hd, Bousso:2015mqa, Sanches:2016sxy}. In the contracting phase of the cosmologies, on the other hand, the screens can be pushed farther into the region between the apparent horizons, which is a region of trapped spheres, as long as they remain in the causal patch of the corresponding observer, respectively. Indeed, as shown in the Penrose diagrams of \Fig{fig:introFRW}, in the contracting phase of the cosmologies, each holographic screen can follow an arbitrary timelike (locally possibly lightlike) trajectory in the purple region, which is the overlap of the region of trapped spheres with the causal patch of the corresponding observer. This trajectory has the same ends with the corresponding apparent horizon in the purple region.
So in contrast to the de Sitter case, there is an infinite number of possible screen trajectories in the contracting phase of the cosmology, leading to different holographic constructions in terms of two-screen quantum mechanical systems. As we will argue, however, these holographic constructions can be grouped in equivalence classes. In each class, the holographic effective theory on the two screens can be derived from the theory on a pair of ``parent'' screens on the apparent horizons. All equivalent screen configurations lead to the same predictions for the fine grained entropy of certain gravitational subsystems, which emerge holographically and are associated to single screen subsystems.  

The two screen trajectories bound the left and right interior regions in the causal patches of the observers, respectively. A third exterior region between the trajectories provides an effective 
%{\blue DISTINGUISH "bridge and exterior, like screens and %horizons"} 
bridge between the single screen subsystems, signaling quantum entanglement, in accordance with the ER=EPR conjecture \cite{Maldacena:2013xja}. Unlike the de Sitter case, the two screens are not always out of causal contact in these more generic cases. Indeed, the screens can exchange energy and information for a certain period of time when they lie in the overlap region of the two causal patches. Therefore, the holographic degrees of freedom on them can interact via time dependent interaction terms. Another crucial difference with the de Sitter case is that the area of the screens does not remain constant as these evolve along their trajectories. This suggests that the evolution of the holographic theory is not unitary, but instead amounts to a sequence of mappings between Hilbert spaces of varying dimensionality -- see \cite{Cotler:2022weg, Cotler:2023xku} for similar discussions. 

\begin{figure}[h!]
	%\centering
	\begin{subfigure}[t]{0.48\linewidth}
		\centering
		\begin{tikzpicture}
			
			\path
			+(2,4)  coordinate (IItopright)
			+(-2,4) coordinate (IItopleft)
			+(2,-4) coordinate (IIbotright)
			+(-2,-4) coordinate(IIbotleft)
			
			;
			
			\fill[fill=red!20] (-0.8,1.6) -- (0.6,3) -- (1.2,2.4) -- (-0.2,1) -- cycle;
			
			\fill[fill=blue!50] (-2,2.8) -- (-0.8,1.6) -- (-2,0.4) --  cycle;
			
			\fill[fill=red!20] (2,3.2) -- (1.2,2.4) -- (2,1.6) --  cycle;
\fill[fill=blue!50] (-3/4,3/2+0.05) -- (-0.225,0.975+0.05) -- (-0.225,0.975-0.05) -- (-3/4,3/2-0.05) -- cycle;

\node at (-0.225,0.975) [circle,fill,inner sep=1.5pt, red, label = right:$M$]{};
			
			\draw (IItopleft) --
          node[midway, above, sloped]    {}
      (IItopright) --
          node[midway, above, sloped] {}
      (IIbotright) -- 
          node[midway, below, sloped] {}
      (IIbotleft) --
          node[midway, above , sloped] {}
      (IItopleft) -- cycle;
			
			\draw (IItopleft) -- (IIbotright)
			(IItopright) -- (IIbotleft) ;
			
			\node at (-0.8,1.6) [circle,fill,inner sep=1.5pt, label = left:$\S_\Le$]{};
			\node at (1.2,2.4) [circle,fill,inner sep=1.5pt, label = right:$\S_\Ri$]{};
			%\node at (0.5,2) [label = right:$\S_2$]{};
			
		\end{tikzpicture}
		\caption{\footnotesize Bouncing cosmology.}
	\end{subfigure}%\hfill
    \quad \,
	\begin{subfigure}[t]{0.48\linewidth}
	\centering
	\begin{tikzpicture}
		
		\path
		+(2,4)  coordinate (IItopright)
		+(-2,4) coordinate (IItopleft)
		+(2,-4) coordinate (IIbotright)
		+(-2,-4) coordinate(IIbotleft)
		
		+(-0.7,2) coordinate (SL)
		+(1.1,2.7) coordinate (SR)
		
		;
		
		\fill[fill=red!20] (SL) -- (0.55,3.25) -- (SR) -- (-0.15,1.45) -- (SL);
		
		\fill[fill=blue!50] (-2,3.3) -- (SL) -- (-2,0.7) --  cycle;
		
		\fill[fill=red!20] (2,3.6) -- (1.1,2.7) -- (2,1.8) --  cycle;

		%\fill[fill=blue!50] (2-0.2,3.2+0.28) -- (1.2-0.2,2.4+0.28) -- (2-0.2,1.6+0.28) --  cycle;
		
		\fill[fill=blue!50] (-0.7,2-0.05) -- (-0.7,2+0.05) -- (0.55,3.25+0.05) -- (0.55,3.25-0.05) -- cycle;
		
		\draw[dashed,gray] (IItopleft) -- (2/3,4/3);
		\draw[dashed,gray] (IItopright) -- (-2/3,4/3);
		
		%\draw[dashed,gray] (IIbotleft) -- (2/3,-4/3);
		%\draw[dashed,gray] (IIbotright) -- (-2/3,-4/3);
		
		\draw[decorate,decoration=snake] (IItopleft) --
		node[midway, above, sloped]    {}
		(IItopright);
		
		\draw (IItopright) --
		node[midway, above, sloped] {}
		(IIbotright);
		
		\draw[decorate,decoration=snake]  (IIbotright) -- 
		node[midway, below, sloped] {}
		(IIbotleft);
		
		\draw (IIbotleft) --
		node[midway, above , sloped] {}
		(IItopleft);
		
		\draw (IItopleft) -- (IIbotright)
		(IItopright) -- (IIbotleft) ;
		
		%\draw (-0.2,3.2+0.25) -- (0,3+0.25) -- (0.2,3.2+0.25);
		
		\node at (SL) [circle,fill,inner sep=1.5pt]{};
		\node at (-0.77,2) [label = above:$\S_\Le$]{};
		\node at (SR) [circle,fill,inner sep=1.5pt]{};
		\node at (1.15,2.7) [label = right:$\S_\Ri$]{};
		\node at (0.55,3.25) [circle,fill,inner sep=1.5pt, red, label = above:$M$]{};
		\node at (0,4) [label = above:$\phantom{M}$]{};
		
	\end{tikzpicture}
	\caption{\footnotesize Big Bang/Big Crunch cosmology.}
\end{subfigure}
	\caption{\footnotesize Examples of applications of the bilayer proposal to bouncing (a) and Big Bang/Big Crunch (b) cosmologies. For a given global Cauchy slice $\rm{\Sigma}$, the holographic screens $\S_\Le$ and $\S_\Ri$ are located at the intersection of $\rm{\Sigma}$ and the purple timelike trajectories of \Fig{fig:introFRW}. In both cases, the fine grained entropy of $\S_\Le\cup\S_\Ri$ vanishes and the entanglement wedge is the union of the three shaded regions, which is spanned by complete Cauchy slices. The minimal extremal surface associated with $\S_\Le$ is depicted by the red dot $M$, corresponding to the trapped or anti-trapped surface with the smallest area in the causal diamond connecting $\S_\Le$ and $\S_\Ri$. The entanglement wedge of $\S_\Le$ is depicted by the blue region. As in de Sitter space, this exterior causal diamond emerges from the entanglement between the $\S_\Le$ and $\S_\Ri$ subsystems, entailing a realization of the ER=EPR correspondence. At late times, $M$ reaches a sphere of finite area in the bouncing cases, while it hits the singularity after a finite time in the singular cases.
	\label{fig:introEW}}
\end{figure}
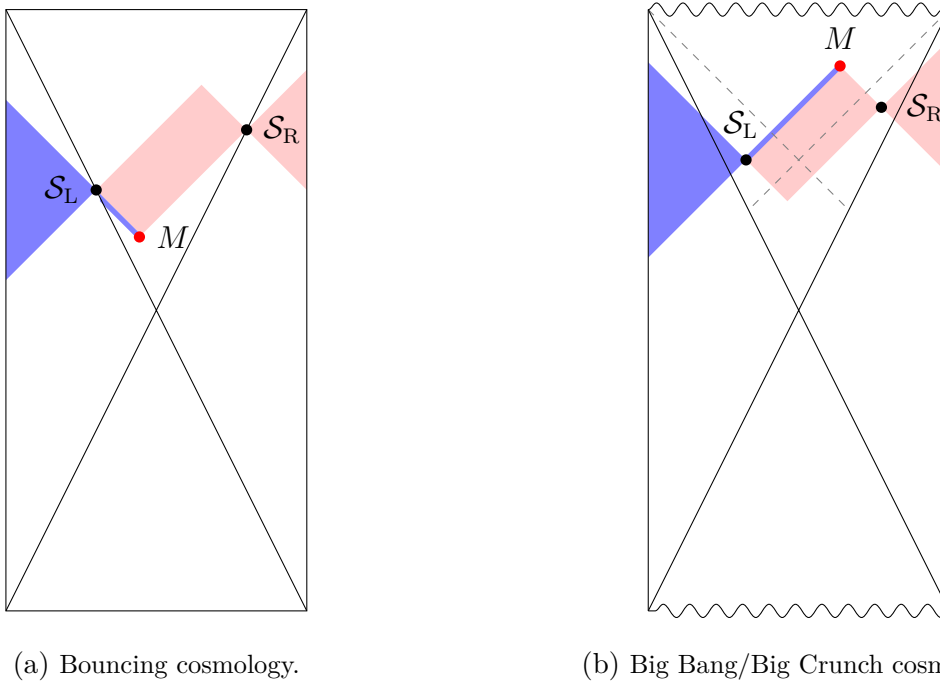

We show how the bilayer prescription for holographic entanglement entropy calculations can be beautifully extended to these general closed FRW cosmologies and adapted to incorporate quantum corrections at the semiclassical level. We then proceed to compute the fine grained entropy of the two-screen system and the single screen subsystems to leading geometrical order, and determine the entanglement wedges. For the two-screen system, we find that the entropy vanishes to leading order, and the entanglement wedge comprises of complete bulk Cauchy slices passing through the screens. Based on this result, we argue that the entropy vanishes to all orders, in the cases where the bulk state on full Cauchy slices is pure. The entropy of the single screen subsystems receives geometrical contributions from the exterior region only. In the contracting (expanding) phase of the cosmology, we find that the minimal extremal homologous surface is the minimal area trapped (anti-trapped) sphere, which lies on the boundary of the causal diamond in the exterior region -- see the red dot $M$ in Figure \ref{fig:introEW}.
This minimal area sphere solves a constrained extremization problem \cite{Franken:2023pni}, where Lagrange multipliers and auxiliary fields are introduced enforcing all extremal surfaces to lie in the exterior causal diamond. Our investigations lead us to conjecture that the entire cosmological spacetime can be holographically encoded on the two screens and reconstructed from the holographic data. Moreover, the computations of the entropies provide a nice time-dependent realization of the ER=EPR conjecture. Quantum entanglement creates an effective geometrical bridge via the minimal trapped or anti-trapped sphere in the exterior causal diamond. Indeed, the entanglement wedges of the single screen subsystems extend in the exterior region via lightlike segments joining the screen with the minimal extremal sphere, as shown in Figure \ref{fig:introEW}. Therefore, the latter is nothing but the intersection of the two entanglement wedges. In the contracting phase of the Big Bang/Big Crunch cosmologies, the minimal extremal sphere hits the Big Crunch singularity at a certain time before the screens reach the Big Crunch singularity. Just at this moment the bridge closes off (since the minimal area sphere collapses to zero size), leading to disentanglement. In the expanding phase of the bouncing cosmologies, the entropy of a single screen saturates an upper bound despite the fact that the areas of the screens grow to infinity in the far future. 

On the other hand for Big Bang/Big Crunch cosmologies with $4/n-1<w\leq 1$, the observers at the antipodal points are separated by particle and/or event horizons and remain out of causal contact for the entire cosmological evolution. We argue that the screens in these cases must be placed on the boundaries of the causal patches, and they never coincide. The two-screen system is always in a mixed state, and the entropies of subsystems do not receive exterior geometrical contributions. Only the interior causal patch regions can be encoded holographically in these cases, on each screen, respectively. Two observers do not suffice to reconstruct the full bulk spacetime. 

We then proceed to study semiclassical and thermal corrections at the 1-loop level, focusing on radiation dominated closed Big Bang/Big Crunch cosmologies. We show how to incorporate the $(G\hbar)^0$ corrections due to the coarse-grained, thermal entropy carried by the gas of particles in the bulk. The first corrected entropy of a single screen subsystem behaves similarly to the leading geometrical entropy. 
In the contracting phase of the cosmology, the first order corrected entropy decreases and eventually gets dominated by the quantum and thermal corrections when the classical extremal surface reaches the Big Crunch singularity.  

Let us stress that in this work we extend and apply the bilayer proposal for holographic entropy computations and study the consequences for the closed FRW cosmological spacetimes under study without providing a detailed proof or derivation. It would be interesting to derive direct evidence for this proposal by studying gravitational replica path integrals in the cosmological context.  

%%%%%%%%%%%%%%%%%%%%%%%%%%%%%%%%%%%%%%%%%%%%%%%%%%%%%%%%%%%%%%%%%%%%%%%%%%%%%%%%%%

The plan of the paper is as follows. In \Sect{basics_FRW}, we review some basic properties of closed FRW cosmologies and their horizon structures. We then proceed with a proposal to extend the static patch holographic proposal of de Sitter space to closed FRW cosmologies in arbitrary dimensions in \Sect{holo_proposal}. We discuss both bouncing and Big Bang/Big Crunch cosmologies for which the Penrose diagrams are taller than wide. In the expanding phase of the cosmologies, we argue that the holographic screens must be placed on the two apparent horizons, in the causal patches of the comoving observers at two antipodal points of the spatial sphere. In the contracting phase of the cosmology, the screens can be pushed farther into the region of trapped spheres between the apparent horizons, leading to different holographic constructions. The latter can be grouped in equivalence classes, with the effective holographic theory on them derived from the theory on a ``parent'' two-screen system at the apparent horizons. We then proceed to generalize the bilayer entropy prescription to these more general cases and specify the extremization problems enforcing all extremal surfaces to lie in the two interior and exterior causal diamond regions. In \Sect{time_dep_EREPR}, we compute the leading geometrical fine grained entropy of the two-screen system and the single screen subsystems. These lead to an interesting interpretation of the behavior of the fine grained entropy of a single screen in terms of a time-dependent version of the ``ER=EPR'' relation. For Big Bang/Big Crunch cosmologies with Penrose diagrams wider than tall, we argue that the screens must be placed on the cosmological horizons in \Sect{BBBC}, and argue that the two-screen system cannot encode the full spacetime in these cases. In \Sect{semicla}, we incorporate semiclassical, thermal corrections in radiation dominated cosmologies, applying our prescription in \Sect{holo_proposal}. Our conclusions are presented in \Sect{conclu}. In Appendix~\ref{FRW_cosmo}, we derive the expression of the scale factor of closed FRW cosmologies in arbitrary dimensions as a function of the conformal time and in Appendix~\ref{thermo}, we review how thermodynamical relations in radiation dominated FRW cosmologies can be derived from an effective action corrected at the 1-loop level. In Appendix~\ref{Lagrange_multipliers}, we specify the area extremization problem in the causal diamond in the exterior region and determine the minimal extremal sphere homologous to a single screen.  

%%%%%%%%%%%%%%%%%%%%%%%%%%%%%%%%%%%%%%%%%%%%%%
%%%%%%%%%%%%%%%%%%%%%%%%%%%%%%%%%%%%%%%%%%%%%%

\section{Basics of closed FRW cosmology}
\label{basics_FRW}

In order to motivate in the next section our proposals for holographic systems dual to closed FRW cosmological universes, we begin our discussion with a brief survey of basic features of closed FRW cosmologies, the associated Penrose diagrams and horizon structures. 

%%%%%%%%%%%%%%%%%%%

\subsection{Cosmological evolutions} 

We consider an $(n+1)$-dimensional closed FRW cosmology, where $n\ge 2$. In conformal gauge, the spacetime metric can be written as 
\begin{equation}
\label{meeta}
    \d s^2= a^2(\eta) \left( -\d\eta^2 + \d\Omega^2_{n}\right)\!,
\end{equation}
where $\eta$ is the conformal time, $a(\eta)$ is the scale factor and $\d\Omega^2_{n}$ is the metric of the unit $n$-dimensional sphere S$^{n}$. In this work, we restrict to an FRW cosmology induced by a single perfect fluid of energy density $\rho$ and pressure $p$, satisfying the state equation
\be
p=w\rho, 
\label{seq}
\ee
where the constant $w\in[-1,1]$ is the perfect fluid index. As reviewed in Appendix~\ref{FRW_cosmo}, the energy density satisfies 
\be
\rho=\frac{C}{a^{n(1+w)}},
\ee
where $C> 0$ is a constant. The evolution of the scale factor depends drastically on $w$, which admits a critical value 
\be
w_{\rm c}=-1+{2\over n}\in(-1,0].
\ee
Throughout this work, we will focus on the generic case where $w\neq w_{\rm c}$, which yields\footnote{The scale factor evolution for $w=w_{\rm c}$ can be found in \Eq{acri}.} 
\be
\label{a(eta)}
a(\eta)=a_0\!\left(\sin{\eta\over |\gamma|}\right)^\gamma, ~~\quad \eta\in\big[0,|\gamma|\pi\big],
\ee
where we have defined
\be
a_0=\left({n(n-1)\over 16\pi C}\right)^{1\over n(w_{\rm c}-w)}, \qquad \gamma={2\over n(w-w_{\rm c})}={2\over n(1+w)-2}.
\label{a0gamma}
\ee
Qualitatively, the cosmological evolution is as follows:

\noindent $\bullet$ When $w$ increases from $-1$ to $w_{\rm c}$, the parameter $\gamma$ decreases from $-1$ to $-\infty$,
 \be
 -1\le w<w_{\rm c} \qquad \Longrightarrow \qquad -1\ge \gamma >-\infty.
 \label{g<-1}
 \ee 
Since $\gamma<0$, the cosmological evolution bounces. The scale factor decreases from an infinite value at $\eta=0$ and reaches its minimum $a_0$ at $\eta=|\gamma|\pi/2$. It then expands and becomes infinite at finite conformal time $\eta=|\gamma|\pi$. The evolution is nowhere singular. The particular case $w=-1$ ($\gamma=-1$) corresponds to a cosmological evolution induced by a positive cosmological constant, \ie a de Sitter spacetime. 

\noindent $\bullet$ When $w$ increases from $w_{\rm c}$ to 1, the parameter $\gamma$ decreases from $+\infty$ to $1/(n-1)$,
 \be
 w_{\rm c}< w\le 1 \qquad \Longrightarrow \qquad +\infty> \gamma \ge{1\over n-1}.
 \ee 
Since $\gamma>0$, the scale factor increases from a Big Bang singularity at $\eta=0$ where it vanishes, and reaches its maximum $a_0$ at $\eta=|\gamma|\pi/2$. It then decreases up to a Big Crunch singularity at $\eta=|\gamma|\pi$, where it vanishes. Semiclassical bulk computations can be trusted for a great part of the cosmological evolution as long as $a_0 >> l_p$. However, near the singularities, the geometrical description breaks down, and so obtaining a holographic dual picture seems necessary to understand the physics. The case $w=1$ ($\gamma=1/(n-1)$) corresponds to a cosmological evolution induced by moduli fields. $w=1/n$ ($\gamma=2/(n-1)$) corresponds to an evolution induced by radiation. Finally, $w=0$ ($\gamma=2/(n-2)$ for $n\ge 3$) corresponds to an evolution induced by massive non-relativistic matter.

%%%%%%%%%%%%%%%%%%%

\subsection{Penrose diagrams} 

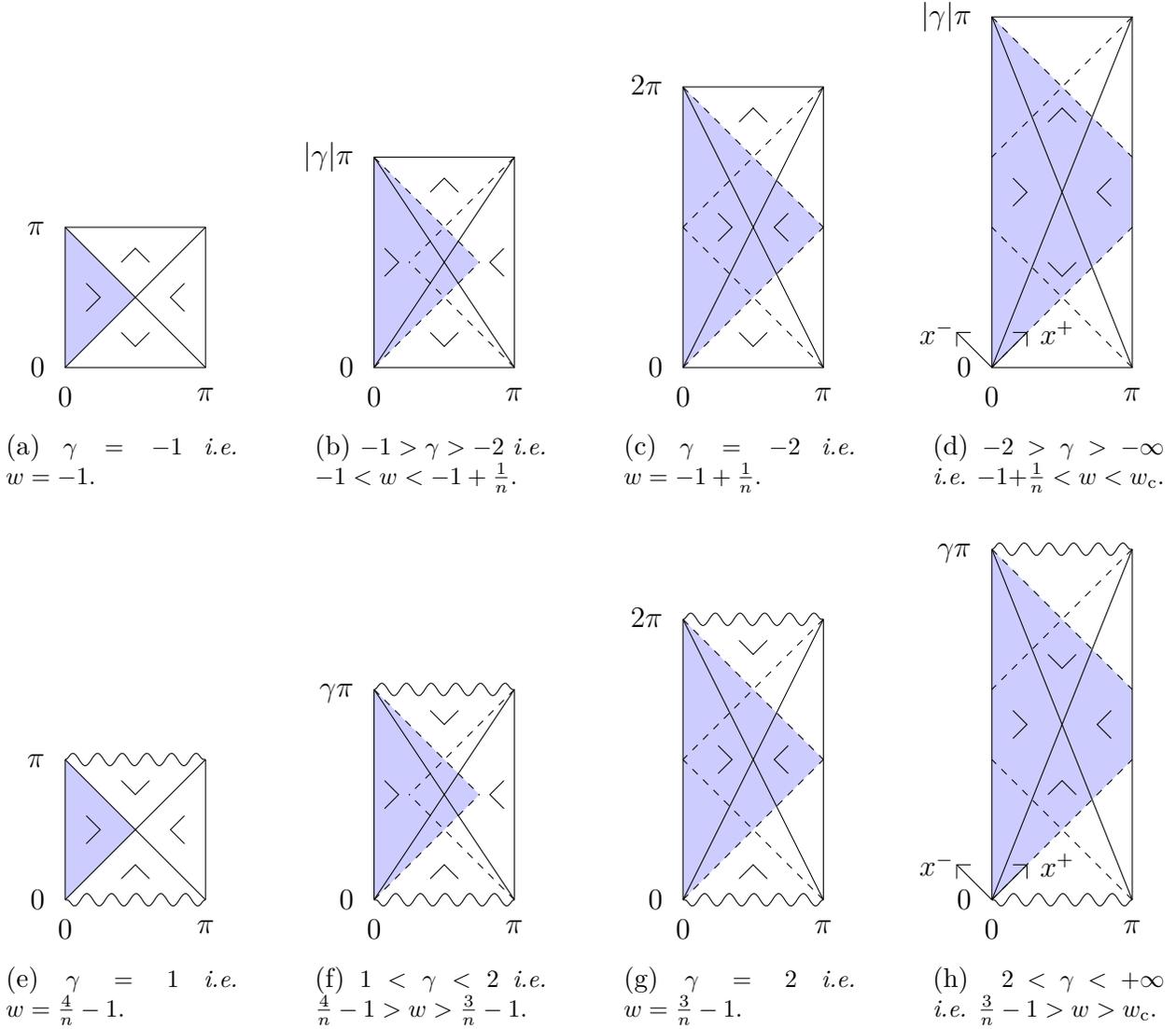
\begin{figure}[h!]
\centering
\begin{subfigure}[b]{0.2\linewidth}\centering
\!\!\!\!\!\!\!\!\!\!\!\!
\begin{tikzpicture}
\path
       +(1,1) coordinate (IItopright)
       +(-1,1) coordinate (IItopleft)
       +(1,-1) coordinate (IIbotright)
       +(-1,-1) coordinate(IIbotleft)
      
       ;
       
       \fill[fill=blue!20] (-1,1) -- (-1,-1) -- (0,0) -- cycle;
       
\draw (IItopleft) --
          node[midway, above, sloped]    {}
      (IItopright) --
          node[midway, above, sloped] {}
      (IIbotright) -- 
          node[midway, below, sloped] {}
      (IIbotleft) --
          node[midway, above , sloped] {}
      (IItopleft) -- cycle;

\draw (IItopleft) -- (IIbotright)
        (IItopright) -- (IIbotleft) ;

\draw (-0.2,1/2) -- (0,1/2+0.2) -- (0.2,1/2);
\draw (-1/2-0.2,-0.2) -- (-1/2,0) -- (-1/2-0.2,0.2);
\draw (1/2+0.2,-0.2) -- (1/2,0) -- (1/2+0.2,0.2);
\draw (-0.2,-1/2) -- (0,-1/2-0.2) -- (0.2,-1/2);

\node at (-1,1) [label = left:$\phantom{2|\gamma|}\pi$]{};
\node at (-1,-1) [label = left:$0$]{};

\node at (-1,-1) [label = below:$0$]{};
\node at (1,-1) [label = below:$\pi$]{};
\end{tikzpicture}
    \caption{\footnotesize$\gamma=-1$ \ie \mbox{$w=-1$}. \label{Pena}}
\end{subfigure}\hfill
\begin{subfigure}[b]{0.2\linewidth}\centering
\!\!\!\!\!\!\!\!\!\!\!\!
\begin{tikzpicture}
\path
       +(1,3/2) coordinate (IItopright)
       +(-1,3/2) coordinate (IItopleft)
       +(1,-3/2) coordinate (IIbotright)
       +(-1,-3/2) coordinate(IIbotleft)
      
       ;
       
       \fill[fill=blue!20] (-1,3/2) -- (1/2,0) -- (-1,-3/2) -- cycle;
       
\draw (IItopleft) --
          node[midway, above, sloped]    {}
      (IItopright) --
          node[midway, above, sloped] {}
      (IIbotright) -- 
          node[midway, below, sloped] {}
      (IIbotleft) --
          node[midway, above , sloped] {}
      (IItopleft) -- cycle;

\draw (IItopleft) -- (IIbotright)
        (IItopright) -- (IIbotleft) ;
\draw[dashed] (IItopleft) -- (1/2,0)
                (IItopright) -- (-1/2,0)
                (IIbotright) -- (-1/2,0)
                (IIbotleft) -- (1/2,0);

\draw (-0.2,1/2+1/2) -- (0,1/2+1/2+0.2) -- (0.2,1/2+1/2);
\draw (-1/2-0.2-0.15,-0.2) -- (-1/2-0.15,0) -- (-1/2-0.2-0.15,0.2);
\draw (1/2+0.2+0.15,-0.2) -- (1/2+0.15,0) -- (1/2+0.2+0.15,0.2);
\draw (-0.2,-1/2-1/2) -- (0,-1/2-1/2-0.2) -- (0.2,-1/2-1/2);

\node at (-1,3/2) [label = left:$\phantom{2}|\gamma|\pi$]{};
\node at (-1,-3/2) [label = left:$0$]{};

\node at (-1,-3/2) [label = below:$0$]{};
\node at (1,-3/2) [label = below:$\pi$]{};
\end{tikzpicture}
    \caption{\footnotesize$-1>\gamma>-2$ \ie $ -1<w<-1+{1\over n}$.\label{Penb}}
\end{subfigure}\hfill
\begin{subfigure}[b]{0.2\linewidth}\centering
\!\!\!\!\!\!\!\!\!\!\!\!
\begin{tikzpicture}
\path
       +(1,2) coordinate (IItopright)
       +(-1,2) coordinate (IItopleft)
       +(1,-2) coordinate (IIbotright)
       +(-1,-2) coordinate(IIbotleft)
      
       ;
       
       \fill[fill=blue!20] (-1,2) -- (-1,-2) -- (1,0) -- cycle;
       
\draw (IItopleft) --
          node[midway, above, sloped]    {}
      (IItopright) --
          node[midway, above, sloped] {}
      (IIbotright) -- 
          node[midway, below, sloped] {}
      (IIbotleft) --
          node[midway, above , sloped] {}
      (IItopleft) -- cycle;

\draw (IItopleft) -- (IIbotright)
        (IItopright) -- (IIbotleft) ;
\draw[dashed] (IItopleft) -- (1,0)
                (IItopright) -- (-1,0)
                (IIbotright) -- (-1,0)
                (IIbotleft) -- (1,0);

\draw (-0.2,1+1/2) -- (0,1+1/2+0.2) -- (0.2,1+1/2);
\draw (-1/2-0.2+0.2,-0.2) -- (-1/2+0.2,0) -- (-1/2-0.2+0.2,0.2);
\draw (1/2+0.2-0.2,-0.2) -- (1/2-0.2,0) -- (1/2+0.2-0.2,0.2);
\draw (-0.2,-1-1/2) -- (0,-1-1/2-0.2) -- (0.2,-1-1/2);

\node at (-1,2) [label = left:$\phantom{|\gamma|}2\pi$]{};
\node at (-1,-2) [label = left:$0$]{};

\node at (-1,-2) [label = below:$0$]{};
\node at (1,-2) [label = below:$\pi$]{};
\end{tikzpicture}
    \caption{\footnotesize$\gamma=-2$ \ie  \mbox{$w=-1+{1\over n}$}.\label{Penc}}
\end{subfigure}\hfill
\begin{subfigure}[b]{0.2\linewidth}\centering
\!\!\!\!\!\!\!\!\!\!\!\!
\begin{tikzpicture}
\path
       +(1,5/2) coordinate (IItopright)
       +(-1,5/2) coordinate (IItopleft)
       +(1,-5/2) coordinate (IIbotright)
       +(-1,-5/2) coordinate(IIbotleft)
      
       ;
       
       \fill[fill=blue!20] (-1,5/2) -- (1,1/2) -- (1,-1/2) -- (-1,-5/2) -- cycle;
       
\draw (IItopleft) --
          node[midway, above, sloped]    {}
      (IItopright) --
          node[midway, above, sloped] {}
      (IIbotright) -- 
          node[midway, below, sloped] {}
      (IIbotleft) --
          node[midway, above , sloped] {}
      (IItopleft) -- cycle;

\draw (IItopleft) -- (IIbotright)
        (IItopright) -- (IIbotleft) ;
\draw[dashed] (IItopleft) -- (1,1/2)
                (IItopright) -- (-1,1/2)
                (IIbotright) -- (-1,-1/2)
                (IIbotleft) -- (1,-1/2);

\draw (-1-0.5,-5/2+0.5) -- (-1,-5/2);
\draw (-1+0.5,-5/2+0.5) -- (-1,-5/2);
\draw (-1-0.5,-5/2+0.3) -- (-1-0.5,-5/2+0.5) -- node[midway, left, sloped] {$x^-$} (-1-0.3,-5/2+0.5) ;
\draw (-1+0.5,-5/2+0.3) -- (-1+0.5,-5/2+0.5) -- (-1+0.3,-5/2+0.5) ;
\node at (-0.6,-5/2+0.5) [label = right:$x^+$]{};

\draw (-0.2,3/2+1/2-1) -- (0,3/2+1/2+0.2-1) -- (0.2,3/2+1/2-1);
\draw (-1/2-0.2,-0.2) -- (-1/2,0) -- (-1/2-0.2,0.2);
\draw (1/2+0.2,-0.2) -- (1/2,0) -- (1/2+0.2,0.2);
\draw (-0.2,-3/2-1/2+1) -- (0,-3/2-1/2-0.2+1) -- (0.2,-3/2-1/2+1);

\node at (-1,5/2) [label = left:$\phantom{2}|\gamma|\pi$]{};
\node at (-1,-5/2) [label = left:$0$]{};

\node at (-1,-5/2) [label = below:$0$]{};
\node at (1,-5/2) [label = below:$\pi$]{};
\end{tikzpicture}
\caption{\footnotesize$-2>\gamma>-\infty$  \ie $-1+{1\over n}<w<w_{\rm c}$.\label{Pend}}
\end{subfigure}\bigskip

\begin{subfigure}[b]{0.2\linewidth}\centering
\!\!\!\!\!\!\!\!\!\!\!\!
\begin{tikzpicture}
\path
       +(1,1) coordinate (IItopright)
       +(-1,1) coordinate (IItopleft)
       +(1,-1) coordinate (IIbotright)
       +(-1,-1) coordinate(IIbotleft)
      
       ;
       
       \fill[fill=blue!20] (-1,1) -- (-1,-1) -- (0,0) -- cycle;
       
\draw[decorate,decoration=snake] (IItopleft) --
          node[midway, above, sloped]    {}
      (IItopright);
      
\draw (IItopright) --
          node[midway, above, sloped] {}
      (IIbotright);
      
\draw[decorate,decoration=snake]  (IIbotright) -- 
          node[midway, below, sloped] {}
      (IIbotleft);
      
\draw (IIbotleft) --
          node[midway, above , sloped] {}
      (IItopleft);

\draw (IItopleft) -- (IIbotright)
        (IItopright) -- (IIbotleft) ;

\draw (-0.2,1/2+0.2) -- (0,1/2) -- (0.2,1/2+0.2);
\draw (-1/2-0.2,-0.2) -- (-1/2,0) -- (-1/2-0.2,0.2);
\draw (1/2+0.2,-0.2) -- (1/2,0) -- (1/2+0.2,0.2);
\draw (-0.2,-1/2-0.2) -- (0,-1/2) -- (0.2,-1/2-0.2);

\node at (-1,1) [label = left:$\phantom{2|\gamma|}\pi$]{};
\node at (-1,-1) [label = left:$0$]{};

\node at (-1,-1) [label = below:$0$]{};
\node at (1,-1) [label = below:$\pi$]{};
\end{tikzpicture}
    \centering \caption{\footnotesize$\gamma=1$ \ie \mbox{$w={4\over n}-1$}.\label{Pene}}
\end{subfigure}\hfill
\begin{subfigure}[b]{0.2\linewidth}\centering
\!\!\!\!\!\!\!\!\!\!\!\!
\begin{tikzpicture}
\path
       +(1,3/2) coordinate (IItopright)
       +(-1,3/2) coordinate (IItopleft)
       +(1,-3/2) coordinate (IIbotright)
       +(-1,-3/2) coordinate(IIbotleft)
      
       ;
       
       \fill[fill=blue!20] (-1,3/2) -- (-1,-3/2) -- (1/2,0) -- cycle;
       
\draw[decorate,decoration=snake] (IItopleft) --
          node[midway, above, sloped]    {}
      (IItopright);
      
\draw (IItopright) --
          node[midway, above, sloped] {}
      (IIbotright);
      
\draw[decorate,decoration=snake]  (IIbotright) -- 
          node[midway, below, sloped] {}
      (IIbotleft);
      
\draw (IIbotleft) --
          node[midway, above , sloped] {}
      (IItopleft);

\draw (IItopleft) -- (IIbotright)
        (IItopright) -- (IIbotleft) ;
\draw[dashed] (IItopleft) -- (1/2,0)
                (IItopright) -- (-1/2,0)
                (IIbotright) -- (-1/2,0)
                (IIbotleft) -- (1/2,0);

\draw (-0.2,1/2+1/2+0.2) -- (0,1/2+1/2) -- (0.2,1/2+1/2+0.2);
\draw (-1/2-0.2-0.15,-0.2) -- (-1/2-0.15,0) -- (-1/2-0.2-0.15,0.2);
\draw (1/2+0.2+0.15,-0.2) -- (1/2+0.15,0) -- (1/2+0.2+0.15,0.2);
\draw (-0.2,-1/2-1/2-0.2) -- (0,-1/2-1/2) -- (0.2,-1/2-1/2-0.2);

\node at (-1,3/2) [label = left:$\phantom{2} \phantom{||}\gamma\pi$]{};
\node at (-1,-3/2) [label = left:$0$]{};

\node at (-1,-3/2) [label = below:$0$]{};
\node at (1,-3/2) [label = below:$\pi$]{};
\end{tikzpicture}
    \caption{\footnotesize$1<\gamma<2$ \ie \mbox{${4\over n}-1>w>{3\over n}-1$}.\label{Penf}}
\end{subfigure}\hfill
\begin{subfigure}[b]{0.2\linewidth}\centering
\!\!\!\!\!\!\!\!\!\!\!\!
\begin{tikzpicture}
\path
       +(1,2) coordinate (IItopright)
       +(-1,2) coordinate (IItopleft)
       +(1,-2) coordinate (IIbotright)
       +(-1,-2) coordinate(IIbotleft)
      
       ;
       
       \fill[fill=blue!20] (-1,2) -- (-1,-2) -- (1,0) -- cycle;
       
\draw[decorate,decoration=snake] (IItopleft) --
          node[midway, above, sloped]    {}
      (IItopright);
      
\draw (IItopright) --
          node[midway, above, sloped] {}
      (IIbotright);
      
\draw[decorate,decoration=snake]  (IIbotright) -- 
          node[midway, below, sloped] {}
      (IIbotleft);
      
\draw (IIbotleft) --
          node[midway, above , sloped] {}
      (IItopleft);

\draw (IItopleft) -- (IIbotright)
        (IItopright) -- (IIbotleft) ;

\draw (-0.2,1+1/2+0.2) -- (0,1+1/2) -- (0.2,1+1/2+0.2);
\draw (-1/2-0.2+0.2,-0.2) -- (-1/2+0.2,0) -- (-1/2-0.2+0.2,0.2);
\draw (1/2+0.2-0.2,-0.2) -- (1/2-0.2,0) -- (1/2+0.2-0.2,0.2);
\draw (-0.2,-1-1/2-0.2) -- (0,-1-1/2) -- (0.2,-1-1/2-0.2);

\draw[dashed] (IItopleft) -- (1,0)
                (IItopright) -- (-1,0)
                (IIbotright) -- (-1,0)
                (IIbotleft) -- (1,0);

\node at (-1,2) [label = left:$\phantom{|\gamma|}2\pi$]{};
\node at (-1,-2) [label = left:$0$]{};

\node at (-1,-2) [label = below:$0$]{};
\node at (1,-2) [label = below:$\pi$]{};
\end{tikzpicture}
    \caption{\footnotesize$\gamma=2$ \ie \mbox{$w={3\over n}-1$}.\label{Peng}}
\end{subfigure}\hfill
\begin{subfigure}[b]{0.2\linewidth}\centering
\!\!\!\!\!\!\!\!\!\!\!\!
\begin{tikzpicture}
\path
       +(1,5/2) coordinate (IItopright)
       +(-1,5/2) coordinate (IItopleft)
       +(1,-5/2) coordinate (IIbotright)
       +(-1,-5/2) coordinate(IIbotleft)
      
       ;
       
       \fill[fill=blue!20] (-1,5/2) -- (1,1/2) -- (1,-1/2) -- (-1,-5/2) -- cycle;
       
\draw[decorate,decoration=snake] (IItopleft) --
          node[midway, above, sloped]    {}
      (IItopright);
      
\draw (IItopright) --
          node[midway, above, sloped] {}
      (IIbotright);
      
\draw[decorate,decoration=snake]  (IIbotright) -- 
          node[midway, below, sloped] {}
      (IIbotleft);
      
\draw (IIbotleft) --
          node[midway, above , sloped] {}
      (IItopleft);
      
\draw (IItopleft) -- (IIbotright)
        (IItopright) -- (IIbotleft) ;

\draw (-1-0.5,-5/2+0.5) -- (-1,-5/2);
\draw (-1+0.5,-5/2+0.5) -- (-1,-5/2);
\draw (-1-0.5,-5/2+0.3) -- (-1-0.5,-5/2+0.5) -- node[midway, left, sloped] {$x^-$} (-1-0.3,-5/2+0.5) ;
\draw (-1+0.5,-5/2+0.3) -- (-1+0.5,-5/2+0.5) -- (-1+0.3,-5/2+0.5) ;
\node at (-0.6,-5/2+0.5) [label = right:$x^+$]{};

\draw (-0.2,3/2+1/2+0.2-1.2) -- (0,3/2+1/2-1.2) -- (0.2,3/2+1/2+0.2-1.2);
\draw (-1/2-0.2,-0.2) -- (-1/2,0) -- (-1/2-0.2,0.2);
\draw (1/2+0.2,-0.2) -- (1/2,0) -- (1/2+0.2,0.2);
\draw (-0.2,-3/2-1/2-0.2+1.2) -- (0,-3/2-1/2+1.2) -- (0.2,-3/2-1/2-0.2+1.2);

\draw[dashed] (IItopleft) -- (1,1/2)
                (IItopright) -- (-1,1/2)
                (IIbotright) -- (-1,-1/2)
                (IIbotleft) -- (1,-1/2);

\node at (-1,5/2) [label = left:$\phantom{2}\phantom{||}\gamma\pi$]{};
\node at (-1,-5/2) [label = left:$0$]{};

\node at (-1,-5/2) [label = below:$0$]{};
\node at (1,-5/2) [label = below:$\pi$]{};
\end{tikzpicture}
    \caption{\footnotesize$\phantom{-}2<\gamma<+\infty$ \ie ${3\over n}-1>w>w_{\rm c}$.\label{Penh}}
\end{subfigure}%\hfill
\caption{\footnotesize Penrose diagrams of closed FRW spacetimes, when $|\gamma|\ge 1$, with the corresponding values of the fluid index~$w\in[-1,w_{\rm c})\cup(w_{\rm c},4/n-1]$. Cases (a) to (d) correspond to bouncing cosmologies, while cases (e) to (h) correspond to Big Bang/Big Crunch cosmologies.  The causal patches of the pode and antipode are delimited by their cosmological horizons, shown in dashed lines. The causal patch of the pode appears in blue. The apparent horizons of the pode and antipode are the diagonal line segments. They divide the Penrose diagrams in four triangular domains in which the Bousso wedges are indicated. The axes of the light-cone coordinates $x^+$, $x^-$ are shown in cases~(d) and~(h). \label{Pen}}
\end{figure}

The spacetime metric~(\ref{meeta}) can be written in the explicitly SO($n$)-symmetric form
\be
\d s^2=a^2(\eta) \left( -\d\eta^2 +\d\theta^2+\sin^2(\theta)\d\Omega_{n-1}^2\right)\!,
\label{so met}
\ee
where $\theta \in [0,\pi]$ is a polar angle and $\d\Omega^2_{n-1}$ is the metric of the unit $(n-1)$-dimensional sphere~S$^{n-1}$. The causal structure of spacetime can be clarified by drawing a Penrose diagram, which is the rectangle parametrized by the coordinates $(\theta,\eta)$. Slices of constant conformal time $\eta$ are spheres S$^n$. Moreover, every generic point $(\theta,\eta)$ of the diagram corresponds to a sphere S$^{n-1}$. In the particular cases where $\theta=0$ or $\theta=\pi$, S$^{n-1}$ reduces to a point, which is the north pole or the south pole of  S$^n$, respectively. Since $\theta$ ranges from $0$ to $\pi$ and $\eta$ ranges from $0$ to $|\gamma|\pi$, the diagrams are taller than wide when $|\gamma|\ge 1$. In the bouncing case, they are shown in \Figs{Pena} to~\ref{Pend}
for various values of $\gamma$ in the range $-1\ge \gamma >-\infty$, \ie $-1\le w<w_{\rm c}$. The diagrams are identical in the Big Bang/Big Crunch case when
$1\le \gamma<+\infty$, \ie $-1+4/n\ge w> w_{\rm c}$, up to wavy lines indicating the singularities in \Figs{Pene} to~\ref{Penh}. When $1/(n-1)\le \gamma<1$, \ie $1\ge w>-1+4/n$, the Penrose diagram is wider than tall, as shown in \Fig{Pen2}.
\begin{figure}[h!]
\centering
\begin{tikzpicture}
\path
       +(5/2,1) coordinate (IItopright)
       +(-5/2,1) coordinate (IItopleft)
       +(5/2,-1) coordinate (IIbotright)
       +(-5/2,-1) coordinate(IIbotleft)
      
       ;
       
       \fill[fill=blue!20] (IItopleft) -- (-3/2,0) -- (IIbotleft) -- cycle;
       
\draw[decorate,decoration=snake] (IItopleft) --
          node[midway, above, sloped]    {}
      (IItopright);
      
\draw (IItopright) --
          node[midway, above, sloped] {}
      (IIbotright);
      
\draw[decorate,decoration=snake]  (IIbotright) -- 
          node[midway, below, sloped] {}
      (IIbotleft);
      
\draw (IIbotleft) --
          node[midway, above , sloped] {}
      (IItopleft);

\draw (IItopleft) -- (IIbotright)
        (IItopright) -- (IIbotleft) ;

\draw (-0.2,1/2+0.2) -- (0,1/2) -- (0.2,1/2+0.2);
\draw (3/2+1/2+0.2-1.2,-0.2) -- (3/2+1/2-1.2,0) -- (3/2+1/2+0.2-1.2,0.2);
\draw (-3/2-1/2-0.2+1.2,-0.2) -- (-3/2-1/2+1.2,0) -- (-3/2-1/2-0.2+1.2,0.2);
\draw (-0.2,-1/2-0.2) -- (0,-1/2) -- (0.2,-1/2-0.2);

\draw[dashed] (IItopleft) -- (-3/2,0) -- (IIbotleft);
%\draw[dashed] (-1/2,1) -- (-3/2,0) -- (-1/2,-1);
\draw[dashed] (IItopright) -- (3/2,0) -- (IIbotright);
%\draw[dashed] (1/2,1) -- (3/2,0) -- (1/2,-1);

\node at (IItopleft) [label = left:$\gamma\pi$]{};
\node at (IIbotleft) [label = left:$0$]{};

\node at (IIbotleft) [label = below:$0$]{};
\node at (IIbotright) [label = below:$\pi$]{};
\end{tikzpicture}
    \caption{\footnotesize Penrose diagram of a closed FRW spacetime, when $1/(n-1)\le \gamma<1$. The cosmological evolution is of Big Bang/Big Crunch type and the fluid index satisfies $1\ge w> -1+{4\over n}$. The causal patches and apparent horizons of the pode and antipode, as well as the Bousso wedges, are shown as explained in \Fig{Pen}.\label{Pen2}}
\end{figure}
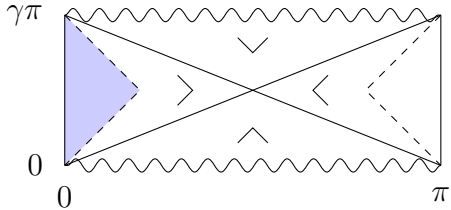
%

%%%%%%%%%%%%%%%%%%%

\subsection{Horizons}

In the following, we will refer to the north and south poles of S$^n$, which are located at $\theta=0$ and $\theta=\pi$, as the pode and the antipode~\cite{Susskind:2021omt}. Let us consider a pair of comoving observers sitting respectively at the pode and the antipode. Their worldlines on the Penrose diagrams are the left and right vertical lines. Each observer has a ``particle horizon,'' which delimits the part of spacetime that he/she will ever be able to send a signal to. He/she also has an ``event horizon,'' which delimits the part of spacetime that he/she will ever be able to receive a signal from. In the Penrose diagrams, these horizons for the observers at the pode and the antipode correspond to segments on the lines:
\be
\begin{tabular} {c|c|c|}\cline{2-3}
& pode &antipode\\ 
\hline
\!\!\!\vline ~~particle horizon & \!\!\!\!\!\!\!\!\!\!\!\!\!\!\!\!\!\!$\eta=\theta$ & \!\!\!\!\!\!\!\!\!\!\!\!\!\!\!\!\!\!$\eta=\pi-\theta$ \\ \hline
\!\!\!\!\!\!\!\!\!\!\:\vline ~~event horizon & $\eta=|\gamma|\pi-\theta$ & $\eta=|\gamma|\pi-\pi+\theta$ \\ \hline
\end{tabular}
\ee
The particle and event horizons of the observers delimit their respective causal patches. Their boundaries, or cosmological horizons, are shown in dashed lines in Figs~\ref{Pen} and~\ref{Pen2}, while the causal patch of the pode is depicted in blue. When $|\gamma|\ge1$, the causal patches of the pode and antipode overlap, while for $1/(n-1)\le\gamma<1$ they are disconnected. 
%In this case, the cosmological horizon of the causal patch of the pode satisfies
%\be
%\left\{\!\begin{array}{ll}
%\dis \eta=\gamma\pi-\theta , &~~\mbox{if}\quad \dis \gamma\, {\pi\over 2}\le \eta \le \gamma\pi , \\
%\dis \eta=\theta , &~~\mbox{if}\quad\dis 0\le \eta \le \gamma\, {\pi\over 2} ,\esps
%\end{array}\right.
%\label{dH1}
%\ee
%while that of the antipode corresponds to 
%\be
%\left\{\!\begin{array}{ll}
%\dis \eta=\theta , &\quad\mbox{if}\quad \dis \gamma\, {\pi\over 2}\le \eta \le \gamma\pi , \\
%\dis \eta=\gamma\pi-\theta , &\quad\mbox{if}\quad \dis 0\le \eta \le \gamma\, {\pi\over 2} .\esps
%\end{array}\right.
%\label{dH2}
%\ee 

As will be seen in the next section, another notion, which is that of ``apparent horizons'' of the observers, enters naturally in the discussion of holographic dual descriptions of bulk cosmological evolutions. Strictly speaking, apparent horizons are the boundaries of the union of all trapped surfaces. However, in the present work, we extend the definition of apparent horizons to also include the boundaries of the union of all anti-trapped surfaces. Trapped (anti-trapped) surfaces are spacelike codimension~2 surfaces whose areas decrease (increase) locally along any future timelike direction. In the case of an FRW cosmology, thanks to the SO($n$) symmetry of the metric~(\ref{so met}), the apparent horizons can be determined by looking for the domains of the Penrose diagram where all points $(\theta,\eta)$ correspond to trapped or anti-trapped spheres~S$^{n-1}$. 

From \Eq{so met}, the area of S$^{n-1}$ located at $(\theta,\eta)$ is given by 
\be
\A(\theta,\eta)=\omega_{n-1}\,a_0^{n-1}\left(\sin{\eta\over |\gamma|}\right)^{\gamma(n-1)}(\sin\theta)^{n-1},
\label{AS}
\ee
where $\omega_{n-1}$ is the area of the unit S$^{n-1}$. 
To discuss how it varies locally along any future timelike direction starting from $(\theta,\eta)$, it is convenient to define light-cone coordinates
\be
x^+=\eta+\theta,\qquad x^-=\eta-\theta.
\label{x+-}
\ee
A sphere S$^{n-1}$ is trapped  if its area decreases when both $x^+$ and~$x^-$ increase. Since  
\be
\label{dAplusminus}
{\partial \A\over \partial x^{\pm}}=\sign(\gamma)\,{n-1\over 2}\,\omega_{n-1}\,a_0^{n-1}\left(\sin{\eta\over |\gamma|}\right)^{\gamma(n-1)-1}(\sin\theta)^{n-2}\,\sin\!\Big(\theta\pm {\eta\over \gamma}\Big),
\ee
this is the case when both inequalities
\be
\sign(\gamma)\sin\!\Big(\theta\pm {\eta\over \gamma}\Big)\le 0
\ee
are satisfied. Similarly, the sphere S$^{n-1}$ is anti-trapped if its area increases when both $x^+$ and $x^-$ increase, which is true when the reversed inequalities are met. The apparent horizons are the sets of points that saturate either of these bounds. As a result, they are the two diagonal lines of the Penrose diagrams,
\be
{\eta\over |\gamma|}=\theta,\qquad {\eta\over |\gamma|}=\pi-\theta.
\label{app}
\ee
The rectangular Penrose diagrams are thus divided in four triangular regions, as shown  in Figures~\ref{Pen} and~\ref{Pen2}. In each triangle, the constant signs of $\partial \A/ \partial x^+$ and $\partial \A/ \partial x^-$ are symbolized by so-called Bousso wedges~\cite{Bousso:1999xy}. The latter are 90$^{\circ}$ wedges, whose sides indicate in which directions along the $x^+$ and $x^-$ axes the area of the S$^{n-1}$ located at the tip decreases. For the bouncing (Big Bang/Big Crunch) cosmologies, the lower triangle is the region of trapped (anti-trapped) surfaces, while the upper triangle is the region of anti-trapped (trapped) surfaces. We define the ``apparent horizon of the pode'' as the boundary of the left triangle,
\be
 \left\{\!\begin{array}{ll}
\dis {\eta\over |\gamma|}=\pi-\theta , &~~\mbox{if}\quad\dis {\pi\over 2}\le {\eta\over |\gamma|} \le \pi , \\
\dis {\eta\over |\gamma|}=\theta , &~~\mbox{if}\quad\dis 0\le {\eta\over |\gamma|} \le {\pi\over 2} ,\esp
\end{array}\right.
\label{a1}
\ee 
and the apparent horizon of the antipode as the boundary of the right triangle,
\be
 \left\{\!\begin{array}{ll}
\dis {\eta\over |\gamma|}=\theta , &\quad\mbox{if}\quad\dis {\pi\over 2}\le {\eta\over |\gamma|} \le \pi , \\
\dis  {\eta\over |\gamma|}=\pi-\theta, &\quad\mbox{if}\quad \dis 0\le {\eta\over |\gamma|} \le {\pi\over 2} .\esp
\end{array}\right.
\label{a2}
\ee 
They are timelike when  $|\gamma|> 1$, spacelike when $1/(n-1)\le \gamma<1$ and lightlike when  $|\gamma|= 1$. In the latter case, they coincide with the cosmological horizons of the pode and antipode. In all cases, we define the bifurcate horizon as the intersection of the apparent horizon of the pode and antipode. It is the S$^{n-1}$ at $(\theta,\eta)=(\pi/2,|\gamma|\pi/2)$.

%%%%%%%%%%%%%%%%%%%%%%%%%%%%%%%%%%%%%%%%%

\section{Holographic proposal for cosmologies with \bm  $|\gamma| \geq 1$}
\label{holo_proposal}

In this section, we motivate our proposals for generalizing the de Sitter static-patch holography conjecture~\cite{Susskind:2021omt} to closed FRW cosmologies, when $|\gamma|\ge 1$. We argue that one possibility is to place a holographic screen on the apparent horizon of the observer at the pode and another screen on the apparent horizon of the antipode. An infinite number of other choices can be considered in the contracting phases of the cosmological evolutions, which may be equivalent. In Section~\ref{BBBC}, we will argue why, in the case where $1/(n-1)\le \gamma < 1$, we do not expect that two observers at the pode and antipode and associated screens suffice to describe holographically the bulk cosmological evolution. 

We will reach our conclusions by considering an arbitrary foliation $\F$ of spacetime, with SO$(n)$-symmetric Cauchy slices.\footnote{
%{\blue Expand and/or comment on that.} 
We focus on SO$(n)$-symmetric slices, exploiting the symmetries of the FRW metric, but more general slices may also be considered. In these cases, the holographic screens on the apparent horizons will not be spherical, but the Bousso bound can still be applied.} $\Sigma$ will denote a generic slice of $\F$. In this section, we will assume that the state on $\Sigma$ is pure, and thus with vanishing entropy. As will be seen in Section~\ref{semicla}, our considerations also apply to the case of a universe filled with radiation, for which the state on $\Sigma$ is taken to be a mixed thermal state, and thus with entropy vanishing at leading order~$(G\hbar)^{-1}$ only. Our proposal, which relies on the Bousso covariant entropy conjecture~\cite{Bousso:1999xy} as well as the bilayer prescription of \Refs{Susskind:2021esx,Shaghoulian:2021cef,
Shaghoulian:2022fop}, will be tested in the forthcoming sections. 

%%%%
\newpage
\subsection{Bousso bound} 
\label{Bb}

\begin{figure}[h!]
   % \centering
\begin{subfigure}[t]{0.48\linewidth}
\centering
\begin{tikzpicture}

\path
       +(2,4)  coordinate (IItopright)
       +(-2,4) coordinate (IItopleft)
       +(2,-4) coordinate (IIbotright)
       +(-2,-4) coordinate(IIbotleft)
      
       ;
       
\draw (IItopleft) --
          node[midway, above, sloped]    {}
      (IItopright) --
          node[midway, above, sloped] {}
      (IIbotright) -- 
          node[midway, below, sloped] {}
      (IIbotleft) --
          node[midway, above , sloped] {}
      (IItopleft) -- cycle;
      
\draw (IItopleft) -- (IIbotright)
              (IItopright) -- (IIbotleft) ;

\draw (-2-0.5,-4+0.5) -- (-2,-4);
\draw (-2+0.5,-4+0.5) -- (-2,-4);
\draw (-2-0.5,-4+0.3) -- (-2-0.5,-4+0.5) -- node[midway, left, sloped] {$x^-$} (-2-0.3,-4+0.5) ;
\draw (-2+0.5,-4+0.3) -- (-2+0.5,-4+0.5) -- (-2+0.3,-4+0.5) ;
\node at (-1.6,-4+0.5) [label = right:$x^+$]{};

\draw (-0.2,2.8) -- (0,3) -- (0.2,2.8);
\draw (-1-0.2,-0.2) -- (-1,0) -- (-1-0.2,0.2);
\draw (1+0.2,-0.2) -- (1,0) -- (1+0.2,0.2);
              
\draw[domain=-2:-0.75, smooth, variable=\x, red,line width=0.8mm] plot ({\x}, {sin(deg((\x/2-1)))+2.5});
\draw[domain=-0.75:2, smooth, variable=\x, red] plot ({\x}, {sin(deg((\x/2-1)))+2.5});

\draw[blue] (-0.75,1.53) -- (-2,2.78) ;

\node at (-0.75,1.53) [circle, fill, inner sep=1.5 pt]{};
\node at (-0.5,1.4) [label=above:$\S_1$]{};

\node at (-1.4,1.6) [label=below:$\red \Sigma_1$]{};
\node at (1.5,2.3) [label=below:$\red \Sigma$]{};

\end{tikzpicture} 
\caption{\footnotesize Case of a bouncing cosmology, $\gamma\le -1$.\label{screenA}}
\end{subfigure}
\quad \,
\begin{subfigure}[t]{0.48\linewidth}
\centering
\begin{tikzpicture}

\path
       +(2,4)  coordinate (IItopright)
       +(-2,4) coordinate (IItopleft)
       +(2,-4) coordinate (IIbotright)
       +(-2,-4) coordinate(IIbotleft)
      
       ;

\draw[decorate,decoration=snake] (IItopleft) --
          node[midway, above, sloped]    {}
      (IItopright);
      
\draw (IItopright) --
          node[midway, above, sloped] {}
      (IIbotright);
      
\draw[decorate,decoration=snake]  (IIbotright) -- 
          node[midway, below, sloped] {}
      (IIbotleft);
      
\draw (IIbotleft) --
          node[midway, above , sloped] {}
      (IItopleft);
      
\draw (IItopleft) -- (IIbotright)
              (IItopright) -- (IIbotleft) ;

\draw (-2-0.5,-4+0.5) -- (-2,-4);
\draw (-2+0.5,-4+0.5) -- (-2,-4);
\draw (-2-0.5,-4+0.3) -- (-2-0.5,-4+0.5) -- node[midway, left, sloped] {$x^-$} (-2-0.3,-4+0.5) ;
\draw (-2+0.5,-4+0.3) -- (-2+0.5,-4+0.5) -- (-2+0.3,-4+0.5) ;
\node at (-1.6,-4+0.5) [label = right:$x^+$]{};

\draw (-0.2,-3.2) -- (0,-3) -- (0.2,-3.2);
\draw (-1-0.2,-0.2) -- (-1,0) -- (-1-0.2,0.2);
\draw (1+0.2,-0.2) -- (1,0) -- (1+0.2,0.2);
              
\draw[domain=-2:-0.85, smooth, variable=\x, red, line width=0.8mm] plot ({\x}, {sin(deg((\x/2-1)))-0.7});
\draw[domain=-0.85:2, smooth, variable=\x, red] plot ({\x}, {sin(deg((\x/2-1)))-0.7});

\draw[blue] (-0.85,-1.7) -- (-2,-0.55) ;

\node at (-0.85,-1.7) [circle, fill, inner sep=1.5 pt]{};

\node at (-0.6,-2.5) [label=above:$\S_1$]{};

\node at (-1.5,-1.6) [label=below:$\red \Sigma_1$]{};
\node at (1.5,-0.9) [label=below:$\red \Sigma$]{};

\end{tikzpicture}     
\caption{\footnotesize Case of a Big Bang/Big Crunch cosmology, \mbox{$\gamma\ge 1$}.\label{screenB}}
\end{subfigure}
    \caption{\footnotesize An SO$(n)$-symmetric Cauchy slice $\Sigma$. The intersection of $\Sigma$ with the apparent horizon of the pode is an S$^{n-1}$ sphere where the screen $\S_1$ is located. $\Sigma_1$ is the part of $\Sigma$ to the left of $\S_1$. The coarse-grained entropy on $\Sigma_1$ is less than or equal to the coarse-grained entropy on the future-directed light-sheet of $\S_1$, which is shown in blue. When the part of $\Sigma$ between the apparent horizons of the pode and antipode lies in the region of anti-trapped surfaces, the screen $\S_1$ must be retained on the apparent horizon of the pode.   
     \label{screens}}
\end{figure}
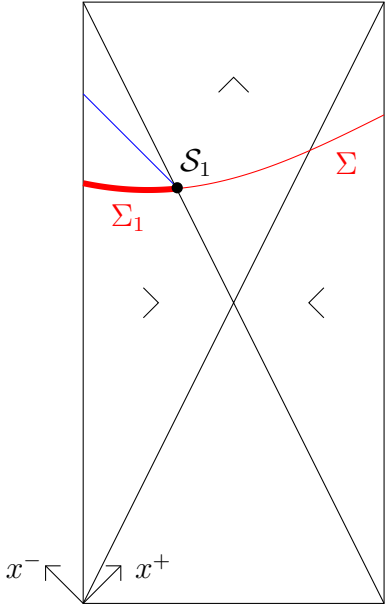
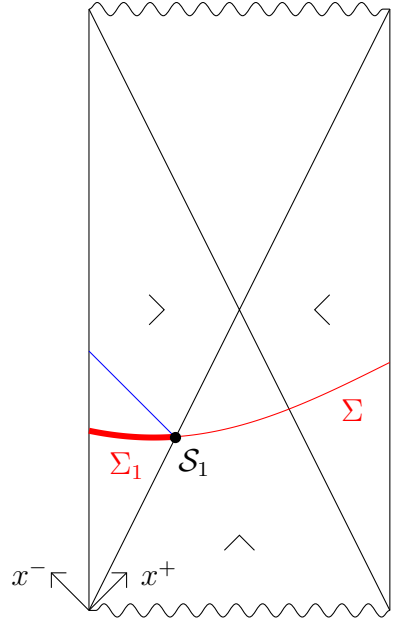

Pick a Cauchy slice $\Sigma$ in $\F$ and consider placing a holographic $(n-1)$-dimensional spherical screen on its part lying in the causal patch of the pode. We will denote this screen $\S_1$. According to the holographic principle, the area of the screen in Planck units is a measure of the number of the holographic degrees of freedom on the screen. In particular, we would like $\S_1$ to be large enough so as to encode the state on $\Sigma_1$, the part of $\Sigma$ located to the left of the screen and lying entirely in the causal patch of the pode. This slice $\Sigma_1$, which has the topology of an $n$-dimensional spherical cap bounded by $\S_1$ and having the pode as its pole, is shown in bold red in the Penrose diagrams of \Fig{screens} (for the particular cases when $\S_1$ is placed at the apparent horizon of the pode). The precise location of $\S_1$ on $\Sigma$ will be determined so as to maximize as much as possible the extent of $\Sigma_1$, which is to be described holographically in accordance with the covariant entropy conjecture~\cite{Bousso:1999xy}.\footnote{See also comments at the end of this subsection.} The screen $\S_1$ must be kept in the causal patch of the observer at the pode in order for the information associated with it to be accessible to this observer. Let us also remark that the bilayer entropy prescription may imply in certain cases that the entanglement wedge of $\S_1$ extends to the right of the screen -- see \Refs{Susskind:2021esx,Shaghoulian:2021cef,
Shaghoulian:2022fop,Franken:2023pni} for the de Sitter case -- thus effectively covering a larger region than $\Sigma_1$. In \Sect{time_dep_EREPR}, we will argue that this is indeed the case for the generic closed FRW cosmologies with $|\gamma|\ge1$.

%$\Sigma_1$ is shown in bold red in the Penrose diagram in \Fig{screens}.
%\sout{Define $\S_1$ the S$^{n-1}$ sphere at the intersection of $\Sigma$ and the apparent horizon of the pode, which is given in \Eq{a1}. We denote $\Sigma_1$ the part of $\Sigma$ located to the left of $\S_1$ and shown in bold red in the Penrose diagram in \Fig{screens}.}
%
Let us initially place the screen $\S_1$  
%Consider the S$^{n-1}$ sphere 
at the intersection of $\Sigma$ with the apparent horizon of the pode, which is given in \Eq{a1}. See \Fig{screens}. Due to SO$(n)$-symmetry, this intersection of $\Sigma$ and the apparent horizon is an S$^{n-1}$ sphere. We would like to show that the holographic degrees of freedom on $\S_1$ suffice so as to encode the state on the left part $\Sigma_1$. 
%(the part of $\Sigma$ located to the left of $\S_1$). 
To this end, we first argue that the coarse-grained entropy on $\Sigma_1$ is expected to be bounded from above by the area of (the sphere where) $\S_1$ (is located) divided by $4G\hbar$. Now consider a codimension~2 spacelike surface in spacetime and define a light-sheet as a codimension~1 surface generated by light-like rays that begin at the surface, extend orthogonally from it and are of non-positive expansion~\cite{Bousso:1999xy}. Applying this definition to $\S_1$, we see from the orientation of the Bousso wedge in the left triangular region of the Penrose diagram that the future-directed light-sheet of $\S_1$ corresponds to the blue line segment. See Figure~\ref{screens}. The latter is parallel to the $x^-$ axis, starts from $\S_1$ and ends on the cosmic line of the pode. In particular, it lies on the future boundary of the causal diamond of $\Sigma_1$. Thanks to the second law of thermodynamics, the coarse-grained entropy on $\Sigma_1$ is bounded from above by the coarse-grained entropy passing through this future-directed light-sheet.  
Then, applying the Bousso covariant-entropy conjecture~\cite{Bousso:1999xy}, this light-sheet being of non-positive expansion, the entropy passing through it is expected to be bounded from above by the area of $\S_1$ divided by $4G\hbar$. Combining the two statements above, we conclude that the coarse-grained entropy on $\Sigma_1$ is bounded from above by the area of $\S_1$ divided by $4G\hbar$. Moreover, as we already remarked, $\Sigma_1$ and $\S_1$ in particular are located in the causal patch of the pode. Hence, in order to describe the state on $\Sigma_1$ in terms of a holographic dual theory, it is natural to place the screen $\S_1$ at the intersection of $\Sigma$ with the apparent horizon of the pode. Notice, though, that the screen may encode more information than that contained on $\Sigma_1$.

Unlike the de Sitter case $\gamma=-1$ and the Big Bang/Big Crunch case $\gamma=1$, when $|\gamma|>1$, the causal patch of the pode extends in all four triangular regions of the Penrose diagram, as seen in \Figs{Penb} to~\ref{Pend} and~\ref{Penf} to~\ref{Penh}. Since a screen associated with the observer at the pode must be retained in the causal patch of the pode, one may wonder whether Bousso's rule allows $\S_1$ to be pushed farther towards the right along $\Sigma$, in order to describe holographically a greater part of it (because then the left part $\Sigma_1$ becomes bigger). Placing $\S_1$ in the right triangular region containing the cosmic line of the antipode does not yield a bound on the coarse-grained entropy on the left part $\Sigma_1$. Specifically, as shown in Figures~\ref{Pen} and \ref{screens}, all future directed light-sheets of negative expansion in this region are parallel to the positive $x^+$ axis. So the future-directed light-sheet ending on $\S_1$ (if this were to be placed in the right triangular region) would lie outside the causal diamond of $\Sigma_1$ and the coarse-grained entropy passing through this light-sheet would not bound the coarse-grained entropy on $\Sigma_1$. Therefore, we need only investigate whether $\S_1$ can be pushed farther and placed on the part of $\Sigma$ that lies between the apparent horizons of the pode and antipode. Now since $\Sigma$ is a spacelike slice, its part located between the apparent horizons lies entirely either in the top or bottom triangular region of the Penrose diagram. It is either in the region of expansion or contraction of the cosmology. If it lies in the expanding phase\footnote{The top triangular region is in the expanding phase for the bouncing cases while the bottom triangular region is in the expanding phase for the Big Bang/Big Crunch cosmologies.}, 
%(the region of anti-trapped surfaces) 
the orientation of the Bousso wedges shown in \Fig{screens} does not allow $\S_1$ to be pushed towards the right and placed on it. In particular in the expanding phase, the region between the apparent horizons of the pode and antipode is a region of anti-trapped surfaces and, therefore, any sphere in it does not have any future-directed light-sheet of negative expansion. Instead, the light-sheets of negative expansion are past-directed, as shown in Figures~\ref{screens} and \ref{Pen}. As a result, the coarse-grained entropy on $\Sigma_1$ is not bounded in general by the entropy passing through the past-directed light-sheet ending on the sphere at its right endpoint. Therefore, if the screen $\S_1$ were to be pushed and placed on the part of $\Sigma$ in the region of anti-trapped surfaces, we would not get a bound for the coarse-grained entropy on $\Sigma_1$ in terms of the area of $\S_1$ in Planck units. 
%preventing the application of Bousso's spacelike projection theorem.
Hence in the expanding phase, we must keep $\S_1$ on the apparent horizon of the pode.

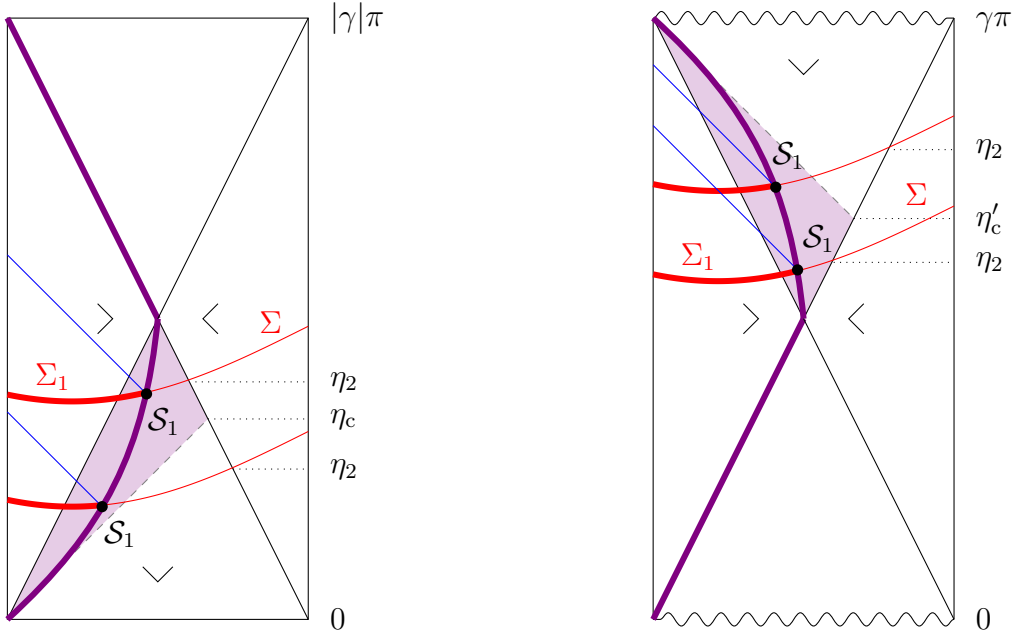
\begin{figure}[h!]
   % \centering
\begin{subfigure}[t]{0.48\linewidth}
\centering
\begin{tikzpicture}

\path
       +(2,4)  coordinate (IItopright)
       +(-2,4) coordinate (IItopleft)
       +(2,-4) coordinate (IIbotright)
       +(-2,-4) coordinate(IIbotleft)
      
       ;

\fill[fill=violet!20] (-2,-4) -- (2/3,-4/3) -- (0,0) -- cycle;       
       
\draw (IItopleft) --
          node[midway, above, sloped]    {}
      (IItopright) --
          node[midway, above, sloped] {}
      (IIbotright) -- 
          node[midway, below, sloped] {}
      (IIbotleft) --
          node[midway, above , sloped] {}
      (IItopleft) -- cycle;
      
\draw[violet,line width=0.8mm] (IItopleft) -- (0,0);
\draw (0,0) -- (IIbotright);
\draw (IItopright) -- (IIbotleft) ;

\draw[dotted] (2/3,-4/3)--(2,-4/3);   
\node at (2,-4/3) [label = right:$\eta_{\rm c}$]{};

\draw[dotted] (1,-2)--(2,-2);  
\node at (2,-2) [label = right:$\eta_2$]{};

\draw[dotted] (0.5,-0.84)--(2,-0.84);  
\node at (2,-0.84) [label = right:$\eta_2$]{};

\draw (-0.2,-2.8-0.5) -- (0,-3-0.5) -- (0.2,-2.8-0.5);
\draw (-1-0.2+0.4,-0.2) -- (-1+0.4,0) -- (-1-0.2+0.4,0.2);
\draw (1+0.2-0.4,-0.2) -- (1-0.4,0) -- (1+0.2-0.4,0.2);

\node at (IItopright) [label = right:$|\gamma|\pi$]{};
\node at (IIbotright) [label = right:$0$]{};
              
\draw[dashed,gray] (IIbotleft) -- (2/3,-4/3);
%\draw[dashed,gray] (IIbotright) -- (-2/3,-4/3);

%\draw[domain=-2:0, smooth, variable=\x, violet,line width=0.8mm] plot ({\x}, {(\x)^2/2+3*\x});

\draw[violet,line width=0.8mm] (-2,-4) to [bend right=21] (0,0) ;

\draw[domain=-2:-0.15, smooth, variable=\x, red,line width=0.8mm] plot ({\x}, {sin(deg((\x/2-1)))-0.1});
\draw[domain=-0.15:2, smooth, variable=\x, red] plot ({\x}, {sin(deg((\x/2-1)))-0.1});

\draw[domain=-2:-0.74, smooth, variable=\x, red,line width=0.8mm] plot ({\x}, {sin(deg((\x/2-1)))-1.5});
\draw[domain=-0.74:2, smooth, variable=\x, red] plot ({\x}, {sin(deg((\x/2-1)))-1.5});

\draw[blue] (-0.74,-2.5) -- (-2,-1.24) ;
\draw[blue] (-0.15,-1) -- (-2,0.85) ;

\node at (-0.74,-2.5) [circle, fill, inner sep=1.5 pt]{};
\node at (-0.5,-2.4) [label=below:$\S_1$]{};

\node at (-0.15,-1) [circle, fill, inner sep=1.5 pt]{};
\node at (0.08,-0.9) [label=below:$\S_1$]{};

\node at (-1.4,-0.3) [label=below:$\red \Sigma_1$]{};
\node at (1.5,0.4) [label=below:$\red \Sigma$]{};

\end{tikzpicture} 
\caption{\footnotesize Case of a bouncing cosmology, $\gamma\le -1$. \label{pushA}}
\end{subfigure}
\quad \,
\begin{subfigure}[t]{0.48\linewidth}
\centering
\begin{tikzpicture}

\path
       +(2,4)  coordinate (IItopright)
       +(-2,4) coordinate (IItopleft)
       +(2,-4) coordinate (IIbotright)
       +(-2,-4) coordinate(IIbotleft)
      
       ;

\fill[fill=violet!20] (-2,4) -- (2/3,4/3) -- (0,0) -- cycle; 
       
\draw[decorate,decoration=snake] (IItopleft) --
          node[midway, above, sloped]    {}
      (IItopright);
      
\draw (IItopright) --
          node[midway, above, sloped] {}
      (IIbotright);
      
\draw[decorate,decoration=snake]  (IIbotright) -- 
          node[midway, below, sloped] {}
      (IIbotleft);
      
\draw (IIbotleft) --
          node[midway, above , sloped] {}
      (IItopleft);
      
\draw[violet,line width=0.8mm] (IIbotleft) -- (0,0);
\draw (0,0) -- (IItopright);
\draw (IItopleft) -- (IIbotright) ;

\draw[dotted] (2/3,4/3)--(2,4/3);   
\node at (2,4/3) [label = right:$\eta'_{\rm c}$]{};

\draw[dotted] (0.4,0.75)--(2,0.75);   
\node at (2,0.75) [label = right:$\eta_2$]{};

\draw[dotted] (1.2,2.25)--(2,2.25);   
\node at (2,2.25) [label = right:$\eta_2$]{};

\draw (-0.2,3.2+0.25) -- (0,3+0.25) -- (0.2,3.2+0.25);
\draw (-1-0.2+0.4,-0.2) -- (-1+0.4,0) -- (-1-0.2+0.4,0.2);
\draw (1+0.2-0.4,-0.2) -- (1-0.4,0) -- (1+0.2-0.4,0.2);

\node at (IItopright) [label = right:$\gamma\pi\phantom{||}$]{};
\node at (IIbotright) [label = right:$0$]{};
              
\draw[dashed,gray] (IItopleft) -- (2/3,4/3);
%\draw[dashed,gray] (IItopright) -- (-2/3,4/3);

%\draw[domain=-2:0, smooth, variable=\x, violet,line width=0.8mm] plot ({\x}, {-(\x)^2/2-3*\x});

\draw[violet,line width=0.8mm] (0,0) to [bend right=21] (-2,4) ;

\draw[domain=-2:-0.37, smooth, variable=\x, red,line width=0.8mm] plot ({\x}, {sin(deg((\x/2-1)))+2.7});
\draw[domain=-0.37:2, smooth, variable=\x, red] plot ({\x}, {sin(deg((\x/2-1)))+2.7});

\draw[domain=-2:-0.08, smooth, variable=\x, red,line width=0.8mm] plot ({\x}, {sin(deg((\x/2-1)))+1.5});
\draw[domain=-0.08:2, smooth, variable=\x, red] plot ({\x}, {sin(deg((\x/2-1)))+1.5});

\draw[blue] (-0.08,0.65) -- (-2,2.57) ;
\draw[blue] (-0.37,1.75) -- (-2,3.38) ;

\node at (-0.08,0.65) [circle, fill, inner sep=1.5 pt]{};
\node at (0.2,0.6) [label=above:$\S_1$]{};

\node at (-0.37,1.75) [circle, fill, inner sep=1.5 pt]{};
\node at (-0.2,1.7) [label=above:$\S_1$]{};

\node at (-1.4,1.3) [label=below:$\red \Sigma_1$]{};
\node at (1.5,2.1) [label=below:$\red \Sigma$]{};

\end{tikzpicture}     
\caption{\footnotesize Case of a Big Bang/Big Crunch cosmology, \mbox{$\gamma\ge 1$}.\label{pushB}}
\end{subfigure}
    \caption{\footnotesize When the part of $\Sigma$ between the apparent horizons of the pode and antipode lies in the region of trapped surfaces, the screen $\S_1$ can follow any timelike (possibly locally lightlike) trajectory in the purple region, with its end points fixed. \label{push}}
\end{figure}

The situation when the part of $\Sigma$ between the apparent horizons lies in the contracting phase (the region of trapped surfaces) is more involved. Indeed, in this case, we can push $\S_1$ farther towards the right and place it on this part of $\Sigma$. In particular, consider the overlap of the causal patch of the pode with the region of trapped spheres -- see the purple region in the Penrose diagrams of Figure~\ref{push} -- and deform the segment of the apparent horizon of the pode in this region to any timelike (possibly locally lightlike) trajectory, while keeping the two endpoints fixed. Such a timelike trajectory of the Penrose diagram represents actually a codimension 1, spherically symmetric, timelike surface of topology $\mathbb{R} \times {\rm S}^{n-1}$. Notice that the timelike trajectories we consider in the purple region are continuous but not necessarily differentiable. Indeed, the apparent horizon of the pode, given in \Eq{a1}, is continuous but not differentiable at the bifurcation point. As we will explain in more detail below, we can place $\S_1$ at the intersection of any such timelike trajectory in the purple region with $\Sigma$. See Figure~\ref{push}. To this end, it turns out to be relevant to define $\eta_2$ as the conformal time at which $\Sigma$ intersects the apparent horizon of the antipode. Let us discuss the 2 types of cosmologies separately:

\noindent $\bullet$ Bouncing case $\gamma<-1$: Define $\eta_{\rm c}$ the conformal time at which the apparent horizon of the antipode intersects the particle horizon of the pode. We have 
\be
\eta_{\rm c}={|\gamma|\over 1+|\gamma|}\, \pi\, ,
\ee
as indicated in \Fig{pushA}.
When $\eta_2\in[\eta_{\rm c},|\gamma|\pi/2]$, the orientation of the Bousso wedge in the bottom triangular region shows that $\S_1$ may be pushed up to any point (actually sphere S$^{n-1}$) of $\Sigma$ between the two apparent horizons.\footnote{However, as we already explained they cannot be pushed farther along $\Sigma$, in the right triangular region, due to the orientation of the Bousso wedge in this domain.} Indeed, the coarse-grained entropy on $\Sigma_{1}$ is still expected to be bounded from above by the area of $\S_1$ divided by $4G\hbar$, while $\Sigma_1$ and $\S_1$ still remain in the causal patch of the pode. Hence, the state on the entire $\Sigma_1$ may be reconstructed from the state of the screen $\S_1$. We arrive at the same conclusions when $\eta_2\in[0,\eta_{\rm c}]$, except that $\S_1$ can be pushed at most up to the particle horizon of the pode, in order for the screen and the whole $\Sigma_1$ to stay in the causal patch of the pode.  In short, $\S_1$ can be placed along $\Sigma$ in the purple region in \Fig{pushA}. Therefore, letting the Cauchy slice $\Sigma$ evolve throughout the foliation $\F$, we have infinitely many choices of trajectories for the screen associated with the observer at the pode. We take these trajectories to be continuous and timelike (locally possibly lightlike). In particular, $\S_1$ first follows any timelike (possibly locally lightlike) curve that starts from the lower left corner of the Penrose diagram in \Fig{pushA}, ends at the bifurcate horizon and lies in the purple region. Subsequently in the expanding phase, it follows the apparent horizon of the pode. Notice that given any other foliation of the whole spacetime with SO$(n)$-symmetric Cauchy slices, each slice intersects once and only once the trajectory of the screen. This follows from the fact that the trajectory is nowhere spacelike.

\noindent $\bullet$ Big Bang/Big Crunch case $\gamma>1$: The analysis is similar in every respect to the previous case. Let $\eta'_{\rm c}$ be the conformal time at which the apparent horizon of the antipode intersects the event horizon of the pode. This definition leads to  
\be
\eta'_{\rm c}={\gamma^2\over 1+|\gamma|}\, \pi=|\gamma|\pi-\eta_{\rm c}, 
\ee
as indicated in the Penrose diagram in \Fig{pushB}. When $\eta_2\in[\gamma\pi/2,\eta'_{\rm c}]$, the Bousso wedge in the upper triangular region allows $\S_1$ to be pushed up to any point (actually sphere S$^{n-1}$) along $\Sigma$ between the two apparent horizons. They cannot be pushed farther, in the right triangular region. For the same reasons as before, the state on $\Sigma_1$ may be described holographically from the state of the screen $\S_1$. When $\eta_2\in[\eta'_{\rm c},\gamma \pi]$, the conclusion remains the same except that $\S_1$ can be pushed at most up to the event horizon of the pode. This has to be the case for $\Sigma_1$ to stay in the causal patch of the pode. Another reason is that the light-sheet of $\S_1$ in the direction of the positive $x^-$ axis would otherwise terminate at the Big Crunch singularity (instead of the cosmic line of the pode), and so part of the coarse-grained entropy on $\Sigma_1$ would end up at the singularity, without passing through the light-sheet. In summary, $\S_1$ can be placed on the part of $\Sigma$ in the purple region in \Fig{pushB}. Again we take the screen trajectory to be timelike (locally possibly lightlike). In particular, when $\Sigma$ varies throughout~$\F$, the screen first follows the apparent horizon of the pode up to the bifurcate horizon in the expanding phase, and then any timelike (possibly locally lightlike) curve in the purple region, up to the upper left corner of the Penrose diagram. Also in this case, the screen trajectory being nowhere spacelike, an SO$(n)$-symmetric Cauchy slice of any other foliation of the whole spacetime will intersect once and only once the trajectory of the screen. %$^{\ref{foot}}$

Everything said so far for the observables accessible from the pode, as well as the holographic screen $\S_1$, can be readily adapted to the case of the observer at the antipode and a second holographic screen $\S_2$. This screen $\S_2$ is initially localized on the S$^{n-1}$ sphere at the intersection of $\Sigma$ and the apparent horizon of the antipode, which is defined in \Eq{a2}. Denote $\Sigma_2$ the part of $\Sigma$ located to the right of $\S_2$ in the Penrose diagram. The state on $\Sigma_2$ may be reconstructed from the state of the screen $\S_2$. Moreover, $\S_2$ can be pushed towards the left along $\Sigma$ only in the region of trapped surfaces, provided it remains in the causal patch of the observer at the antipode. In this way, it may follow in the overlap of the causal patch of the antipode with the region of trapped surfaces any timelike (possibly locally lightlike) curve connecting the bifurcate horizon and the origin (in the bouncing case) or end (in the Big Bang/Big Crunch case) of the cosmic line of the observer at the antipode. 

Notice that the trajectories of $\S_1$ and $\S_2$ intersect at the bifurcate horizon and possibly elsewhere in the region of trapped surfaces. When this occurs on a Cauchy slice $\Sigma$, the union of the coincident screens should describe holographically the entire slice $\Sigma$, since the latter is nothing but the union $\Sigma_1\cup\Sigma_2$. More generally, {\it when $\S_1$ and thus $\S_2$ on $\Sigma$ are in the region of trapped surfaces, two cases can occur:}

\noindent $\bullet$  If $\S_1$ is to the right of $\S_2$ in the Penrose diagram, $\Sigma$ is still the union of $\Sigma_1$ and $\Sigma_2$ (with some overlap) and thus the two-screen system should describe the state on the entire slice~$\Sigma$. Notice that for this to happen, the two screens must be in the overlap region of the causal patches of the pode and antipode.

\noindent $\bullet$  If $\S_1$ is to the left of $\S_2$, the two-screen system may still have the capacity to describe more than the states on $\Sigma_1$ and $\Sigma_2$. To see why, let us apply Bousso's rule to the complement of $\Sigma_1$ in $\Sigma$. From the Bousso wedges in the region of trapped surfaces and the right triangular region in \Fig{push}, we see that the coarse-grained entropy on the complement of $\Sigma_1$ is bounded by the area of $\S_1$ divided by $4G\hbar$. The same remark applies to the complement of $\Sigma_2$ in $\Sigma$ and $\S_2$. Since $\Sigma$ is now equal to the union of these complementary slices (with some overlap), additivity of coarse-grained entropy guaranties that the two-screen system has enough number of degrees of freedom to describe the state on the entire $\Sigma$. This motivates us to consider arbitrary timelike trajectories for the screens in the region of trapped surfaces, as long as they remain in the causal patches of the respective observers and have the same endpoints as the segments of the respective apparent horizons in this region.

As can be seen from the Bousso wedges in \Fig{screens}, the above argument involving the complement of $\Sigma_1$ in $\Sigma$ and the screen on $\S_1$ does not apply when $\S_1$ is in the left triangular region. Hence, when this is the case, in order to describe the state on the greatest part of the Cauchy slice $\Sigma$, the screen should be placed on the apparent horizon of the pode so that $\Sigma_1$ is the largest. The argument involving the complement of $\Sigma_1$ in $\Sigma$ does not apply either if one tries to push $\S_1$ in the region of anti-trapped surfaces.

%We have raised the fact that the Bousso wedge in the region of trapped surface allows $\S_1$ to be pushed from the apparent horizon of the pode towards the interior of the purple region, as shown in \Fig{push}. From the orientation of the Bousso wedge in the left triangular region of the Penrose diagram, one sees that $\S_1$ may also be located in the interior of this region. The reason why we have not considered this possibility is the following. We will present in \Sect{time_dep_EREPR} arguments based on entanglement wedges indicating that the two-screen system does have the capacity of describing the state on any entire Cauchy slice $\Sigma$ of the foliation $\F$. It then turns out that placing the two screens respectively in the left and right triangular regions  yields paradoxical behavior.  Indeed, letting the screens approach the worldlines of the observers at the pode and antipode, the number of degrees of freedom on the two-screen system would become arbitrarily small, while still being enough to describe the state on the whole Cauchy slice $\Sigma$. 

%%%%%%%%%%%%%%%%%%%

\subsection{Equivalent configurations} 
\label{equivconf}

In order to investigate the possibility for various configurations of the screens to be equivalent (in a sense that needs to be specified), let us begin with a few remarks. The observer at the pode and the screen $\S_1$ evolve along their own trajectories, which start and end at the same points in the Penrose diagram in \Fig{push}. As a result, they share the same causal patches shown in \Fig{Pen}. The same applies to the observer at the antipode and the screen $\S_2$. When one of the screens is in the causal patch of the other, they can exchange energy and information, possibly \via time dependent interaction terms. The exchanges can be reciprocal when both screens are in each other's causal patch.\footnote{This is in contrast to the example of the eternal AdS black hole, for which there are no interactions between the two copies of the dual CFT. 
%\blue FOOTNOTE TO BE DISCUSSED: In the particular case of a de Sitter cosmological evolution, $\gamma=-1$, INTERSECT AT BIFURCASE XXXXX\cite{Franken:2023pni}, we argued that the existence of phase transitions regarding the entanglement wedges of single-screen systems at the quantum level implies the existence of non-local interactions between the holographic degrees of freedom of the two screens, which are not obvious from the geometrical picture. Presumably, these interactions concern the degrees of freedom describing the bulk region between the screens.
}

As we remarked in the previous section, during the contracting phase and in particular in the overlap region of the causal patches of the pode and antipode, there can be trajectories for which the antipode screen $\S_2$ lies to the left of the pode screen $\S_1$ for some Cauchy slices $\Sigma$. In the overlap of the causal patches, the screens can exchange information and both observers at the pode and antipode have access to both screens. We will argue below that in such cases the roles of the screens can be swapped with $\S_2$ serving as the pode screen and $\S_1$ as the antipode screen.  

So, on each slice $\Sigma$ of $\F$, let us denote $\S_\Le$ the leftmost screen among $\S_1$ and $\S_2$. We also define $\Sigma_\Le$ the part of $\Sigma$ on the left of $\S_\Le$. Likewise, $\S_\Ri$ is the rightmost screen among $\S_1$ and $\S_2$, while $\Sigma_\Ri$ is the part of $\Sigma$ on the right of $\S_\Ri$. We claim that the states on $\Sigma_\Le$ and $\Sigma_\Ri$ may always be reconstructed from the states of the screens on $\S_\Le$ and $\S_\Ri$, respectively. This is a trivial statement when $\S_1$ is to the left of $\S_2$. Moreover, when $\S_1$ is to the right of $\S_2$, $\S_2$ is located on $\Sigma_1$, implying the second screen to be in the causal patch of the pode, and thus in the causal patch of the first screen. In fact, both screens are in the causal patch of the other. Since they are in causal contact, they can exchange information, enabling $\S_2$ to serve as an alternative screen, $\tilde{\cal S}_1$, associated with the observer at the pode. In this case, the screen $\tilde{\cal S}_1$ is expected to contain the information necessary to reconstruct the state on $\tilde\Sigma_1$, the complement of $\Sigma_2$ in $\Sigma$. We thus have $\S_\Le=\S_2$, while $\Sigma_\Le$ is the complement of $\Sigma_2$ in $\Sigma$. Similarly, $\S_\Ri=\S_1$ and $\Sigma_\Ri$ is the complement of $\Sigma_1$ in $\Sigma$. Another way of arguing that the roles of the screens can be swapped is as follows. If $\S_2$ is to the left of $\S_1$ on some slices in the overlap of the causal patches, we must have $\S_1=\S_2$ on an earlier Cauchy slice of the foliation. On this earlier slice, the degrees of freedom of the coincident screens can be exchanged with each other. Hence, it is permitted to impose that, subsequently, the screen associated with the pode is the leftmost, while that associated with the antipode is the rightmost. In the following, we will adopt this convention.

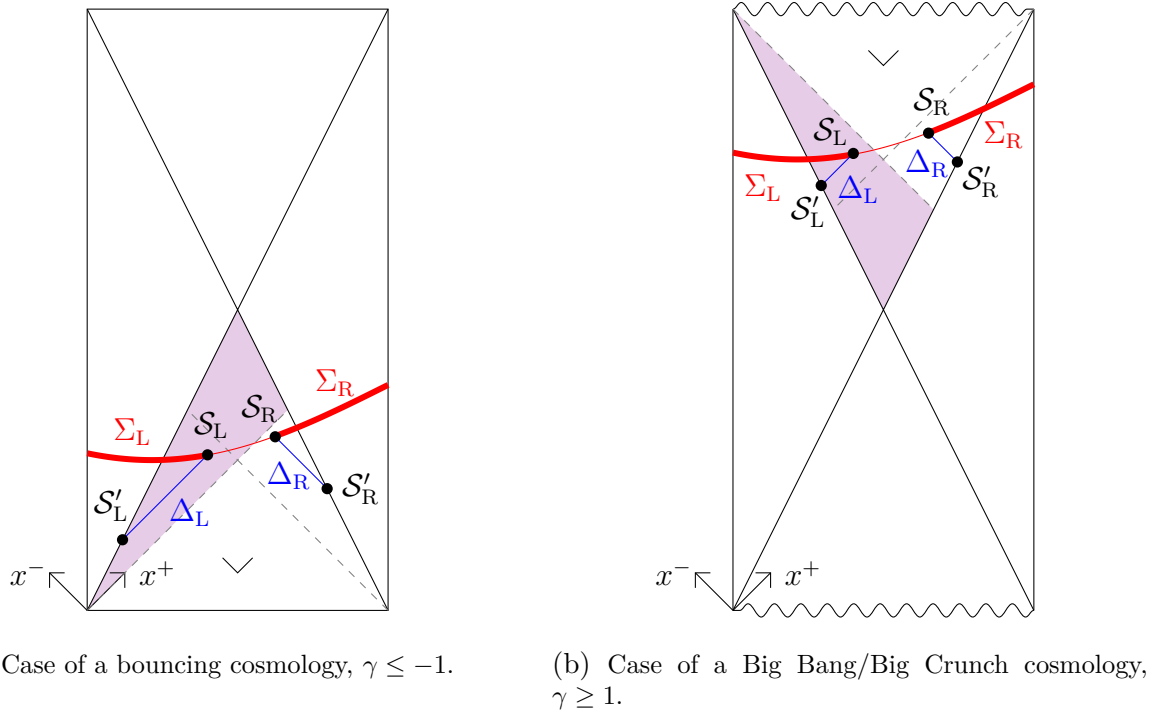
\begin{figure}[h!]
   % \centering
\begin{subfigure}[t]{0.48\linewidth}
\centering
\begin{tikzpicture}

\path
       +(2,4)  coordinate (IItopright)
       +(-2,4) coordinate (IItopleft)
       +(2,-4) coordinate (IIbotright)
       +(-2,-4) coordinate(IIbotleft)
      
       ;

%\begin{scope}[transparency group]
\begin{scope}[blend mode=multiply]
\fill[fill=violet!20] (-2,-4) -- (2/3,-4/3) -- (0,0) -- cycle;
%\fill[fill=orange!20] (2,-4) -- (-2/3,-4/3) -- (0,0) -- cycle;
\end{scope}
%\end{scope}
       
\draw (IItopleft) --
          node[midway, above, sloped]    {}
      (IItopright) --
          node[midway, above, sloped] {}
      (IIbotright) -- 
          node[midway, below, sloped] {}
      (IIbotleft) --
          node[midway, above , sloped] {}
      (IItopleft) -- cycle;
      
\draw (IItopleft) -- (IIbotright)
              (IItopright) -- (IIbotleft) ;

\draw (-2-0.5,-4+0.5) -- (-2,-4);
\draw (-2+0.5,-4+0.5) -- (-2,-4);
\draw (-2-0.5,-4+0.3) -- (-2-0.5,-4+0.5) -- node[midway, left, sloped] {$x^-$} (-2-0.3,-4+0.5) ;
\draw (-2+0.5,-4+0.3) -- (-2+0.5,-4+0.5) -- (-2+0.3,-4+0.5) ;
\node at (-1.6,-4+0.5) [label = right:$x^+$]{};

\draw (-0.2,-2.8-0.5) -- (0,-3-0.5) -- (0.2,-2.8-0.5);
              
\draw[dashed,gray] (IIbotleft) -- (2/3,-4/3);
\draw[dashed,gray] (IIbotright) -- (-2/3,-4/3);

\draw[domain=-2:-0.4, smooth, variable=\x, red, line width=0.8mm] plot ({\x}, {sin(deg((\x/2-1)))-1});
\draw[domain=-0.4:0.5, smooth, variable=\x, red] plot ({\x}, {sin(deg((\x/2-1)))-1});
\draw[domain=0.5:2, smooth, variable=\x, red, line width=0.8mm] plot ({\x}, {sin(deg((\x/2-1)))-1});

\draw[blue] (-0.4,-1.93) -- (-1.53,-3.06);
\node at (-0.4,-1.93) [circle, fill, inner sep=1.5 pt]{};
\node at (-0.35,-2) [label=above:$\S_\Le$]{};
\node at (-1.53,-3.06) [circle, fill, inner sep=1.5 pt]{};
\node at (-2.2,-2.6) [label=right:$\S_\Le'$]{};
\node at (-1.2,-2.7) [label=right:$\blue \Delta_\Le$]{};

\draw[blue] (0.5,-1.69) -- (1.19,-2.38);
\node at (0.5,-1.69) [circle, fill, inner sep=1.5 pt]{};
\node at (0.3,-1.78) [label=above:$\S_\Ri$]{};
\node at (1.19,-2.38) [circle, fill, inner sep=1.5 pt]{};
\node at (1.1,-2.38) [label=right:$\S_\Ri'$]{};
\node at (0.1,-2.2) [label=right:$\blue \Delta_\Ri$]{};

\node at (-1.4,-1.15) [label=below:$\red \Sigma_\Le$]{};
\node at (1.3,-0.5) [label=below:$\red \Sigma_\Ri$]{};

\end{tikzpicture} 
\caption{\footnotesize Case of a bouncing cosmology, $\gamma\le -1$.\label{puschbackA}}
\end{subfigure}
\quad \,
\begin{subfigure}[t]{0.48\linewidth}
\centering
\begin{tikzpicture}

\path
       +(2,4)  coordinate (IItopright)
       +(-2,4) coordinate (IItopleft)
       +(2,-4) coordinate (IIbotright)
       +(-2,-4) coordinate(IIbotleft)
      
       ;

%\begin{scope}[transparency group]
\begin{scope}[blend mode=multiply]
\fill[fill=violet!20] (-2,4) -- (2/3,4/3) -- (0,0) -- cycle;
%\fill[fill=orange!20] (2,4) -- (-2/3,4/3) -- (0,0) -- cycle;
\end{scope}
%\end{scope}
       
\draw[decorate,decoration=snake] (IItopleft) --
          node[midway, above, sloped]    {}
      (IItopright);
      
\draw (IItopright) --
          node[midway, above, sloped] {}
      (IIbotright);
      
\draw[decorate,decoration=snake]  (IIbotright) -- 
          node[midway, below, sloped] {}
      (IIbotleft);
      
\draw (IIbotleft) --
          node[midway, above , sloped] {}
      (IItopleft);
      
\draw (IItopleft) -- (IIbotright)
              (IItopright) -- (IIbotleft) ;

\draw (-2-0.5,-4+0.5) -- (-2,-4);
\draw (-2+0.5,-4+0.5) -- (-2,-4);
\draw (-2-0.5,-4+0.3) -- (-2-0.5,-4+0.5) -- node[midway, left, sloped] {$x^-$} (-2-0.3,-4+0.5) ;
\draw (-2+0.5,-4+0.3) -- (-2+0.5,-4+0.5) -- (-2+0.3,-4+0.5) ;
\node at (-1.6,-4+0.5) [label = right:$x^+$]{};

\draw (-0.2,3.2+0.25) -- (0,3+0.25) -- (0.2,3.2+0.25);
              
\draw[dashed,gray] (IItopleft) -- (2/3,4/3);
\draw[dashed,gray] (IItopright) -- (-2/3,4/3);

\draw[domain=-2:-0.4, smooth, variable=\x, red, line width=0.8mm] plot ({\x}, {sin(deg((\x/2-1)))+3});
\draw[domain=-0.4:0.6, smooth, variable=\x, red] plot ({\x}, {sin(deg((\x/2-1)))+3});
\draw[domain=0.6:2, smooth, variable=\x, red, line width=0.8mm] plot ({\x}, {sin(deg((\x/2-1)))+3});

\draw[blue] (-0.4,2.08) -- (-2.48/3,2/3*2.48);
\node at (-0.4,2.08) [circle, fill, inner sep=1.5 pt]{};
\node at (-0.7,1.9) [label=above:$\S_\Le$]{};
\node at (-2.48/3,2/3*2.48) [circle, fill, inner sep=1.5 pt]{};
\node at (-1,1.8) [label=below:$\S_\Le'$]{};
\node at (-0.9,1.6) [label=right:$\blue \Delta_\Le$]{};

\draw[blue] (0.6,2.35) -- (2.95/3,2/3*2.95);
\node at (0.6,2.35) [circle, fill, inner sep=1.5 pt]{};
\node at (0.65,2.3) [label=above:$\S_\Ri$]{};
\node at (2.95/3,2/3*2.95) [circle, fill, inner sep=1.5 pt]{};
\node at (1.3,2.2) [label=below:$\S_\Ri'$]{};
\node at (0,1.95) [label=right:$\blue \Delta_\Ri$]{};

\node at (-1.6,2.1) [label=below:$\red \Sigma_\Le$]{};
\node at (1.6,2.8) [label=below:$\red \Sigma_\Ri$]{};

\end{tikzpicture}     
\caption{\footnotesize Case of a Big Bang/Big Crunch cosmology, \mbox{$\gamma\ge 1$}.\label{puschbackB}}
\end{subfigure}
    \caption{\footnotesize  The screens $\S_\Le$ and $\S_\Ri$ can be moved along the lightlike slices $\Delta_\Le$ and $\Delta_\Ri$, which are parallel to the $x^+$ and $x^-$ axes in the region of trapped surfaces. The ``parent screens'' $\S'_\Le$, $\S'_\Ri$ are located on the spheres of greatest areas along $\Delta_\Le$, $\Delta_\Ri$, respectively, which lie on the apparent horizons.\label{pushback}}
   
\end{figure}

At this stage, we have infinitely many possible choices of foliations $\F$ and trajectories for $\S_\Le$ and $\S_\Ri$. Anticipating with results presented and explained in more detail in \Sect{equivconf2}, we now argue that these configurations can be grouped in equivalence classes. In each class, the effective holographic theory on the screens may be derived from a ``parent'' setup, where both screens evolve along the apparent horizons, with the slicing of spacetime corresponding to a new foliation. To arrive to these conclusions, consider a Cauchy slice $\Sigma$ of $\F$, where $\S_\Le$ is in the purple region in \Fig{pushback}.
Define $\Delta_\Le$ the blue line segment that runs parallel to the $x^+$ axis from $\S_\Le$ to the apparent horizon of the pode. Likewise, $\Delta_\Ri$ is parallel to the $x^-$ axis from $\S_\Ri$ to the apparent horizon of the antipode. All ``points'' of $\Delta_\Le$ and $\Delta_\Ri$ are spheres S$^{n-1}$, with different areas. Let us displace the screens $\S_\Le$ and $\S_\Ri$ to any spheres of $\Delta_\Le$ and $\Delta_\Ri$, respectively. Notice that the region between these two spheres and lying on any Cauchy slice of spacetime has the topology of a barrel. In the spirit of the ER=EPR relation~\cite{Maldacena:2013xja}, we expect that 
the existence of this region is a manifestation of entanglement between the two boundary screens in the holographic picture. Moreover, we will see in \Sect{equivconf2} that applying the bilayer prescription presented below, one always obtains the same result for the fine-grained entanglement entropy between the two screens, irrespective of their locations on $\Delta_\Le$ and $\Delta_\Ri$. In addition, we will see in Section~\ref{equivconf2} that each such choice leads to the same predictions concerning the fine-grained entropy of certain gravitational bulk slices, which may be reconstructed from the holographic theory on the screens. However, from the Bousso wedge in the region of trapped surfaces, we see that the spheres on $\Delta_\Le$ and $\Delta_\Ri$ having the greatest areas, on which we can place alternative screens $\S_\Le'$ and $\S_\Ri'$ respectively, are at the boundaries of $\Delta_\Le$ and $\Delta_\Ri$ on the apparent horizons. As a result, the screens $\S_\Le'$ and $\S_\Ri'$ must have more degrees of freedom than their counterparts $\S_\Le$ and $\S_\Ri$. This suggests that the effective holographic theory on $\S_\Le\cup \S_\Ri$ can be obtained by integrating out some degrees of freedom of the ``parent'' system $\S_\Le'\cup\S_\Ri'$. This is reminiscent of an RG flow in local quantum field theories.\footnote{However, as in the de Sitter case $\gamma=-1$, we do not expect the holographic dual theory to these closed FRW cosmologies to be local \cite{Shaghoulian:2021cef, Shaghoulian:2022fop, Franken:2023pni}. The leading entanglement entropies computed from a bilayer-like prescription do not always obey an area law as they should in a local quantum field theory. In particular, computing holographic entropies of subsystems of the horizons to leading order produces a volume-law for the growth of fine-grained entropy.} Since the construction of $\S_\Le'$ and $\S_\Ri'$ guarantees that they are spacelike separated, we can choose a Cauchy slice $\Sigma'$ of spacetime that passes through them. Repeating all steps for every slice $\Sigma$ belonging to $\F$, we generate a new foliation $\F'$ of slices $\Sigma'$ intersecting the apparent horizons of the pode and antipode at $\S_\Le'$ and $\S_\Ri'$, respectively. It is always possible to choose $\F'$ so that $\Sigma'=\Sigma$ for every slice $\Sigma$ where $\S_\Le$ and $\S_\Ri$ are on the apparent horizons, \ie $\S_\Le=\S_\Le'$ and $\S_\Ri=\S_\Ri'$. This is in particular the case in the expanding phase of the cosmological evolution. However, in general, the other slices $\Sigma'$ of the foliation $\F'$ do not belong to $\F$. 

The state of the two-screen system $\S_\Le\cup \S_\Ri$ can be found by evolving the screens along their arbitrary trajectories, slice after slice in $\F$. Alternatively, we may evolve the parent screens $\S_\Le'$ and $\S_\Ri'$ along the apparent horizons, slice after slice in $\F'$, up to a certain Cauchy slice in $\F'$, and then along $\Delta_\Le$ and $\Delta_\Ri$ up to $\S_\Le$ and $\S_\Ri$, while maintaining them spacelike separated.\footnote{In this second stage, the screens do not lie in general on Cauchy slices of $\F$ or $\F'$.} This procedure is independent of the choice of the timelike trajectory we chose for the screens. Note that it requires to study the evolution along a trajectory that is non-differentiable, but continuous, at the intersection of $\Delta_\Le$ and $\Delta_\Ri$ with the apparent horizons of the pode and antipode, respectively. The final states of the two-screen system in these two cases should be unitarily equivalent, with the same entanglement entropies between the two screens.

Let us focus on a parent configuration. For each slice $\Sigma'$ of the foliation $\F'$, let us denote $\eta_\Le'$ and $\eta_\Ri'$ the conformal times of the screens on the apparent horizons defined in \Eqs{a1} and~(\ref{a2}). The areas of $\S_\Le'$ and $\S_\Ri'$ are given by  
\be
    \A(\S_\Le')=\omega_{n-1}\,a_0^{n-1}\left(\sin{\eta_\Le'\over |\gamma|}\right)^{(\gamma+1)(n-1)},\qquad   \A(\S_\Ri')=\omega_{n-1}\,a_0^{n-1}\left(\sin{\eta_\Ri'\over |\gamma|}\right)^{(\gamma+1)(n-1)}.
 \ee
They are constant in the de Sitter case $\gamma=-1$, which means that there is a perfect compensation between the fact that the areas decrease as we approach the worldlines of the observers, and the fact that they increase when the conformal time approaches 0 or $|\gamma|\pi$. In the generic bouncing case $\gamma<-1$, the increasing effect dominates: the area of $\S_\Le'$ and $\S_\Ri'$ start infinite at $\eta_\Le'=\eta_\Ri'=0$, reach their minimum values at $\eta_\Le'=\eta_\Ri'=|\gamma|\pi/2$ and then grow to infinity at $\eta_\Le'=\eta_\Ri'=|\gamma|\pi$. In the Big Bang/Big Crunch case $\gamma\ge 1$, the areas of $\S_\Le'$ and $\S_\Ri'$ grow from zero at $\eta_\Le'=\eta_\Ri'=0$, reach their maximum values at $\eta_\Le'=\eta_\Ri'=\gamma\pi/2$ and then decrease until they vanish at $\eta_\Le'=\eta_\Ri'=\gamma\pi$. In all cases, the total area of the screens in Planck units should be a measure of the number of the degrees of freedom on them, and the exponential of this a measure of the dimensionality of the Hilbert space of the dual holographic theory. See \Refs{Banks:2000fe,Witten:2001kn} for such proposals in the context of de Sitter holography. As a result, the evolution of the holographic theory is not unitary, in the sense that it corresponds to a sequence of maps between Hilbert spaces of different dimensionalities. For the Big Bang/Big Crunch case, the dimensionality of the Hilbert space is maximum  when both screens are at the bifurcation point. Near the singularities the dimensionality of the Hilbert space is very small, with the holographic dual having very few degrees of freedom. In the bouncing cases, the maximum dimensionality occurs when the screens are at the future or past null infinity. 
%\footnote{\blue DANGEROUS... TO CORRECT/CHANGE?? It is possible that the contracting (expanding) phase of the cosmology gets mapped to an RG flow (inverse RG flow) towards low (high) energy scales  in the dual holographic theory, as in \Refe{Strominger:2001gp}.} 

%{\blue ADD: the causal diamonds of the pode trajectory and screen trajectory are the same: they are the envelope of each trajectory, as said in footnote 1 of the last paper of witten. }

%%%%%%%%%%%%%%%%%%%

\subsection{Bilayer proposal}
\label{bilayer}

We are ready to present our proposal for computing entanglement entropies between complementary subsystems of the pair of screens, as well as the bulk regions that may be reconstructed from such holographic subsystems. 

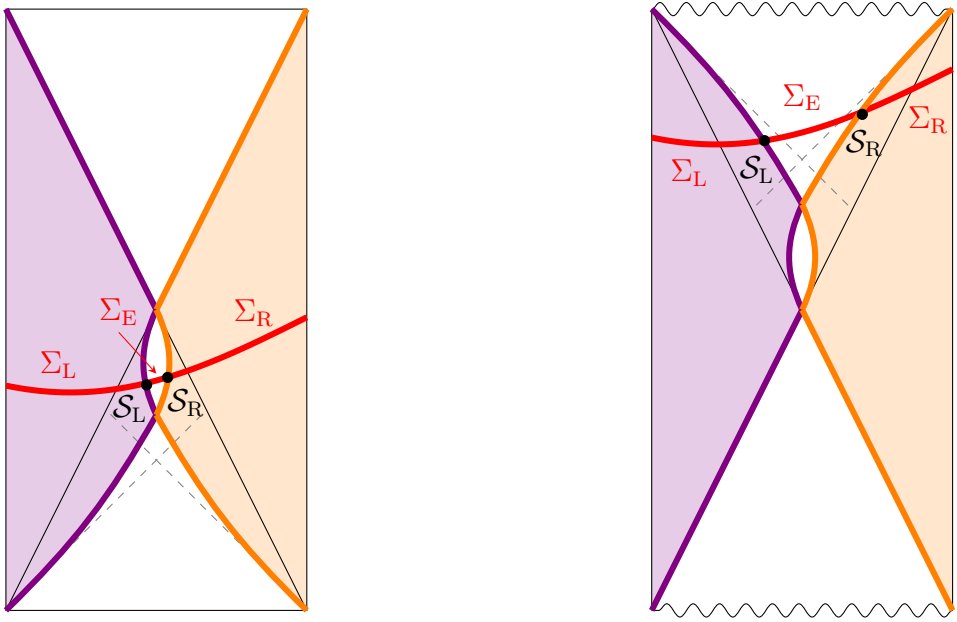
\begin{figure}[h!]
   % \centering
\begin{subfigure}[t]{0.48\linewidth}
\centering
\begin{tikzpicture}

\path
       +(2,4)  coordinate (IItopright)
       +(-2,4) coordinate (IItopleft)
       +(2,-4) coordinate (IIbotright)
       +(-2,-4) coordinate(IIbotleft)
      
       ;

%\begin{scope}[transparency group]
\begin{scope}[blend mode=multiply]
\fill[fill=violet!20] (IIbotleft) to [bend right=8] (0,-1.4) to [bend left=25] (0,0) -- (IItopleft) -- cycle;
\fill[fill=orange!20] (IIbotright) to [bend left=8] (0,-1.4) to [bend right=25] (0,0) -- (IItopright) -- cycle;
\end{scope}
%\end{scope}
       
\draw (IItopleft) --
          node[midway, above, sloped]    {}
      (IItopright) --
          node[midway, above, sloped] {}
      (IIbotright) -- 
          node[midway, below, sloped] {}
      (IIbotleft) --
          node[midway, above , sloped] {}
      (IItopleft) -- cycle;
      
\draw[violet,line width=0.8mm] (IItopleft) -- (0,0);
\draw (0,0) -- (IIbotright);
\draw[orange,line width=0.8mm] (IItopright) -- (0,0);
\draw (0,0) -- (IIbotleft) ;
              
\draw[dashed,gray] (IIbotleft) -- (2/3,-4/3);
\draw[dashed,gray] (IIbotright) -- (-2/3,-4/3);

%\draw[domain=-2:2, smooth, variable=\x, red] plot ({\x}, {sin(deg((\x/2-1)))-1.5});

\draw[domain=-2:2, smooth, variable=\x, red, line width=0.8mm] plot ({\x}, {sin(deg((\x/2-1)))-0.1});
%\draw[domain=-0.13:0.15, smooth, variable=\x, red] plot ({\x}, {sin(deg((\x/2-1)))-0.1});
%\draw[domain=0.15:2, smooth, variable=\x, red, line width=0.8mm] plot ({\x}, {sin(deg((\x/2-1)))-0.1});

\draw[violet,line width=0.8mm] (IIbotleft) to [bend right=8] (0,-1.4) ;
\draw[violet,line width=0.8mm] (0,-1.4) to [bend left=25] (0,0) ;

\draw[orange,line width=0.8mm] (IIbotright) to [bend left=8] (0,-1.4) ;
\draw[orange,line width=0.8mm] (0,-1.4) to [bend right=25] (0,0) ;

\node at (-0.13,-1) [circle, fill, inner sep=1.5 pt]{};
\node at (-0.35,-0.85) [label=below:$\S_\Le$]{};

\node at (0.15,-0.9) [circle, fill, inner sep=1.5 pt]{};
\node at (0.4,-0.77) [label=below:$\S_\Ri$]{};

\node at (-0.5,0.45) [label=below:$\red \Sigma_\Ex$]{};
\draw[red,-stealth] (-0.5,-0.3) -- (0,-0.85);
\node at (-1.3,-1.2) [label=above:$\red \Sigma_\Le$]{};
\node at (1.3,-0.5) [label=above:$\red \Sigma_\Ri$]{};

%\node at (-0.65,-2.45) [circle, fill, inner sep=1.5 pt]{};
%\node at (-0.78,-1.6) [label=below:$\S_\Le$]{};

%\node at (0.48,-2.17) [circle, fill, inner sep=1.5 pt]{};
%\node at (0.6,-1.36) [label=below:$\S_\Ri$]{};

%\node at (0,-2.3) [label=below:$\red \Sigma_\Ex$]{};
%\node at (-1.6,-2.5) [label=above:$\red \Sigma_\Le$]{};
%\node at (1.5,-1.8) [label=above:$\red \Sigma_\Ri$]{};

\end{tikzpicture} 
\caption{\footnotesize Case of a bouncing cosmology, $\gamma\le -1$. \label{lerA}}
\end{subfigure}
\quad \,
\begin{subfigure}[t]{0.48\linewidth}
\centering
\begin{tikzpicture}

\path
       +(2,4)  coordinate (IItopright)
       +(-2,4) coordinate (IItopleft)
       +(2,-4) coordinate (IIbotright)
       +(-2,-4) coordinate(IIbotleft)
      
       ;

%\begin{scope}[transparency group]
\begin{scope}[blend mode=multiply]
\fill[fill=violet!20] (IItopleft) to [bend left=8] (0,1.4) to [bend right=25] (0,0) -- (IIbotleft) -- cycle;
\fill[fill=orange!20] (IItopright) to [bend right=8] (0,1.4) to [bend left=25] (0,0) -- (IIbotright) -- cycle;
\end{scope}
%\end{scope}
      
\draw[decorate,decoration=snake] (IItopleft) --
          node[midway, above, sloped]    {}
      (IItopright);
      
\draw (IItopright) --
          node[midway, above, sloped] {}
      (IIbotright);
      
\draw[decorate,decoration=snake]  (IIbotright) -- 
          node[midway, below, sloped] {}
      (IIbotleft);
      
\draw (IIbotleft) --
          node[midway, above , sloped] {}
      (IItopleft);
      
\draw[violet,line width=0.8mm] (IIbotleft) -- (0,0);
\draw (0,0) -- (IItopright);
\draw[orange,line width=0.8mm] (IIbotright) -- (0,0);
\draw (0,0) -- (IItopleft) ;
              
\draw[dashed,gray] (IItopleft) -- (2/3,4/3);
\draw[dashed,gray] (IItopright) -- (-2/3,4/3);

\draw[domain=-2:2, smooth, variable=\x, red, line width=0.8mm] plot ({\x}, {sin(deg((\x/2-1)))+3.2});
%\draw[domain=-0.5:0.8, smooth, variable=\x, red] plot ({\x}, {sin(deg((\x/2-1)))+3.2});
%\draw[domain=0.8:2, smooth, variable=\x, red, line width=0.8mm] plot ({\x}, {sin(deg((\x/2-1)))+3.2});

\draw[violet,line width=0.8mm] (IItopleft) to [bend left=8] (0,1.4) ;
\draw[violet,line width=0.8mm] (0,1.4) to [bend right=25] (0,0) ;

\draw[orange,line width=0.8mm] (IItopright) to [bend right=8] (0,1.4) ;
\draw[orange,line width=0.8mm] (0,1.4) to [bend left=25] (0,0) ;

\node at (-0.5,2.25) [circle, fill, inner sep=1.5 pt]{};
\node at (0.8,2.6) [circle, fill, inner sep=1.5 pt]{};

\node at (-0.6,1.4) [label=above:$\S_\Le$]{};
\node at (0.83,1.75) [label=above:$\S_\Ri$]{};

\node at (0,3.3) [label=below:$\red \Sigma_\Ex$]{};
\node at (-1.5,2.3) [label=below:$\red \Sigma_\Le$]{};
\node at (1.68,3) [label=below:$\red \Sigma_\Ri$]{};

\end{tikzpicture}     
\caption{\footnotesize Case of a Big Bang/Big Crunch cosmology, \mbox{$\gamma\ge 1$}. \label{lerB}}
\end{subfigure}
    \caption{\footnotesize Trajectories of left screen $\S_\Le$ (purple) and right screen $\S_\Ri$ (orange). The interior region of the pode is purple shaded, while the interior region of the antipode is orange shaded. The exterior region is white. \label{ler}}

\end{figure}

Let us first repeat our notations and complete them. For any cosmological evolution satisfying $|\gamma|\ge 1$, consider a foliation $\F$ of spacetime, with SO$(n)$-symmetric Cauchy slices. In the Penrose diagram in \Fig{push}, choose a nowhere spacelike curve that starts at the bottom left corner, ends at the top left corner, is located in the purple region of the contracting era, and follows the apparent horizon of the pode in the expanding era. Similarly, choose a second curve, swapping the roles of left and right. As shown in \Fig{ler}, 
each Cauchy slice $\Sigma$ of $\F$ intersects these trajectories at two S$^{n-1}$ spheres, where the left screen $\S_\Le$ and the right screen $\S_\Ri$ are located. We denote $\Sigma_\Le$ the part of $\Sigma$ to the left of $\S_\Le$, $\Sigma_\Ri$ the part of $\Sigma$ to the right of $\S_\Ri$ and $\Sigma_\Ex$ the part between $\Sigma_\Le$ and $\Sigma_\Ri$. $\Sigma_\Le$ and $\Sigma_\Ri$ have the topologies of spherical caps and their boundaries are respectively $\S_\Le$ and $\S_\Ri$. The topology of $\Sigma_\Ex$ is that of a barrel and its boundary is $\S_\Le\cup\S_\Ri$. We refer to the ``interior region of the pode,'' which is shown in purple in \Fig{ler}, the domain spanned by all Cauchy slices $\Sigma_\Le$. Likewise the ``interior region of the antipode,'' which is shown in orange, is covered by all slices $\Sigma_\Ri$. The remaining domain of spacetime, which we call the ``exterior region,'' is spanned by all slices $\Sigma_\Ex$ and appears in white in \Fig{ler}. We will label these three regions as $\Le$, $\Ex$ and $\Ri$, respectively. Note that given $\Sigma\in\F$, all Cauchy slices $\hat \Sigma$ of spacetime that pass through $\S_\Le$ and $\S_\Ri$ have equivalent causal structures: denoting $\hat \Sigma_i$, $i\in\{\Le,\Ex,\Ri\}$, the part of $\hat \Sigma$ located in region $i$, the causal diamonds of $\Sigma_i$ and $\hat \Sigma_i$ are the same. 

\noindent {\bf Classical entropy:} {\it For any subregion $A$ of $\S_\Le\cup \S_\Ri$ on the Cauchy slice $\Sigma$, denote 
\be
A_\Le=A\cap \S_\Le, \qquad A_\Ex=A, \qquad A_\Ri=A\cap \S_\Ri.
\ee  
For $i\in\{\Le,\Ex,\Ri\}$, let $\chi_i$ be a codimension-2 surface of minimal extremal area  that is homologous to $A_i$ and lies on a Cauchy slice $\hat \Sigma_i$.\footnote{Notice that the homology constraint implies that $\chi_i$ is anchored on the boundaries of $A_i$, namely \mbox{$\partial \chi_i=\partial A_i$.}\label{anch}} At leading order in $G\hbar$, the von Neumann entropy $S$ of the subsystem on $A$, \ie the entanglement entropy between $A$ and its complement in $\S_\Le\cup \S_\Ri$, satisfies 
\be
S(A)={\A(\chi_\Le)+\A(\chi_\Ex)+\A(\chi_\Ri)\over 4G\hbar}+\O((G\hbar)^0),
\ee
where $\A(\chi_i)$ is the area of $\chi_i$.} 

\noindent Let us stress that since $\chi_i$ must lie on some Cauchy slice $\hat\Sigma_i$, one must look for it within the causal diamond of $\Sigma_i$. Since this region has a boundary, extremizing the area functional of surfaces homologous to $A_i$ requires the introduction of Lagrange multipliers and auxiliary fields enforcing all extremal surfaces to lie in the diamond, including its boundary~\cite{Franken:2023pni}. If there exists a single extremal-area surface in region $i$, it is $\chi_i$. If several exist, $\chi_i$ is one with the smallest area. 

\noindent  {\bf Entanglement wedge:} {\it Let $\hat \C_i$ be the codimension-1 surface\footnote{All codimensions are with respect to the whole $(n+1)$-dimensional spacetime.} on $\hat \Sigma_i$ bounded by $\chi_i$ and $A_i$, \ie satisfying $\partial\hat \C_i=\chi_i\cup A_i$. Assuming entanglement wedge reconstruction~\cite{Dong:2016eik}, we have the following:

\noindent $\bullet$ The state on $\hat \C_\Le\cup\hat \C_\Ex\cup\hat \C_\Ri$ is dual to the state of the holographic subsystem on $A$. Generically, they  are mixed states. 

\noindent $\bullet$ In particular, the von Neumann entropy of the state on $\hat \C_\Le\cup\hat \C_\Ex\cup\hat \C_\Ri$, \ie the entanglement entropy between $\hat \C_\Le\cup\hat \C_\Ex\cup\hat \C_\Ri$ and its complement in $\hat \Sigma$, equals the von Neumann entropy of the holographic subsystem on $A$, \ie the entanglement entropy between $A$ and its complement in $\S_\Le\cup \S_\Ri$. 

\noindent $\bullet$ The ``entanglement wedge'' of $A$, which is a spacetime region reconstructible from the dual holographic subsystem on $A$, is the union of the three causal diamonds of $\hat\C_\Le$, $\hat\C_\Ex$, $\hat\C_\Ri$.\footnote{Notice that the union of causal diamonds is not the causal diamond of a union. The former is a subset of the latter.} }
%{\blue H: 1) I will add a comment saying the the true region than can be reconstructed is the diamond of the union. The wedges are the regions reconstructed without redundancies since they are equivalent to those on cauch slices etc... 2) change order: put wedge before classical story.}

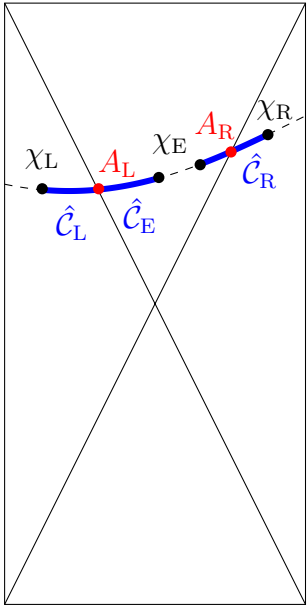
\begin{figure}[h!]
\centering
\begin{tikzpicture}

\path
       +(2,4)  coordinate (IItopright)
       +(-2,4) coordinate (IItopleft)
       +(2,-4) coordinate (IIbotright)
       +(-2,-4) coordinate(IIbotleft)
      
       ;
       
\draw (IItopleft) --
          node[midway, above, sloped]    {}
      (IItopright) --
          node[midway, above, sloped] {}
      (IIbotright) -- 
          node[midway, below, sloped] {}
      (IIbotleft) --
          node[midway, above , sloped] {}
      (IItopleft) -- cycle;
      
\draw (IItopleft) -- (IIbotright)
              (IItopright) -- (IIbotleft) ;

\draw[domain=-2:-1.5, smooth, variable=\x,dashed] plot ({\x}, {sin(deg((\x/2-1)))+2.5});
\draw[domain=-1.5:-0.75, smooth, variable=\x, blue, line width=0.8mm] plot ({\x}, {sin(deg((\x/2-1)))+2.5});
\draw[domain=-0.75:0, smooth, variable=\x, blue, line width=0.8mm] plot ({\x}, {sin(deg((\x/2-1)))+2.5});
\draw[domain=0:0.6, smooth, variable=\x, dashed] plot ({\x}, {sin(deg((\x/2-1)))+2.5});
\draw[domain=0.6:1.5, smooth, variable=\x, blue, line width=0.8mm] plot ({\x}, {sin(deg((\x/2-1)))+2.5});
\draw[domain=1.5:2, smooth, variable=\x, dashed] plot ({\x}, {sin(deg((\x/2-1)))+2.5});

\node at (-0.75,1.53) [circle, fill, inner sep=1.5 pt,red]{};
\node at (-0.5,1.4) [label=above:$\red A_\Le$]{};

\node at (1.01,2.02) [circle, fill, inner sep=1.5 pt,red]{};
\node at (0.8,1.9) [label=above:$\red A_\Ri$]{};

\node at (-1.5,1.53) [circle, fill, inner sep=1.5 pt]{};
\node at (0.6,1.85) [circle, fill, inner sep=1.5 pt]{};
\node at (-1.5,1.5) [label=above:$\chi_\Le$]{};

\node at (0.05,1.68) [circle, fill, inner sep=1.5 pt]{};
\node at (0.2,1.7) [label=above:$\chi_\Ex$]{};

\node at (1.5,2.25) [circle, fill, inner sep=1.5 pt]{};
\node at (1.6,2.15) [label=above:$\chi_\Ri$]{};

\node at (-1.1,1.6) [label=below:$\blue \hat{\mathcal{C}}_\Le$]{};
\node at (-0.2,1.7) [label=below:$\blue \hat{\mathcal{C}}_\Ex$]{};
\node at (1.4,2.3) [label=below:$\blue \hat{\mathcal{C}}_\Ri$]{};
%\node at (0.4,1.9) [label=below:$\hat \Sigma$]{};

\end{tikzpicture}    
    \caption{\footnotesize We consider a subregion $A$ of $\S_\Le\cup \S_\Ri$. $A_\Le=A\cap \S_\Le$ and $A_\Ri=A\cap \S_\Ri$ are depicted by the red dots, on a Cauchy slice $\hat \Sigma$ (dashed line). For $i\in\{\Le,\Ex,\Ri\}$, $\chi_i$ is a codimension-2 surface of minimal extremal area depicted by a black dot, which is homologous to $A_i$ on $\hat \Sigma_i$. $\hat \C_i$ is the codimension-1 surface on $\hat \Sigma_i$ bounded by $\chi_i$ and $A_i$. In the example depicted on this Penrose diagram, $\chi_E$ is the union of the two black dots in the upper triangular region, while $\hat \C_\Ex$ is the union of the two blue slices in this region. 
     \label{surfaces}}
\end{figure}

\noindent { An example depicting the minimal extremal surfaces $\chi_i$ associated with a subsystem $A$ of $\S_\Le \cup \S_\Ri$, together with the associated codimension-1 surfaces $\hat \C_i$, is presented on the Penrose diagram of \Fig{surfaces}.}
Notice that the choice of Cauchy slice $\hat \Sigma_i$ containing $\chi_i$ is not unique, and so for the $\hat \C_i$'s. However, this does not introduce any ambiguity in the definition of the entanglement wedge, as the causal diamonds of all these surfaces $\hat \C_i$ are the same. On the contrary, when several surfaces of minimal extremal areas exist in some region $i$, meaning that they have equal areas, the causal diamonds of the corresponding surfaces $\hat \C_i$ are in general distinct. In this case, one is led to include quantum corrections to the classical entropy formula, in order to lift the degeneracy. These quantum corrections can produce non-trivial results as we will see and are crucial to understand the escape of information from an evaporating black hole \cite{Penington:2019npb, Almheiri:2019psf}.

\noindent {\bf Semiclassical entropy:} To take into account the first-order 
quantum corrections to the entanglement entropy between $A$ and its complement in $\S_\Le\cup \S_\Ri$, let us define the generalized entropy 
\be
S_{\rm gen}(\chi_\Le,\chi_\Ex,\chi_\Ri;A)= {\A(\chi_\Le)+\A(\chi_\Ex)+\A(\chi_\Ri)\over 4G\hbar} + S_{\rm semicl}\big(\hat\C_\Le \cup\hat \C_\Ex\cup\hat\C_\Ri\big),
    \label{Sgen}
\ee
where $\chi_i$, $i\in\{\Le,\Ex,\Ri\}$, is an arbitrary codimension-2 surface homologous to $A_i$ and lying on a Cauchy slice $\hat\Sigma_i$, while $\hat \C_i\subset \hat \Sigma_i$ is the codimension-1 surface satisfying $\partial\hat \C_i=\chi_i\cup A_i$. In the above expression, $S_{\rm semicl}$ is the von Neumann entropy of quantum fields, including gravitons, on the semiclassical geometry \cite{Almheiri:2020cfm}. More specifically, using methods of quantum field theory on curved backgrounds, one can construct the density matrix $\rho_R$ of the quantum field system on the region $R=\hat\C_\Le \cup\hat \C_\Ex\cup\hat\C_\Ri$ and define $S_{\rm semicl}(\hat\C_\Le \cup\hat \C_\Ex\cup\hat\C_\Ri)=-\tr\left(\rho_R\log\rho_R\right)$. The leading contributions to the semiclassical entropy are of order $(G\hbar)^0$ and appear at the 1-loop level in a bulk replica path integral computation. Both the bulk matter and radiation fields, including gravitons, contribute. As before, for any $i\in\{\Le,\Ex,\Ri\}$, the choice of Cauchy slice $\hat \Sigma_i$ is not unique, and so for $\hat \C_i$. However, since the density matrices on different $\hat \C_i$'s are unitarily related, $S_{\rm semicl}$ is independent of the choice of $\hat \C_i$ and thus the definition of $S_{\rm gen}$ is unambiguous. 

\noindent {\it To obtain the entanglement entropy of $A$ with its complement in $\S_\Le \cup \S_\Ri$ in the holographic dual theory at the semiclassical level, one has to extremize the generalized entropy with respect to $\chi_\Le$, $\chi_\Ex$ and $\chi_\Ri$, and then select among the extrema the configuration for which $S_{\rm gen}$ is minimal. This can be summarized by the expression, }
\be
S(A)=\min {\rm ext}\; S_{\rm gen}(\chi_\Le,\chi_\Ex,\chi_\Ri;A)+\O(G\hbar).
\label{Ssem}
\ee

\noindent Notice that in general the semiclassical entropy associated with bulk quantum fields is not additive. More specifically, the semiclassical entropy on the union of two surfaces is less than or equal to the sum of the entropies on each surface~\cite{VanRaamsdonk:2016exw}. Therefore, it is important to consider the semiclassical entropy on the union of surfaces in \Eq{Sgen}. 

%{\cyan  \noindent {\bf Beyond semiclassical?} TO DISCUSS: I think that replacing $S_{\rm semicl}$ with an exact version calculated for the fields living on $\hat C_i$ is not correct. The reason is that if we calculate these corrections beyond one loop, then we have to take into account their effect on the geometry: the metric will change. On the other hand, if we stay at 1-loop for fields on $\hat C_i$, then this is already taken into account in the FRW geometry. For example, in the thermal case, $\rho$, $p$ are calculated at 1-loop and injected into the Einstein eqs. It is even this effect, in the form of a source, that gives rise to cosmology. In short, the backreaction of these quantum corrections on the background is taken into account at this order. But not at subsequent orders... What is the conjecture beond semiclass????? How to extremize curves and metric together...??? The fulle path integral should tell us. }

%{\blue PUT THIS LATER: The bilayer proposal and its quantum extension imply that the von Neumann entropy of the two screen system is vanishing to all orders in $G\hbar$ -- that is, the two-screen system remains always in a pure state.\footnote{It would be interesting to explore the possibility of an isometric evolution that preserves the inner product on the Hilbert space, as in \cite{Cotler:2022weg, Cotler:2023eza}.}} 

%%%%%%%%%%%%%%%%%%%%%%%%%%%%%%%%%%%%%%%%%

\section{Time-dependent ER=EPR}
\label{time_dep_EREPR}

In this section we compute the entanglement entropies associated with the two-screen system $\S_\Le\cup \S_\Ri$ and a single-screen system $\S_\Le$ or $\S_\Ri$, for cosmologies with $|\gamma|\geq 1$. The screens lie on SO$(n)$-symmetric Cauchy slices of the foliation $\F$ and follow their respective trajectories, as described in detail in \Sect{bilayer} and shown in \Fig{ler}. We denote the conformal coordinates of $\S_\Le$ by $(\theta_\Le, \eta_\Le)$ and those of $\S_\Ri$ by $(\theta_\Ri, \eta_\Ri)$. The corresponding lightcone coordinates are  $(x^+_\Le, x^-_\Le)$ and  $(x^+_\Ri, x^-_\Ri)$.  We mainly focus in this section on the classical, geometrical contributions to the entropies, which are of order $(G\hbar)^{-1}$, and comment on the effects of quantum corrections. In \Sect{semicla}, we proceed to incorporate thermal corrections due to the coarse-grained entropy carried by the thermal gas of particles permeating the bulk space in radiation-dominated cosmologies, which are of order $(G\hbar)^{0}$.

%Let us proceed to compute the von Neumann entropy of the two and the single-screen systems in cosmologies with $|\gamma|\geq 1$, when both screens lie at the conformal times $\eta_1\geq\eta_c$ and $\eta_2\geq\eta_c$ on the apparent horizons. When the full holographic system is in a pure state, the von Neumann entropies of the two screens are equal, and they will be referred to as the entanglement entropy between the screens. 
%We will see here how the connection between the geometrical bridge and quantum %entanglement manifests itself via our computations. 
%In this section, we will discuss the leading geometrical contributions to the generalized entropy arising from classical bulk surfaces, which are of order $(G\hbar)^{-1}$. In the following section, we incorporate thermal corrections due to the entropy carried by the thermal gas of massless particles permeating the bulk space for radiation dominated cosmologies.

\subsection{The two-screen system in cosmologies with \bm $|\gamma| \geq 1$}
\label{twoscreens}

\begin{figure}[h!]
\centering
\begin{subfigure}[c]{0.3\linewidth}
\centering
\begin{tikzpicture}

\path
       +(2,4)  coordinate (IItopright)
       +(-2,4) coordinate (IItopleft)
       +(2,-4) coordinate (IIbotright)
       +(-2,-4) coordinate(IIbotleft)
      
       ;
       
\fill[fill=blue!50] (-1,-2.7) -- (0.1,-1.6) -- (0.7,-2.2) -- (-0.4,-3.3) -- cycle;

\fill[fill=blue!50] (-2,-1.7) -- (-1,-2.7) -- (-2,-3.7) --  cycle;

\fill[fill=blue!50] (2,-0.9) -- (0.7,-2.2) -- (2,-3.5) --  cycle;

%\draw[dashed,gray] (IItopleft) -- (2/3,4/3);
%\draw[dashed,gray] (IItopright) -- (-2/3,4/3);

\draw[dashed,gray] (IIbotleft) -- (2/3,-4/3);
\draw[dashed,gray] (IIbotright) -- (-2/3,-4/3);
       
\draw (IItopleft) --
      (IItopright) --
      (IIbotright) -- 
      (IIbotleft) --
      (IItopleft) -- cycle;

\draw (IItopleft) -- (IIbotright)
              (IItopright) -- (IIbotleft) ;
      
\node at (-1,-2.7) [circle,fill,inner sep=1.5pt, label = below:$\S_\Le$]{};
\node at (0.7,-2.2) [circle,fill,inner sep=1.5pt, label = below:$\S_\Ri$]{};

\end{tikzpicture}
\caption{\footnotesize For $\eta_\Le, \eta_\Ri\in[0,|\gamma|\pi/2)$.}
\end{subfigure}\hfill
\begin{subfigure}[c]{0.3\linewidth}
\centering
\begin{tikzpicture}

\path
       +(2,4)  coordinate (IItopright)
       +(-2,4) coordinate (IItopleft)
       +(2,-4) coordinate (IIbotright)
       +(-2,-4) coordinate(IIbotleft)
      
       ;
       
\fill[fill=blue!50] (-2,2) -- (0,0) -- (-2,-2) --  cycle;

\fill[fill=blue!50] (2,2) -- (0,0) -- (2,-2) --  cycle;

%\draw[dashed,gray] (IItopleft) -- (2/3,4/3);
%\draw[dashed,gray] (IItopright) -- (-2/3,4/3);

%\draw[dashed,gray] (IIbotleft) -- (2/3,-4/3);
%\draw[dashed,gray] (IIbotright) -- (-2/3,-4/3);
       
\draw (IItopleft) --
      (IItopright) --
      (IIbotright) -- 
      (IIbotleft) --
      (IItopleft) -- cycle;

\draw (IItopleft) -- (IIbotright)
              (IItopright) -- (IIbotleft) ;
      
\node at (0,0) [circle,fill,inner sep=1.5pt, label = left:$\S_\Le$]{};
\node at (0,0) [circle,fill,inner sep=1.5pt, label = right:$\S_\Ri$]{};

\end{tikzpicture}    
\caption{\footnotesize For $\eta_\Le=\eta_\Ri=|\gamma|\pi/2$.}
\end{subfigure}\hfill
\begin{subfigure}[c]{0.3\linewidth}
\centering
\begin{tikzpicture}

\path
       +(2,4)  coordinate (IItopright)
       +(-2,4) coordinate (IItopleft)
       +(2,-4) coordinate (IIbotright)
       +(-2,-4) coordinate(IIbotleft)
      
       ;
       
\fill[fill=blue!50] (-0.8,1.6) -- (0.6,3) -- (1.2,2.4) -- (-0.2,1) -- cycle;

\fill[fill=blue!50] (-2,2.8) -- (-0.8,1.6) -- (-2,0.4) --  cycle;

\fill[fill=blue!50] (2,3.2) -- (1.2,2.4) -- (2,1.6) --  cycle;

%\draw[dashed,gray] (IItopleft) -- (2/3,4/3);
%\draw[dashed,gray] (IItopright) -- (-2/3,4/3);

%\draw[dashed,gray] (IIbotleft) -- (2/3,-4/3);
%\draw[dashed,gray] (IIbotright) -- (-2/3,-4/3);
       
\draw (IItopleft) --
      (IItopright) --
      (IIbotright) -- 
      (IIbotleft) --
      (IItopleft) -- cycle;

\draw (IItopleft) -- (IIbotright)
              (IItopright) -- (IIbotleft) ;
      
\node at (-0.8,1.6) [circle,fill,inner sep=1.5pt, label = left:$\S_\Le$]{};
\node at (1.2,2.4) [circle,fill,inner sep=1.5pt, label = right:$\S_\Ri$]{};
%\node at (0.5,2) [label = right:$\S_2$]{};

\end{tikzpicture}
\caption{\footnotesize For $\eta_\Le, \eta_\Ri\in(|\gamma|\pi/2,|\gamma|\pi]$.}
\end{subfigure}\hfill
     \caption{\footnotesize Entanglement wedge (blue shaded region) of the two-screen system $\S_\Le \cup\S_\Ri$, when $\gamma \leq -1$. The entire spacetime is covered by the entanglement wedge as the latter evolves during the whole cosmological evolution. \label{Penrose_diag_FRW_entang_wedge}}
\end{figure}
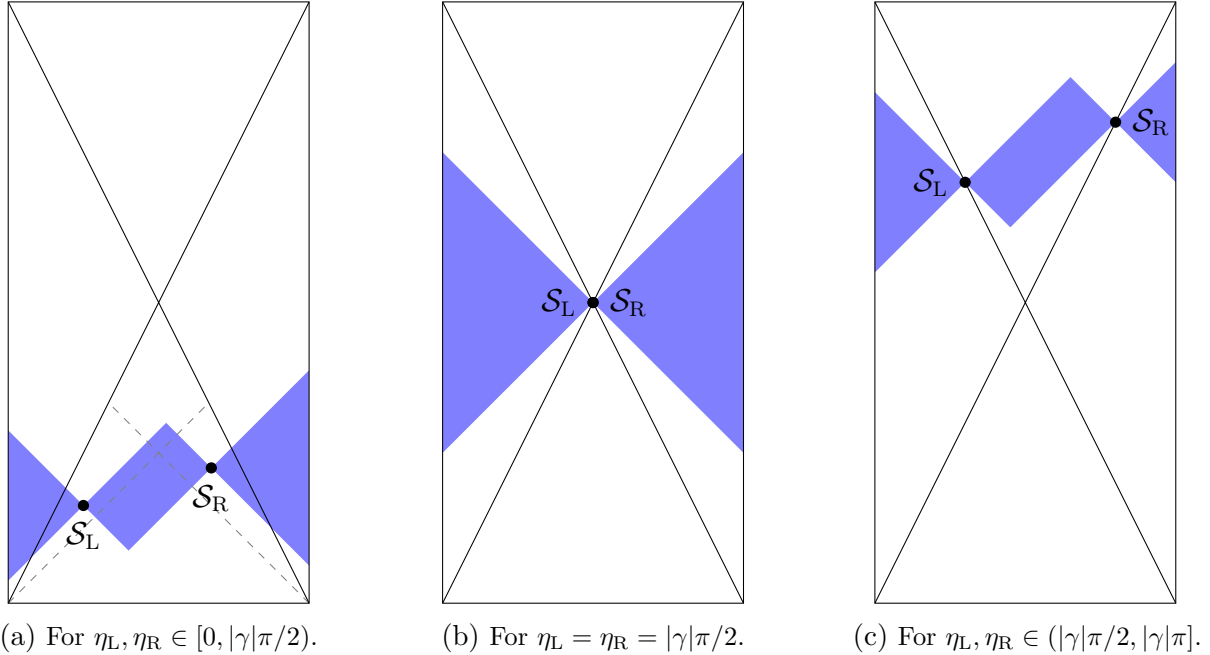

On a given Cauchy slice $\Sigma\in\F$, let us consider the two-screen system $A=\S_\Le \cup \S_\Ri$. We will calculate the geometrical, classical contributions to the fine-grained entropy by applying the rules of \Sect{bilayer}. Following the notations introduced there, we have for this case $A_{\rm L}=\S_\Le$, $A_{\rm E}=\S_\Le \cup \S_\Ri$ and $A_{\rm R}=\S_\Ri$. Now, consider any SO$(n)$-symmetric Cauchy slice $\hat {\Sigma}$ containing the spheres on which the screens $\S_\Le$ and $\S_\Ri$ are located. The parts of $\hat \Sigma$ and $\Sigma$ that lie in the pode interior region, $\hat{\Sigma}_{\rm L}$ and $\Sigma_{\rm L}$, have a common causal diamond corresponding to the left blue triangle in Figure~\ref{Penrose_diag_FRW_entang_wedge},
for the bouncing cases. The situation is completely analogous for the Big Bang/Big Crunch cosmologies. Since any such Cauchy slice $\hat{\Sigma}_{\rm L}$ has the topology of a spherical cap bounded by $A_{\rm L}=\S_\Le$, the surface homologous to $A_{\rm L}$ that is of minimal extremal area is the empty set, $\chi_{\rm L}=\varnothing$. Indeed, the latter is a subset of $\hat{\Sigma}_{\rm L}$ and it is homologous to~$A_{\rm L}$ since the homology condition $\partial \hat{\mathcal{C}}_{\rm L}=\varnothing\cup A_{\rm L}$ is trivially satisfied for $\hat{\mathcal{C}}_{\rm L}=\hat{\Sigma}_{\rm L}$. Moreover, the empty set is a surface of minimal extremal area since it cannot be deformed and has a vanishing area. Hence, the classical contribution from region L to the von Neumann entropy of the two-screen system is null. Since $\hat{\mathcal{C}}_{\rm L}=\hat{\Sigma}_{\rm L}$, the part of the entanglement wedge in L is the full causal diamond of $\hat{\Sigma}_{\rm L}$ and $\Sigma_{\rm L}$, \ie the left blue triangle in Figure~\ref{Penrose_diag_FRW_entang_wedge}. 

The above arguments can be easily adapted to determine the classical contributions to the von Neumann entropy from the interior region of the antipode R. The empty set (which is trivially a subset of (any) $\hat{\Sigma}_{\rm R}$) is the minimal extremal surface homologous to $A_{\rm R}=\S_\Ri$, \mbox{$\chi_{\rm R}=\varnothing$}. In addition, since $\hat{\mathcal{C}}_{\rm R}=\hat{\Sigma}_{\rm R}$, the part of the entanglement wedge in region R is the causal diamond of $\hat{\Sigma}_{\rm R}$ (and $\Sigma_{\rm R}$), which corresponds to the right blue triangle in Figure~\ref{Penrose_diag_FRW_entang_wedge}. Therefore, the classical contribution to the fine-grained entropy of the two-screen system from region R is also vanishing. 

Finally, consider the exterior region, for which $A_{\rm E}=\S_\Le \cup \S_\Ri$. The slices $\hat{\Sigma}_{\rm E}$ in region~E, which are bounded by $S_\Le \cup \S_\Ri$, span a common causal diamond corresponding to the blue rectangle in Figure~\ref{Penrose_diag_FRW_entang_wedge}. The empty set is a subset of minimal extremal area of $\hat{\Sigma}_{\rm E}$ and the homology constraint $\partial \hat{\mathcal{C}}_{\rm E}=\varnothing\cup A_{\rm E}$ is trivially satisfied for $\hat{\mathcal{C}}_{\rm E}=\hat{\Sigma}_{\rm E}$.
%\footnote{\label{fn_hom}The conditions $\partial\hat \C_i=\chi_i\cup A_i$ reduce to $\partial \hat \Sigma_\Le=\varnothing\cup \S_\Le$, $\partial \hat \Sigma_\Ex =\varnothing\cup (\S_\Le\cup\S_\Ri)$ and $\partial\hat\Sigma_\Ri=\varnothing\cup \S_\Ri$, respectively.}
Hence, we obtain $\chi_{\rm E}=\varnothing$, leading to vanishing classical contributions to the von Neumann entropy from the exterior region as well. The entanglement wedge in region E is the full causal diamond of $\hat{\Sigma}_{\rm E}$, {\it i.e.} the blue rectangle in Figure~\ref{Penrose_diag_FRW_entang_wedge}. 

We conclude that the leading classical contributions to the von Neumann entropy of the two-screen system are vanishing: 
\begin{equation}
     S(\S_\Le \cup \S_\Ri) =0 +{\mathcal{O}}(G\hbar)^{0}.
\end{equation}
The full entanglement wedge is the union of the causal diamonds of $\hat{\Sigma}_{\rm L}$, $\hat{\Sigma}_{\rm E}$ and $\hat{\Sigma}_{\rm R}$, \ie the union of the blue triangles and the rectangle in Figure~\ref{Penrose_diag_FRW_entang_wedge}. In particular, the entanglement wedge comprises slices that are complete Cauchy slices, with respect to the bulk cosmology. Assuming entanglement wedge reconstruction \cite{Dong:2016eik}, the state on the whole of $\hat \Sigma$ or $\Sigma$ can be reconstructed from the holographic dual system on $\S_\Le\cup \S_\Ri$. At the end of \Sect{Bb}, we already saw that the two screen-system has an adequate number of degrees of freedom to describe the state on the entire slice $\Sigma$, provided $\Sigma_\Ex$ lies in the region of trapped surfaces. Here, we see that the bilayer proposal leads to the conclusion that the state of the two-screen system is dual to the state on $\Sigma$,  irrespective of the location of the screens. Even if the latter are on the apparent horizons that bound the region of anti-trapped surfaces, the entanglement wedge extends and covers any $\hat \Sigma_\Ex$, leading to full reconstruction of the states on any bulk Cauchy slice $\hat \Sigma$ containing the screens. This result was argued to hold for the de Sitter case $\gamma=-1$ in \Refe{Franken:2023pni}. As in the de Sitter case, each point of the cosmological spacetime will have been inside the entanglement wedge at least once, as the screens evolve with time. Therefore, we expect to be possible to encode holographically on the two-screen system the state of the cosmology on any slice of any complete foliation $\F$. 

To go beyond the vanishing geometrical, classical contributions to the entropy of the two-screen system, we can consider the generalized entropy $S_{\rm gen}$ defined in \Eq{Sgen}. For the classical minimal extremal surfaces $\chi_\Le=\chi_\Ex=\chi_\Ri=\varnothing$, the non-negative, ${\cal{O}}((G\hbar)^0)$ contribution $S_{\rm semicl}$ is evaluated for $\hat \C_\Le\cup \hat \C_\Ex\cup \hat \C_\Ri$, which amounts to the complete Cauchy slice $\hat \Sigma$. So, if the bulk cosmology is in a pure state, $S_{\rm semicl}(\hat \Sigma)$ is zero, implying that the empty surfaces in the regions L, R, E lead to a minimal extremal generalized entropy at the semiclassical level, since the latter vanishes in this case. In fact, the von Neumann entropy of the dual system $\S_\Le\cup \S_\Ri$ is expected to be exactly zero, with the two-screen system being in a pure state at all times $\eta_{\rm L}$, $\eta_{\rm R}$. On the other hand, when the bulk cosmology is in a mixed state, as in the radiation dominated cases, we have $S_{\rm semicl}(\hat \Sigma)>0$. Hence, it is necessary to extremize the generalized entropy $S_{\rm gen}$ in order to find the correct surfaces $\chi_i$, $i\in\{\Le,\Ex,\Ri\}$, that yield the minimal extrema results. In this case, the von Neumann entropy of the two-screen system is non-vanishing and of order $(G\hbar)^0$. 

Finally, notice that the state on the slices of $\F$ remains pure if it starts pure at past infinity. As we remarked at the end of \Sect{equivconf}, the evolution of the two-screen system, along with the bulk cosmology, is not unitary, in the sense that the number of the underlying degrees of freedom and the dimensionality of the Hilbert space changes. It could be that this evolution is isometric~\cite{Cotler:2022weg, Cotler:2023eza} and linear, mapping pure states to pure states. As we have seen, this possibility is indeed consistent with the bilayer proposal. 

%%%%%%%%%

\subsection{The single-screen system in bouncing cosmologies with \bm \mbox{$\gamma < -1$}}
\label{singlescreenbouncing}

We proceed now to consider a single-screen system $\S_\Le$ or $\S_\Ri$, in bouncing cosmologies with  $\gamma< -1$. The de Sitter case $\gamma=-1$ has been analyzed in \Refs{Shaghoulian:2021cef,Shaghoulian:2022fop, Franken:2023pni} in great detail. In the expanding phase, the screens must be placed on the apparent horizons. In the initial contracting phase, they can be pushed farther into the region of trapped surfaces between the apparent horizons, as explained in \Sect{holo_proposal} and shown in \Fig{lerA}. We will analyze the left screen subsystem $\S_\Le$, but all results and conclusions can be adapted to apply equally to the right screen $\S_\Ri$.

\begin{figure}[h!]
\centering
\begin{subfigure}[t]{0.48\linewidth}
\centering
\begin{tikzpicture}

\path
       +(2,4)  coordinate (IItopright)
       +(-2,4) coordinate (IItopleft)
       +(2,-4) coordinate (IIbotright)
       +(-2,-4) coordinate(IIbotleft)

       +(-3/4,3/2) coordinate (SL)
       +(1.2,2.4) coordinate(SR)
      
       ;

\fill[fill=blue!50] (SL) -- (-2,2.75) -- (-2,0.25) --  cycle;       
\fill[fill=red!20] (-3/4,3/2) -- (0.675,2.925) -- (1.2,2.4) -- (-0.225,0.975) -- (-3/4,3/2);
     
\draw (IItopleft) --
      (IItopright) --
      (IIbotright) -- 
      (IIbotleft) --
      (IItopleft) -- cycle;
      
\draw (IItopleft) -- (IIbotright)
              (IItopright) -- (IIbotleft) ;

\fill[fill=blue!50] (-3/4,3/2+0.05) -- (-0.225,0.975+0.05) -- (-0.225,0.975-0.05) -- (-3/4,3/2-0.05) -- cycle;

\node at (SL) [circle,fill,inner sep=1.5pt, label = left:$\S_\Le$]{};
\node at (SR) [circle,fill,inner sep=1.5pt, label = right:$\S_\Ri$]{};
\node at (-0.225,0.975) [circle,fill,inner sep=1.5pt, red, label = right:$M$]{};
%\node at (-0.225,0.975) [label = right:$M$]{};
\end{tikzpicture}
\caption{\footnotesize At early times, the causal diamond in E is a rectangle with $M$ being the lower vertex. The area of $M$ is non-zero and the classical contributions to the entropy of $\S_\Le$ are non-vanishing. As the screens approach future null infinity, the area of $M$ and the von Neumann entropy of $\S_\Le$ increase.}
\end{subfigure}\hfill
\begin{subfigure}[t]{0.48\linewidth}
\centering
\begin{tikzpicture}

\path
       +(2,4)  coordinate (IItopright)
       +(-2,4) coordinate (IItopleft)
       +(2,-4) coordinate (IIbotright)
       +(-2,-4) coordinate(IIbotleft)
      
       ;
       
\fill[fill=red!20] (-2,4) -- (2,4) -- (0,2) -- cycle;
       
\draw (IItopleft) --
      (IItopright) --
      (IIbotright) -- 
      (IIbotleft) --
      (IItopleft) -- cycle;
      
\draw (IItopleft) -- (IIbotright)
              (IItopright) -- (IIbotleft) ;
              
\fill[fill=blue!50] (-2,4+0.05) -- (0,2+0.05) -- (0,2-0.05) -- (-2,4-0.05) -- cycle;

\node at (IItopleft) [circle,fill,inner sep=1.5pt, label = left:$\S_\Le$]{};
\node at (IItopright) [circle,fill,inner sep=1.5pt, label = right:$\S_\Ri$]{};
\node at (0,2) [circle,fill,inner sep=1.5pt, red, label = below:$M$]{};

\end{tikzpicture}
\caption{\footnotesize At large times $\eta_\Le=\eta_\Ri=|\gamma|\pi$, $M$ asymptotes to a fixed point and so the entropy of $\S_\Le$ saturates at a finite value.}
\end{subfigure}\hfill
    \caption{\footnotesize Penrose diagram for a bouncing cosmology satisfying $\gamma<-1$, when the screens are in the expanding phase. The dark blue region corresponds to the entanglement wedge of the single-screen system $\S_\Le$. The causal diamond in the exterior region~E is red shaded. The minimal extremal sphere, denoted by the red dot $M$, corresponds to the lower vertex.}
    \label{Penrose_gamma<-1}
\end{figure}
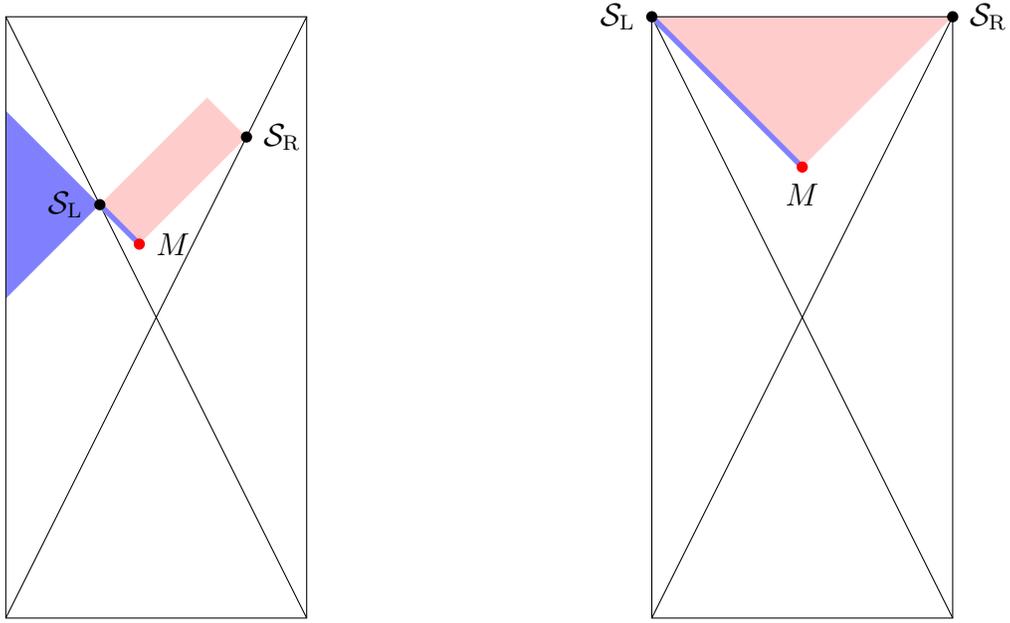

In the pode interior region L, the minimal extremal surface homologous to $A_{\rm L}=\S_\Le$ is the empty set, $\chi_{\rm L}=\varnothing$, which can be taken to lie on any $\hat{\Sigma}_{\rm L}$, for the same reasons explained in \Sect{twoscreens} for the two-screen system. Therefore, the classical contribution from region L to the von Neumman entropy of the screen $\S_\Le$ is zero. Since $\hat{\C}_{\rm L}=\hat{\Sigma}_{\rm L}$, the part of the entanglement wedge in region~L is the common causal diamond of the $\hat{\Sigma}_{\rm L}$'s, corresponding to the blue triangular domain in \Fig{Penrose_gamma<-1}.%\footnote{The condition $\partial\hat{\mathcal{C}}_\Le = \chi_\Le \cup A_\Le$ reduces to $\partial \hat \Sigma_\Le=\varnothing\cup \S_\Le$.}
The classical contribution to the entropy from the antipode interior region R is also null. Indeed, since $A_{\rm R}=\varnothing$, we obtain $\chi_{\rm R}=\varnothing$ as the minimal extremal homologous surface, which ``lies'' on any Cauchy slice $\hat{\Sigma}_{\rm R}$. Indeed, we have $\hat{\mathcal{C}}_{\rm R}=\varnothing$, as it satisfies trivially the condition $\partial \hat{\mathcal{C}}_{\rm R}=\chi_{\rm R}\cup A_{\rm R}=\varnothing$. Since the causal diamond of the empty set $\hat{\mathcal{C}}_{\rm R}$ is empty, the entanglement wedge of the screen $\S_\Le$ does not extend in the antipode interior region R. These conclusions hold regardless whether the screens lie in the expanding or the contracting phase of the cosmology.

The classical contribution to the von Neumann entropy of the single-screen system arises solely from the exterior region E. Let us first discuss the case where the screens are located in the expanding phase of the cosmology and thus lying on the two apparent horizons. %The upper triangular region between the apparent horizons is a region of anti-trapped spheres. 
The Cauchy slices $\hat{\Sigma}_{\rm E}$, which are bounded by $\S_\Le\cup \S_\Ri$, span a common causal diamond corresponding to the red region in Figure~\ref{Penrose_gamma<-1}. Since $A_{\rm E}=\S_\Le$ has no boundary, any homologous surface on a Cauchy slice $\hat{\Sigma}_{\rm E}$ must be closed (see Footnote~\ref{anch}, where we have $\partial \chi_\Ex=\partial A_\Ex=\varnothing$). We will look for minimal extremal homologous surfaces in the causal diamond in region~E that are ${\rm SO}(n)$-symmetric. These are S$^{n-1}$ spheres represented by points on the Penrose diagram, with areas depending on their positions. As remarked in \Sect{bilayer}, since the causal diamond in region~E has a boundary, the extremization of the area functional involves Lagrange multipliers and auxiliary fields, in order to impose all homologous extremal surfaces to lie in this diamond, including its boundary~\cite{Franken:2023pni}.\footnote{It would be interesting to obtain direct evidence for the bilayer proposal and in particular the constrained extremization problem from a replica bulk path-integral approach. This would involve Lorentzian Schwinger-Keldysh path integrals. See \eg  \Refs{Penington:2019kki,Almheiri:2019qdq} for the black hole cases.}

In Appendices~\ref{B3} and \ref{B2}, it is shown that all ${\rm SO}(n)$-symmetric solutions to the constrained-extremization problem are the spheres whose areas are minimal, maximal or saddle points. Moreover, these spheres turn out to lie on the boundary of the causal diamond in region~E. No sphere of extremal surface exists in the interior of this diamond. The upper vertex of the causal diamond is the unique sphere of maximal area, when the diamond has a rectangular shape, as in \Fig{extremaBL}. On the other hand, when $x_{\rm R}^++x_{\rm L}^->|\gamma|\pi$, the causal diamond acquires a fifth edge along future null infinity, as in \Fig{extremaBR}. Since all spheres along this edge have infinite areas, the latter are maximal. For $\gamma < -1$, there is a unique sphere of minimal area corresponding to the lower vertex of the causal diamond. It is depicted by the red dot $M$ in \Fig{Penrose_gamma<-1}, and also in \Figs{extremaBL} and~\ref{extremaBR}. $M$ is indeed the anti-trapped sphere with the smallest area, as can be easily inferred from the Bousso wedges in the upper triangular region (see \Appendix{B2}). Finally, the areas of the spheres on which $\S_\Le$ and $\S_\Ri$ are located are saddle points. %The maximal and minimal area spheres, corresponding to the upper and lower vertices of the causal diamond respectively, and the two saddle points, corresponding to $\S_\Le$ and $\S_\Ri$, are solutions of the constrained extremization problem, which includes Lagrange multipliers. 
Notice that the sphere $M$ of minimal area coincides with the bifurcate horizon only when $\eta_\Le=\eta_\Ri=|\gamma| \pi/2$. As shown in \Appendix{B1}, the bifurcate horizon is also a solution of the {\it unconstrained}-extremization problem for arbitrary $\eta_\Le$, $\eta_\Ri$. However, the latter lies outside the causal diamond in region~E, except in the particular case where $\eta_\Le=\eta_\Ri=|\gamma| \pi/2$.%\footnote{What is meant by this is that the variation of the area of the sphere located in the diamond vanishes as its position varies infinitesimally in the diamond, $\delta\A=0$.}

We conclude that in the expanding phase of the bouncing cosmologies with $\gamma < -1$, the minimal extremal surface $\chi_{\rm ext}$ is the lower vertex of the causal diamond $M$. As a result, $\hat{\mathcal{C}}_\Ex$ is the lightlike slice joining $\S_\Le$ to $M$, which we denote $\hat \Sigma_{\S_\Le\mbox{\scriptsize -} M}$.\footnote{The condition $\partial\hat{\mathcal{C}}_\Ex = \chi_\Ex \cup \S_\Le$ reduces to $\partial \hat \Sigma_{\S_\Le\mbox{\scriptsize -} M}=M\cup \S_\Le$}. The entanglement wedge of $\S_\Le$ then extends in the exterior causal diamond region, acquiring a component consisting of $\hat \Sigma_{\S_\Le\mbox{\scriptsize -} M}$ which is its own causal diamond. Indeed, since $\S_\Le$ and $M$ are lightlike separated, there is a unique $\hat{\C}_{\rm E}$. Since this Cauchy slice is lightlike, its causal diamond consists only of itself.  The conformal coordinates of $M$ are given by
\begin{equation}
    (\theta_M,\eta_M)=\left(\frac{|\gamma|-1}{|\gamma|}\frac{\eta_{\rm L}-\eta_{\rm R}}{2}+\frac{\pi}{2}~,~ \frac{|\gamma|-1}{|\gamma|}\frac{\eta_{\rm L}+\eta_{\rm R}}{2}+\frac{\pi}{2}\right),\,\,\,\, \eta_{\rm L}, \eta_{\rm, R}\geq |\gamma|\pi/2,
\end{equation}
and the minimal extremal area for $\eta_{\rm L}, \eta_{\rm, R}\geq |\gamma|\pi/2$ is equal to
\begin{equation}
    \A_{\rm {min}}(M) = \omega_{n-1}a_0^{n-1}\sin^{\gamma(n-1)}\left(\frac{|\gamma|-1}{|\gamma|^2}\frac{\eta_{\rm L}+\eta_{\rm R}}{2}+\frac{\pi}{2|\gamma|}\right)\cos^{n-1}\left(\frac{|\gamma|-1}{|\gamma|}\frac{\eta_{\rm L}-\eta_{\rm R}}{2}\right). 
\end{equation}
Since the interior contributions vanish at the classical level, we deduce that in the expanding phase, the leading classical entropy of the single-screen subsystem $\S_{\Le}$ is given as a function of $\eta_{\Le}$ and $\eta_\Ri$ by
\begin{equation}\label{sexpanding}
    S_{\S_{\Le}}(\eta_{\rm L},\eta_{\rm R}) = \frac{\A_{\rm {min}}(M)}{4G\hbar}+\mathcal{O}((G\hbar)^0).
\end{equation}

The entropy increases as $\eta_{\rm L}$ and $\eta_{\rm R}$ increase, and eventually saturates a finite upper bound, as $\eta_{\rm L}, \eta_{\rm R} \to |\gamma|\pi$:
\begin{equation}
    S_{\S_\Le}(\eta_{\rm L}=\eta_{\rm R}=|\gamma|\pi) = \frac{1}{4G\hbar}\omega_{n-1}a_0^{n-1}\sin^{\gamma(n-1)}\left(\frac{\pi}{2|\gamma|}\right)+\mathcal{O}((G\hbar)^0).
\end{equation}
Notice that this result holds despite the growing to infinity of the area of the screen $\S_\Le$. At latter times, the entanglement entropy is a small fraction of the maximal possible value. The extra degrees of freedom added to the screens as these evolve along the apparent horizons remain disentangled. This hints towards the fact that the effective bridge manifesting the entanglement between the two screen system is realized by the Cauchy slice with the smallest possible bottleneck, via the sphere $M$. 

%A similar behavior will be observed in the singular cases $\gamma \geq 1$, which however will lead to a decreasing and eventually vanishing entropy for the single-screen subsystem in sharp contrast with the bouncing cosmologies, as we will see in the following section.

Finally, let us discuss the single screen system $\S_\Le$, when the screens are in the contracting phase of the cosmology, $\eta_{\rm L}, \eta_{\rm R} < |\gamma|\pi/2$. In this case, the screens can be pushed farther into the lower triangular region, which is a region of trapped spheres. When $(x^+_{\rm L}+ x^-_{\rm R})/2 \ge 0$, the exterior causal diamond region is a rectangle, as shown in Figure~\ref{extremaBL2}. On the other hand, when $(x^+_{\rm L}+ x^-_{\rm R})/2 < 0$, it has a fifth edge along past lightlike infinity, as in Figure~\ref{extremaBR2}. In both cases, there is a unique minimal extremal surface, homologous to $\S_\Le$, corresponding to the upper vertex of the causal diamond, which is depicted by the red dot $M$ in Figures~\ref{extremaBL2} and \ref{extremaBR2}. This is shown in Appendices~\ref{B2} and \ref{B3}. The entanglement wedge acquires a component in the exterior region E, comprising the lightlike segment joining $\S_\Le$ and $M$, $\hat \Sigma_{\S_\Le\mbox{\scriptsize -} M}$. The lightcone coordinates of $M$ can be expressed in terms of the lightcone coordinates of the screens as follows:
\begin{equation}
(x_M^+,x_M^-)=(x_{\rm R}^+, x_{\rm L}^-).
\end{equation}
The minimal area is now given by
\begin{equation}
    \A_{\rm {min}}(M) = \omega_{n-1}a_0^{n-1}\sin^{\gamma(n-1)}\left(\frac{x_\Ri^++x_\Le^-}{2|\gamma|}\right)\sin^{n-1}\left(\frac{x_\Ri^+-x_\Le^-}{2}\right),
\end{equation}
and the leading geometrical entropy satisfies Eq.~\ref{sexpanding}, where now $\eta_{\rm L}, \eta_{\rm R} \leq \pi |\gamma|/2$.
The entropy increases as the screens approach the lower corners of the Penrose diagram as $\eta_{\rm L}, \eta_{\rm R} \to 0$, saturating an upper bound:
\begin{equation}
    S_{\S_\Le}(\eta_{\rm L}=\eta_{\rm R}=0) = \frac{1}{4G\hbar}\omega_{n-1}a_0^{n-1}\sin^{\gamma(n-1)}\left(\frac{\pi}{2|\gamma|}\right)+\mathcal{O}((G\hbar)^0).
\end{equation}
This maximal value of the fine grained entropy in the contracting phase is equal to the maximal value achieved in the expanding phase, as the screens approach future null infinity.

It can be easily seen that the conclusions above apply to the second screen subsystem $\S_{\Ri}$. The leading geometrical entropy of $\S_{\Ri}$ receives contributions from the exterior region only, with the corresponding minimal extremal homologous surface being the sphere $M$. Therefore, the leading geometrical entropy is equal to that of $\S_{\Le}$:
\begin{equation}\label{sRexpanding}
    S_{\S_{\Ri}}(\eta_{\rm L},\eta_{\rm R}) = \frac{\A_{\rm {min}}(M)}{4G\hbar}+\mathcal{O}((G\hbar)^0).
\end{equation}
As we have argued in subsection~\ref{twoscreens}, the two screen system is in a pure state, when the bulk state on $\Sigma$ is taken to be pure. Therefore, in this case the entropies of the two screens have to be the same at all orders in $G\hbar$.

We conclude this section by comparing with the de Sitter case $\gamma = -1$, analyzed in \cite{Franken:2023pni}. In the de Sitter case there is a degeneracy of minimal area spheres, which also solve the constrained-extremization problem at the classical level. Indeed, for $\gamma=-1$, the apparent horizons become lightlike and coincide with the cosmological horizons delimiting the pode and antipode causal patches. The screens necessarily lie on the cosmological horizons. The segments joining the screens with the bifurcate horizon belong to the boundary of the exterior causal diamond region. All points along these segments correspond to degenerate minimal extremal spheres. As argued in \cite{Franken:2023pni}, the classical degeneracy can be lifted by quantum corrections. The sphere of smallest quantum area, which minimizes the generalized entropy to first order in the quantum corrections, is the screen for which $|\eta - \pi/2|$ is bigger. Here we see that this degeneracy can be also lifted by classical perturbations on the geometry. If we perturb the geometry so that $\gamma < -1$, we find a unique minimal extremal surface controlling the leading geometrical contributions to the entanglement entropy between the screens.

\subsection{The single-screen system in Big Bang/Big crunch cosmologies with \bm $\gamma \geq 1$}
\label{singlescreensingular}

Next, we consider the single-screen subsystem $A=\S_\Le$ in Big Bang/Big Crunch cosmologies with $\gamma\geq 1$. This class includes the matter dominated cases with $w=0$, $\gamma=2/(n-2)$, for $n=3$, $n=4$, as well as the radiation dominated cases, $w=1/n$, $\gamma=2/(n-1)$, for $n=2$, $n=3$. The analysis for the interior regions is completely analogous to the bouncing cases presented in subsection~\ref{singlescreenbouncing}. The minimal extremal surfaces homologous to $A_\Le=\S_\Le$ and $A_\Ri=\varnothing$ are the empty sets, $\chi_\Le=\varnothing$ and $\chi_\Ri=\varnothing$, respectively. Therefore, the classical contributions to the entanglement entropy of $\S_\Le$ with its complement from the interior regions L and R are vanishing. The entanglement wedge has no component in region R, while it contains the full causal diamond region of the $\hat{\Sigma}_{\Le}$'s in the pode interior region L, depicted by the blue triangle in Figure~\ref{Penrose_gamma>1}. 
\begin{figure}[h!]
\centering
\begin{subfigure}[t]{0.3\linewidth}
\centering
\begin{tikzpicture}

\path
       +(2,4)  coordinate (IItopright)
       +(-2,4) coordinate (IItopleft)
       +(2,-4) coordinate (IIbotright)
       +(-2,-4) coordinate(IIbotleft)

       +(-0.7,2) coordinate (SL)
       +(1.1,2.7) coordinate (SR)
      
       ;

\fill[fill=red!20] (SL) -- (0.55,3.25) -- (SR) -- (-0.15,1.45) -- (SL);

\fill[fill=blue!50] (-2,3.3) -- (SL) -- (-2,0.7) --  cycle;

\fill[fill=blue!50] (-0.7,2-0.05) -- (-0.7,2+0.05) -- (0.55,3.25+0.05) -- (0.55,3.25-0.05) -- cycle;

\draw[dashed,gray] (IItopleft) -- (2/3,4/3);
\draw[dashed,gray] (IItopright) -- (-2/3,4/3);

%\draw[dashed,gray] (IIbotleft) -- (2/3,-4/3);
%\draw[dashed,gray] (IIbotright) -- (-2/3,-4/3);
       
\draw[decorate,decoration=snake] (IItopleft) --
          node[midway, above, sloped]    {}
      (IItopright);
      
\draw (IItopright) --
          node[midway, above, sloped] {}
      (IIbotright);
      
\draw[decorate,decoration=snake]  (IIbotright) -- 
          node[midway, below, sloped] {}
      (IIbotleft);
      
\draw (IIbotleft) --
          node[midway, above , sloped] {}
      (IItopleft);
      
\draw (IItopleft) -- (IIbotright)
              (IItopright) -- (IIbotleft) ;

%\draw (-0.2,3.2+0.25) -- (0,3+0.25) -- (0.2,3.2+0.25);

\node at (SL) [circle,fill,inner sep=1.5pt]{};
\node at (-0.77,2) [label = above:$\S_\Le$]{};
\node at (SR) [circle,fill,inner sep=1.5pt]{};
\node at (1.15,2.7) [label = right:$\S_\Ri$]{};
\node at (0.55,3.25) [circle,fill,inner sep=1.5pt, red, label = above:$M$]{};
\node at (0,4) [label = above:$\phantom{M}$]{};

\end{tikzpicture}
\caption{\footnotesize At early times, the causal diamond in E is a rectangle with $M$ being the top vertex. The area of $M$ is non-zero and the classical contributions to the entropy of $\S_\Le$ are non-vanishing. \label{PL}}
\end{subfigure}\hfill
\begin{subfigure}[t]{0.3\linewidth}
\centering
\begin{tikzpicture}

\path
       +(2,4)  coordinate (IItopright)
       +(-2,4) coordinate (IItopleft)
       +(2,-4) coordinate (IIbotright)
       +(-2,-4) coordinate(IIbotleft)

       +(-1.1,2.6) coordinate (SL)
       +(1.3,3) coordinate (SR)
      
       ;

\fill[fill=red!20] (SL) -- (0.3,4) -- (SR) -- (-0.1,1.6) -- (SL);

\fill[fill=blue!50] (-2,3.5) -- (SL) -- (-2,1.7) --  cycle;

\fill[fill=blue!50] (-1.1+0.05,2.6) -- (0.3+0.05,4) -- (0.3-0.05,4) -- (-1.1-0.05,2.6) -- cycle;

\draw[dashed,gray] (IItopleft) -- (2/3,4/3);
\draw[dashed,gray] (IItopright) -- (-2/3,4/3);

%\draw[dashed,gray] (IIbotleft) -- (2/3,-4/3);
%\draw[dashed,gray] (IIbotright) -- (-2/3,-4/3);
       
\draw[decorate,decoration=snake] (IItopleft) --
          node[midway, above, sloped]    {}
      (IItopright);
      
\draw (IItopright) --
          node[midway, above, sloped] {}
      (IIbotright);
      
\draw[decorate,decoration=snake]  (IIbotright) -- 
          node[midway, below, sloped] {}
      (IIbotleft);
      
\draw (IIbotleft) --
          node[midway, above , sloped] {}
      (IItopleft);
      
\draw (IItopleft) -- (IIbotright)
              (IItopright) -- (IIbotleft) ;

%\draw (-0.2,3.2+0.25) -- (0,3+0.25) -- (0.2,3.2+0.25);

\node at (SL) [circle,fill,inner sep=1.5pt]{};
\node at (-1.17,2.6) [label = above:$\S_\Le$]{};
\node at (SR) [circle,fill,inner sep=1.5pt]{};
\node at (1.15,2.7) [label = right:$\S_\Ri$]{};
\node at (0.3,4) [circle,fill,inner sep=1.5pt, red, label = above:$M$]{};

\end{tikzpicture}
\caption{\footnotesize When $(x^+_\Ri+x^-_\Le)/2= \gamma\pi$, $M$ hits the Big Crunch singularity. The area of $M$ is zero and the classical contributions to the entropy of $\S_\Le$ vanish. \label{PC}}
\end{subfigure}\hfill
\begin{subfigure}[t]{0.3\linewidth}
\centering
\begin{tikzpicture}

\path
       +(2,4)  coordinate (IItopright)
       +(-2,4) coordinate (IItopleft)
       +(2,-4) coordinate (IIbotright)
       +(-2,-4) coordinate(IIbotleft)

       +(-1.2,3) coordinate (SL)
       +(1.5,3.3) coordinate (SR)
      
       ;

\fill[fill=red!20] (SL) -- (-0.2,4) -- (0.8,4) -- (SR) -- (0,1.8) -- (SL);

\fill[fill=blue!50] (-2,3.8) -- (SL) -- (-2,2.2) --  cycle;

\fill[fill=blue!50] (-1.2,3-0.05) -- (-1.2,3+0.05) -- (-0.2,4+0.05) -- (-0.2,4-0.05) -- cycle;

\draw[dashed,gray] (IItopleft) -- (2/3,4/3);
\draw[dashed,gray] (IItopright) -- (-2/3,4/3);

%\draw[dashed,gray] (IIbotleft) -- (2/3,-4/3);
%\draw[dashed,gray] (IIbotright) -- (-2/3,-4/3);

\draw[decorate,decoration=snake] (IItopleft) --
          node[midway, above, sloped]    {}
      (IItopright);
      
\draw (IItopright) --
          node[midway, above, sloped] {}
      (IIbotright);
      
\draw[decorate,decoration=snake]  (IIbotright) -- 
          node[midway, below, sloped] {}
      (IIbotleft);
      
\draw (IIbotleft) --
          node[midway, above , sloped] {}
      (IItopleft);
      
\draw (IItopleft) -- (IIbotright)
              (IItopright) -- (IIbotleft) ;

\node at (SL) [circle,fill,inner sep=1.5pt]{};
\node at (-1.27,3) [label = above:$\S_\Le$]{};
\node at (SR) [circle,fill,inner sep=1.5pt]{};
\node at (1.75,3.4) [label = below:$\S_\Ri$]{};
\node at (-0.2,4) [circle,fill,inner sep=1.5pt, red, label = above:$M$]{};

\end{tikzpicture}
\caption{\footnotesize The entropy of $\S_\Le$ remains small of order $(G\hbar)^0$ when $(x^+_\Ri+x^-_\Le)/2> \gamma\pi$, since $M$ is at the Big Crunch singularity. \label{PR}}
\end{subfigure}\hfill
    \caption{\footnotesize Closed FRW cosmologies for $\gamma>1$. The two screens are denoted by the black dots $\S_\Le$ and $\S_\Ri$. We consider the subsystem associated with the single screen $\S_\Le$. The light red shaded region is the causal diamond in the exterior region E, and the dark blue region corresponds to the entanglement wedge of $\S_\Le$. The minimal extremal surface is denoted by the red dot $M$.}
    \label{Penrose_gamma>1}
\end{figure}

The classical geometrical contribution to the entropy of $\S_\Le$ arises from the exterior region between the screens. It is given in terms of the area of a minimal extremal surface, which lies on the boundary of the causal diamond in the exterior region E, spanned by the $\hat\Sigma_\Ex$'s, as explained in Appendices~\ref{B2} and \ref{B3}. 

We will discuss in great detail the contracting phase of the cosmologies and, so, the screen conformal times satisfy $\eta_\Le, \eta_\Ri \geq \gamma\pi/2$. In the contracting phase, the screens can be pushed farther into the upper triangular region, following timelike (and possibly locally lightlike) trajectories, as explained in \Sect{holo_proposal} and illustrated in Figure~\ref{lerB}. The upper triangular region is a region of trapped spheres. For $(x^+_\Ri+x^-_\Le)/2 < \gamma\pi$, the causal diamond in region E is a rectangle with its upper vertex lying below the Big Crunch singularity, as shown in Figure~\ref{PL}. The upper vertex of the rectangle hits the Big Crunch singularity precisely when $(x^+_\Ri+x^-_\Le)/2 = \gamma\pi$ -- see Figure~\ref{PC}. As we show in Appendices~\ref{B2} and \ref{B3}, the minimal extremal surface $\chi_{\rm E}$, homologous to $\S_\Le$, is precisely this top vertex of the causal diamond. We depict this by the red dot $M$ in Figure~\ref{Penrose_gamma>1}. $M$ is indeed the trapped sphere with the smallest area, as can be easily inferred from the structure of the Bousso wedges in the upper triangular region. In addition, it solves the constrained-extremization problem, as demonstrated explicitly in Appendix~\ref{B3}. In terms of the screen lightcone coordinates, the lightcone coordinates of $M$ are given by
\begin{equation}
\label{eta_M_position}
    \left(x_M^+,x_M^-\right)=\left(x_\Ri^+,x_\Le^-\right)
\end{equation}
and the minimal area by
\begin{equation}
    \A_{min}(M) = \omega_{n-1}a_0^{n-1}\sin^{\gamma(n-1)}\left(\frac{x_\Ri^++x_\Le^-}{2\gamma}\right)\sin^{n-1}\left(\frac{x_\Ri^+-x_\Le^-}{2}\right).
\end{equation}
The entanglement wedge acquires a component in the exterior region, comprising of the lightlike segment $\hat \Sigma_{\S_\Le\mbox{\scriptsize -} M}$ joining $\S_\Le$ and $M$ -- see Figure~\ref{PL}. Notice that when $\eta_\Le=\eta_\Ri=\eta_c=\pi|\gamma|/2$, the screens coincide at the bifurcation point of the apparent horizons. The exterior causal diamond becomes trivial, comprising the bifurcation point only. The bifurcate horizon is an extremal of the unconstrained area functional, as shown in Appendix~\ref{B1}. Its area is equal to $\omega_{n-1}a_0^{n-1}$ yielding non-trivial classical contributions to the entropy of $\S_\Le$. 

As the two screens follow their trajectories towards the Big Crunch singularity, the top vertex $M$ of the causal diamond in E approaches the Big Crunch singularity and its area decreases. Consequently the entropy of $\S_\Le$ decreases. $M$ hits the future singularity at $(x^+_\Ri+x^-_\Le)/2= \gamma\pi$ -- see Figure~\ref{PC}. Precisely at this time, the area of $M$ vanishes, so that the classical  contributions to the entropy from the exterior region vanish. 

When $(x^+_\Ri+x^-_\Le)/2>\gamma\pi$, the exterior causal diamond acquires a fifth edge along the Big Crunch singularity slice. Along this edge, the coordinates satisfy $\gamma\pi+\theta_\Le-\eta_\Le\leq\theta\leq -\gamma\pi+\theta_\Ri+\eta_\Ri$ (and $\eta=\gamma\pi$). All spheres on this segment are extremal and of vanishing area. So they yield zero classical contribution to the entropy. To resolve this degeneracy, we pick the left endpoint of this segment so as to minimize the extent of the entanglement wedge in the exterior region \cite{Shaghoulian:2021cef}, as in Figure~\ref{PR}. As we discussed in the case of the entanglement wedge of a single screen system in the de Sitter case \cite{Franken:2023pni}, it is not true that minimizing the extent of the entanglement wedge is always the solution. The correct choice depends on the structure of the quantum corrections. However, we can consider this choice to give a lower bound on the extent of the entanglement wedge since all other choices of extremal surfaces lead to an entanglement wedge that includes this minimal choice.

Since the classical contributions from the interior region are always zero, the bilayer proposal yields for the entropy of the single-screen system $\S_\Le$:
\begin{equation}
\label{entropy_FRW_alpha>0}
    S_{\S_\Le} =   \left\{\begin{array}{cc}
        \frac{\A_{min}(M)}{4G\hbar} + \mathcal{O}((G\hbar)^0) \quad &, (x^+_\Ri+x^-_\Le)/2<\gamma\pi  \\
           \mathcal{O}((G\hbar)^0) \quad &, (x^+_\Ri+x^-_\Le)/2\geq\gamma\pi
    \end{array}\right. .
\end{equation}
Notice that the classical entropy is always smaller or equal to the area of the screen divided by $4G\hbar$ -- the two are equal only at $\eta_\Le=\eta_\Ri=\gamma\pi/2$, when the screens coincide at the bifurcation point. This is a consistency check for the bilayer proposal since the area of the screen divided by $4G\hbar$ sets an upper bound for the von Neumann entropy of the density matrix describing the state of the single screen system.

Therefore, when $(x^+_\Ri+x^-_\Le)/2 < \gamma\pi$, the entanglement entropy is significantly large, of order $(G\hbar)^{-1}$, and decreasing. At $(x^+_\Ri+x^-_\Le)/2= \gamma\pi$ the minimal extremal surface $M$ hits the Big Crunch singularity. As a result, the part of the entanglement wedge of $\S_\Le$ in E is pinched to the singularity. The effective bridge connecting the screens closes off, leading to disentanglement (or a significant decrease in the entanglement entropy), well before the screens reach the Big Crunch singularity.  The entanglement entropy becomes small, of order $(G\hbar)^0$, even if the Universe has not yet collapsed.

The disentanglement phenomenon cannot occur when the two screens are in causal contact, namely when they lie in the overlap region of the pode and antipode causal patches. It is easy to see that in this case $M$ has to lie below the intersection of the event horizons delimiting the pode and antipode causal patches.

The leading classical contributions to the entropy remain vanishing for the rest of the evolution of the screens along their trajectories. So the entropy at latter times is dominated by the semiclassical contributions.
%\footnote{The singular contributions to the semiclassical entanglement entropy are absorbed via renormalization in the geometrical terms, see e.g. \cite{Almheiri:2019hni}.} 
Notice that the slices connecting the screen with the minimal extremal surfaces reach up to the Big Crunch singularity. So quantum corrections to the bulk theory effective action will be important for the determination of the full generalized entropy, including quantum corrections, but we expect the entropy at these latter times to remain much smaller as compared to the value of order $(G\hbar)^{-1}$ at earlier times. Indeed the number of degrees of freedom on the screens decreases as these evolve towards the Big Crunch singularity and the fine grained entropy must be bounded by this. 
Unlike the bouncing cases, the entanglement entropy of $\S_{\Le}$ with its complement becomes very small (of order $(G\hbar)^0$) well before the screens reach the Big Crunch singularity. 

The situation for the right screen $\S_\Ri$ is completely analogous. In particular, the classical entropy of $\S_\Ri$ is equal to that of $\S_\Le$ since this is controlled by the same minimal extremal sphere $M$. Notice also, that assuming the quantum extremal surface is common to the two screens, their entropies, including quantum corrections, will continue being equal assuming the bulk state on $\hat{\Sigma}$ is pure. This is consistent with the fact that the two screen system is in a pure state in this case, as argued in subsection~\ref{twoscreens}. 

Let us now discuss the initial expanding phase of the cosmology, during which $\eta_\Le, \eta_R \le \gamma\pi/2$. In this phase, the screens must lie on the apparent horizons. The region between the apparent horizons is a region of anti-trapped spheres. The exterior causal diamond region is a rectangle for $(x_{\Le}^++x_{\Ri}^-)/2 \ge 0$, and now its lower vertex $M$, the one closer to the Big Bang singularity, is the minimal extremal sphere that leads to non-vanishing contributions to the entropy of $\S_{\Le}$. When $(x_{\Le}^++x_{\Ri}^-)/2=0$, $M$ touches the initial singularity and has vanishing area. On the other hand, when $(x_{\Le}^++x_{\Ri}^-)/2 < 0$, the exterior causal diamond acquires a fifth edge along the Big Bang singularity, with all spheres along this segment being extremal with vanishing area. Therefore, the entropy is initially small of order $(G\hbar)^0$, with the degrees of freedom between the two screens being effectively disentangled. The entanglement entropy is governed by quantum, semiclassical corrections, which are expected to lift the classical degeneracy and determine the precise extent of the entanglement wedge in region E. At some point, the entanglement entropy becomes large, of order $(G\hbar)^{-1}$, when the degrees of freedom on the two screens interact. Then the entropy grows and reaches its maximal value when the screens are at the bifurcate horizon.

\begin{figure}[h!]
\centering
\begin{subfigure}[t]{0.3\linewidth}
\centering
\begin{tikzpicture}

\path
       +(2,2)  coordinate (IItopright)
       +(-2,2) coordinate (IItopleft)
       +(2,-2) coordinate (IIbotright)
       +(-2,-2) coordinate(IIbotleft)
      
       ;
       
\fill[fill=blue!50] (-0.5,0.5) -- (-2,2) -- (-2,-1) --  cycle;       
       
\fill[fill=red!20] (-0.5,0.5) -- (1/2,3/2) -- (1,1) -- (0,0) -- cycle;
       
\draw[decorate,decoration=snake] (IItopleft) --
          node[midway, above, sloped]    {}
      (IItopright);
      
\draw (IItopright) --
          node[midway, above, sloped] {}
      (IIbotright);
      
\draw[decorate,decoration=snake]  (IIbotright) -- 
          node[midway, below, sloped] {}
      (IIbotleft);
      
\draw (IIbotleft) --
          node[midway, above , sloped] {}
      (IItopleft);
      
\draw (IItopleft) -- (IIbotright)
              (IItopright) -- (IIbotleft) ;

\fill[fill=blue!50] (-0.5,0.5-0.05) -- (-0.5,0.5+0.05) -- (1/2,3/2+0.05) -- (1/2,3/2-0.05) -- cycle;

\node at (-0.5,0.5) [circle,fill,inner sep=1.5pt, label = below:$\S_\Le$]{};

\node at (1,1) [circle,fill,inner sep=1.5pt, label = below:$\S_\Ri$]{};
\node at (1/2,3/2) [circle,fill,inner sep=1.5pt, red, label = right:$M$]{};
\end{tikzpicture}
\caption{\footnotesize For $\pi<\eta_\Le+\eta_\Ri<3\pi/2$, the causal diamond region is a rectangle with $M$ being the top vertex.}
\end{subfigure}\hfill
\begin{subfigure}[t]{0.3\linewidth}
\centering
\begin{tikzpicture}

\path
       +(2,2)  coordinate (IItopright)
       +(-2,2) coordinate (IItopleft)
       +(2,-2) coordinate (IIbotright)
       +(-2,-2) coordinate(IIbotleft)
      
       ;
       
\fill[fill=blue!50] (-3/4,3/4) -- (-2,2) -- (-2,-1/2) --  cycle;       
       
\fill[fill=red!20] (-3/4,3/4) -- (1/2,2) -- (5/4,5/4) -- (0,0) -- cycle;
       
\draw[decorate,decoration=snake] (IItopleft) --
          node[midway, above, sloped]    {}
      (IItopright);
      
\draw (IItopright) --
          node[midway, above, sloped] {}
      (IIbotright);
      
\draw[decorate,decoration=snake]  (IIbotright) -- 
          node[midway, below, sloped] {}
      (IIbotleft);
      
\draw (IIbotleft) --
          node[midway, above , sloped] {}
      (IItopleft);
      
\draw (IItopleft) -- (IIbotright)
              (IItopright) -- (IIbotleft) ;

\fill[fill=blue!50] (-3/4,3/4-0.05) -- (-3/4,3/4+0.05) -- (1/2,2+0.05) -- (1/2,2-0.05) -- cycle;

\node at (-3/4,3/4) [circle,fill,inner sep=1.5pt, label = below:$\S_\Le$]{};

\node at (5/4,5/4) [circle,fill,inner sep=1.5pt, label = below:$\S_\Ri$]{};
\node at (1/2,2) [circle,fill,inner sep=1.5pt, red, label = above:$M$]{};

%\draw[dashed] (-2,5/4) -- (2,5/4);
%\node at (2,5/4) [label = right: $\eta_D$]{};

\end{tikzpicture}
\caption{\footnotesize For $\eta_\Le+\eta_\Ri = 3\pi/2$, $M$ hits the Big Crunch singularity and has vanishing area.}
\end{subfigure}\hfill
\begin{subfigure}[t]{0.3\linewidth}
\centering
\begin{tikzpicture}

\path
       +(2,2)  coordinate (IItopright)
       +(-2,2) coordinate (IItopleft)
       +(2,-2) coordinate (IIbotright)
       +(-2,-2) coordinate(IIbotleft)
      
       ;

\fill[fill=blue!50] (-2,2) -- (-1,1) -- (-2,0) -- cycle;     
       
\fill[fill=red!20] (-1,1) -- (0,2) -- (1,2) -- (1.5,1.5) -- (0,0) -- cycle;
       
\draw[decorate,decoration=snake] (IItopleft) --
          node[midway, above, sloped]    {}
      (IItopright);
      
\draw (IItopright) --
          node[midway, above, sloped] {}
      (IIbotright);
      
\draw[decorate,decoration=snake]  (IIbotright) -- 
          node[midway, below, sloped] {}
      (IIbotleft);
      
\draw (IIbotleft) --
          node[midway, above , sloped] {}
      (IItopleft);
      
\draw (IItopleft) -- (IIbotright)
              (IItopright) -- (IIbotleft) ;

\fill[fill=blue!50] (-1,1-0.05) -- (-1,1+0.05) -- (0,2+0.05) -- (0,2-0.05) -- cycle;

\node at (-1,1) [circle,fill,inner sep=1.5pt, label = below:$\S_\Le$]{};
\node at (1.5,1.5) [circle,fill,inner sep=1.5pt, label = below:$\S_\Ri$]{};
\node at (0,2) [circle,fill,inner sep=1.5pt, red, label = above:$M$]{};
\end{tikzpicture}
\caption{\footnotesize For $\eta_\Le+\eta_\Ri\geq3\pi/2$, $M$ remains at the Big Crunch singularity.%, so that the classical geometrical contributions to the entropy of $\S_\Le$ remains zero. 
}
\end{subfigure}\hfill
    \caption{\footnotesize Closed FRW cosmology for $\gamma=1$. The two screens at the cosmological horizons are denoted by the black dots $\S_\Le$ and $\S_\Ri$. The light red shaded region is the causal diamond in the exterior region E. The sphere with the minimal extremal area is denoted by the red dot $M$. The dark blue region is the entanglement wedge associated with $\S_\Le$.}
    \label{Penrose_gamma=1}
\end{figure}
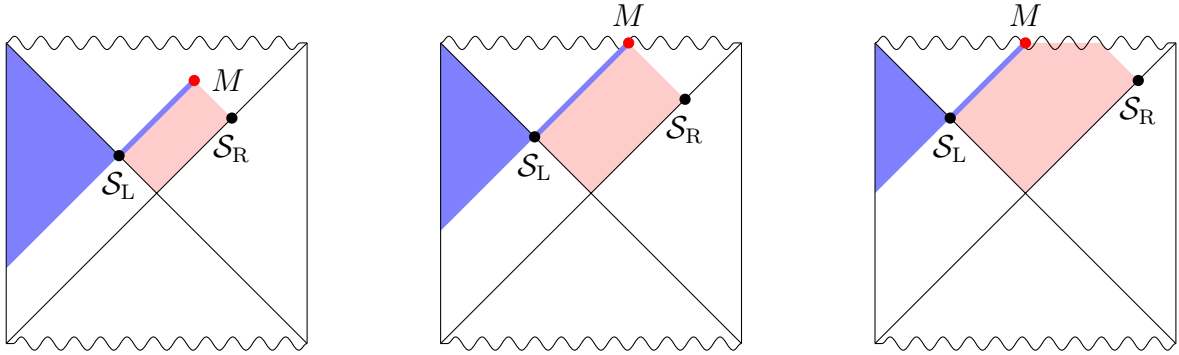

Finally, let us briefly discuss the particular case $\gamma=1$, for which the Penrose diagram is a square. This is the marginal case for the two distinct classes of Big Bang/Big Crunch cosmologies. As in the de Sitter case $\gamma=-1$, the apparent horizons are lightlike and coincide with the cosmological horizons. The screens must be located on the cosmological horizons, in order to remain in the causal patches of the pode and antipode, respectively. The two screens at the horizons remain out of causal contact during the contracting phase, up to the Big Crunch singularity. A sharp difference with the de Sitter case $\gamma=-1$ is that now the areas of the screens on the horizons are no longer constant but decrease with time, and vanish at the Big Crunch singularity. The time evolution of the single screen subsystem $\S_{\Le}$ and the entanglement wedge structure is illustrated in Figure~\ref{Penrose_gamma=1}.
The behavior in the expanding phase can be obtained via time reversal of the behavior in the contracting phase.

\subsection{Maximin spheres}

Let us consider the contracting phase of the  cosmologies with $\gamma< -1$ and $\gamma \ge 1$. When the causal diamond in the exterior region E is rectangular, its four vertices are the only extremal spheres. Let us denote $\S_{\rm small}$ the screen with the smaller area and $\S_{\rm big}$ the screen with the bigger area among the two. Then $\S_{\rm small}$ is a maximin sphere, obtained by first minimizing the area for all Cauchy slices in the causal diamond in region E, and then picking the maximum among these minima. Maximin surfaces were introduced in a seminal paper by Wall \cite{Wall:2012uf} in the context of holographic entanglement entropy computations, showing that a maximin prescription provides an alternative method to find extremal surfaces. Indeed, consider the limiting Cauchy slice consisting of the union of the two lower lightlike edges of the causal rectangle, which join the screens with the bottom vertex, respectively. From the structure of the Bousso wedges, it follows that, as we move along a lightlike edge from the bottom vertex towards the screen at its end, the area of the spheres decreases. Therefore, $\S_{\rm small}$ must be the minimal area sphere on this limiting Cauchy slice. Since any other Cauchy slice in the causal diamond in E ends on the two screens, the minimal area sphere on it necessarily has smaller or equal area to the area of $\S_{\rm small}$. Therefore, $\S_{\rm small}$ must be a maximin sphere: among the set of minimal area spheres from all Cauchy slices, $\S_{\rm small}$ is one with maximal area. On the other hand, $\S_{\rm big}$ is a minimax sphere. Indeed, consider now the limiting Cauchy slice consisting of the two upper edges of the causal rectangle, which join the screens with the top vertex of the rectangle, respectively. $\S_{\rm big}$ is the maximal area sphere on this limiting slice. Since any other Cauchy slice in the exterior causal diamond ends on the screens, the maximal area sphere on it necessarily has bigger or equal area to the area of $\S_{\rm big}$. Therefore, this screen is a minimax sphere. Likewise, it is easy to see that the top vertex, which is the minimal extremal sphere, is a minimin sphere and the bottom vertex is a maximax sphere.

In the expanding phase of these cosmologies, similar arguments can be applied to conclude that now the top vertex is a maximax sphere and the bottom vertex is minimin sphere. The screen with the bigger conformal time, $\S_{\rm big}$, is a minimax sphere, while the screen at the smaller conformal time, $\S_{\rm small}$, is a maximin sphere. 

\subsection{Equivalence of holographic constructions}
\label{equivconf2}
As we already remarked in subsection~\ref{equivconf}, there are infinitely many possible choices of trajectories for the two screens $\S_\Le$ and $\S_\Ri$ in the contracting phase of the cosmologies with $\gamma<-1$, $\gamma > 1$, leading to different holographic constructions. However, by examining the full two-screen and single screen subsystems, evidence can be obtained supporting the fact that these screen configurations can be grouped in equivalence classes. In each class, the effective holographic theory on the screens can be derived from a pair of ``parent'' screens on the apparent horizons of the pode and antipode, yielding identical predictions for the entropy of certain gravitational bulk systems, which can be reconstructed from the single screen subsystems. In the following, we will assume entanglement wedge reconstruction~\cite{Dong:2016eik} and unitary evolution among the Cauchy slices in the entanglement wedge of a single screen subsystem (or any subsystem in general). Recall also that it is consistent to assume unitary evolution among the Cauchy slices $\hat{\Sigma}$ associated with a given two-screen configuration, and also unitary evolution in the three causal diamonds associated with ${\hat\Sigma}_{\rm L}$, ${\hat\Sigma}_{\rm E}$ and ${\hat\Sigma}_{\rm R}$. Indeed, the states and density matrices on these slices are associated with a holographic quantum system of certain dimensionality (determined by the total area of the screens in Planck units), as well as its subsystems.

To determine the ``parent'' screens, we draw the lightlike segments $\Delta_{\rm L}$ and $\Delta_{\rm R}$, which run parallel to the $x^+$ and $x^-$ axes, respectively, and join the screens $\S_\Le$ and $\S_\Ri$ with the apparent horizons of the pode and antipode, as in Figure~\ref{pushback}. The left ``parent'' screen $\S_{\Le}'$ lies at the intersection of $\Delta_{\rm L}$ with the apparent horizon of the pode. Likewise, the right ``parent'' screen $\S_{\Ri}'$ lies at the intersection of $\Delta_{\rm R}$ with the apparent horizon of the antipode. As mentioned in subsection~\ref{equivconf}, all pairs of screens on $\Delta_{\rm L}$ and $\Delta_{\rm R}$ belong to the same equivalence class, and the effective holographic theory on them can be derived from the theory on $\S_{\Le}'\cup \S_{\Ri}'$ by integrating out degrees of freedom. Indeed, $\S_\Le'$ and $\S_\Ri'$ are the maximal area spheres on $\Delta_{\rm L}$ and $\Delta_{\rm R}$, respectively, as follows from the structure of the Bousso wedges in the region between the apparent horizons in the contracting phase. So, as we propagate from $\S_\Le'$ towards $\S_\Le$ along $\Delta_{\rm L}$, the number of degrees of freedom decreases. However, as we will see below, the von Neumann entropy of the screens does not change, suggesting that the extra degrees of freedom that are integrated out are redundant, as they do not participate in the entanglement between the two screens and are not likely to be necessary for the reconstruction of the dual bulk subsystems. The number of degrees of freedom associated with a minimal extremal trapped sphere suffice to reconstruct the state of these dual bulk subsystems.   

We proceed now to argue that the screen configurations associated with $\Delta_{\rm L}$ and $\Delta_{\rm R}$ lead to equivalent predictions, at least for the bulk subsystems that are dual to the single screen subsystems and emerge holographically. These bulk subsystems are associated with field configurations on Cauchy slices in the entanglement wedge of a single screen. First, we observe that the causal diamonds in E associated with pairs of screens on $\Delta_{\rm L}$ and $\Delta_{\rm R}$ form a nested series.\footnote{The importance of causal diamonds in holography was recently emphasised in \cite{A:2023psv}.} Starting with $\S_\Le'$ and $\S_\Ri'$ and propagating along $\Delta_{\rm L}$ and $\Delta_{\rm R}$, the exterior causal diamond of the new screen configuration is contained in the exterior causal diamond of the previous one. Therefore, the exterior causal diamond associated with $\S_{\Le}$ and $\S_{\Ri}$ is contained in the exterior causal diamond of $\S_\Le'$ and $\S_\Ri'$. The same holds for the exterior causal diamond associated with any pair of screens on $\Delta_{\rm L}$ and $\Delta_{\rm R}$. 

If the initial exterior causal diamond of $\S_\Le'$ and $\S_\Ri'$ is rectangular, all subsequent exterior causal diamonds associated with equivalent screens are also rectangular and share the same top vertex $M$. But, $M$ corresponds to the minimal extremal sphere $\chi_{\rm E}$, homologous to the left (and also the right) screen. This holds irrespective of the choice of the pair of screens on $\Delta_{\rm L}$ and $\Delta_{\rm R}$. The area of $M$ divided by $4G\hbar$ is equal to the leading classical entropy of the left (and right) screen subsystem(s). In addition, $M$ determines the extent of the left screen's entanglement wedge in the exterior region. We see that the fine grained entropies of the left and right screen subsystems at the classical level remain constant as the screens move along $\Delta_L$ and $\Delta_R$ and towards $\S_{\Le}$ and $\S_{\Ri}$, respectively. In particular, the fine grained entropies of $\S_{\Le}$ and the parent screen $\S_{\Le}'$ are equal at the classical level. For any choice of pair of equivalent screens, the entanglement wedge of the left screen subsystem consists of the union of the left triangular causal diamond of the associated $\hat{\Sigma}_{\rm L}$ slices and the lightlike segment connecting the left screen and the top vertex $M$ in the exterior region.

One possibility to explain the constancy of the fine grained entropy as we integrate out degrees of freedom is the following. The Hilbert space of the ``daughter'' screen, whose dimensionality is smaller, should be realizable as a vector subspace of the Hilbert space of the ``parent'' screen. Let us now choose a suitable basis in which the density matrix of the ``parent'' screen is diagonal. The non-trivial block of non-zero eigenvalues should correspond to ket-bra operators associated with the subspace of the ``daughter'' screen and should have a smaller size. This non-trivial block, which yields the non-vanishing contributions to the von Neumann entropy, is nothing but the density matrix of the ``daughter'' screen.  

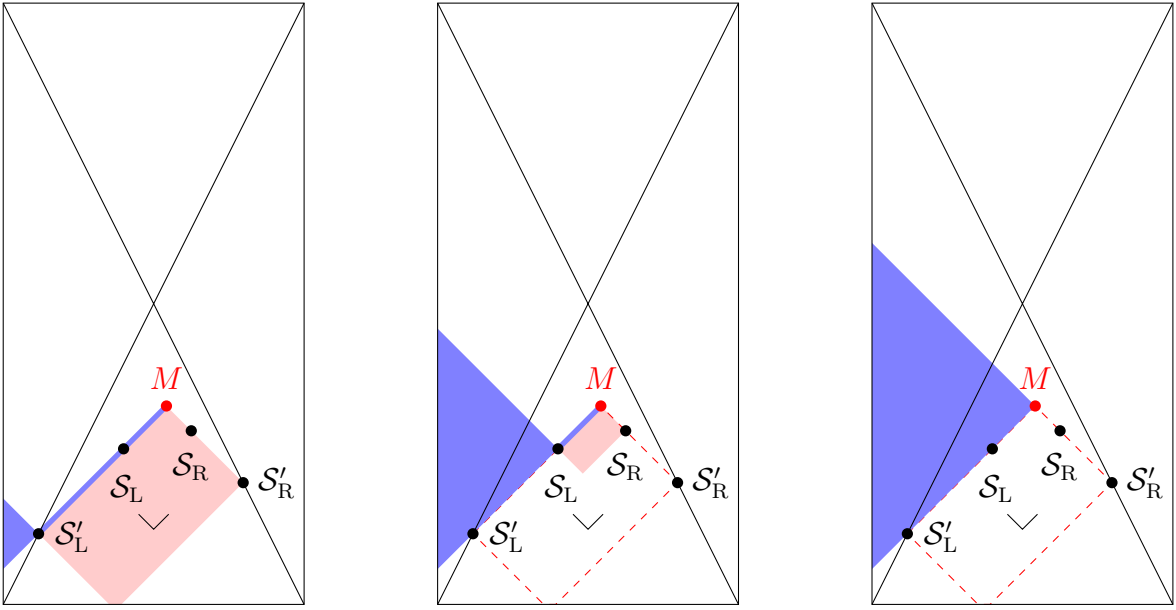
\begin{figure}[h!]
\centering
\begin{subfigure}[t]{0.3\linewidth}
\centering
\begin{tikzpicture}

\path
       +(2,4)  coordinate (IItopright)
       +(-2,4) coordinate (IItopleft)
       +(2,-4) coordinate (IIbotright)
       +(-2,-4) coordinate(IIbotleft)
      
       ;

\fill[fill=red!20] (-1.53,-3.06) -- (0.17,-1.36) -- (1.19,-2.38) -- (-0.43,-4) -- (-0.59,-4);

\fill[fill=blue!50] (-2,-2.59) -- (-1.53,-3.06) -- (-2,-3.53) --  cycle;

\fill[fill=blue!50] (-1.53,-3.06-0.05) -- (-1.53,-3.06+0.05) -- (0.17,-1.36+0.05) -- (0.17,-1.36-0.05) -- cycle;

%\begin{scope}[transparency group]
\begin{scope}[blend mode=multiply]
%\fill[fill=violet!20] (-2,-4) -- (2/3,-4/3) -- (0,0) -- cycle;
%\fill[fill=orange!20] (2,-4) -- (-2/3,-4/3) -- (0,0) -- cycle;
\end{scope}
%\end{scope}
       
\draw (IItopleft) --
          node[midway, above, sloped]    {}
      (IItopright) --
          node[midway, above, sloped] {}
      (IIbotright) -- 
          node[midway, below, sloped] {}
      (IIbotleft) --
          node[midway, above , sloped] {}
      (IItopleft) -- cycle;
      
\draw (IItopleft) -- (IIbotright)
              (IItopright) -- (IIbotleft) ;

%\draw (-2-0.5,-4+0.5) -- (-2,-4);
%\draw (-2+0.5,-4+0.5) -- (-2,-4);
%\draw (-2-0.5,-4+0.3) -- (-2-0.5,-4+0.5) -- node[midway, left, sloped] {$x^-$} (-2-0.3,-4+0.5) ;
%\draw (-2+0.5,-4+0.3) -- (-2+0.5,-4+0.5) -- (-2+0.3,-4+0.5) ;
%\node at (-1.6,-4+0.5) [label = right:$x^+$]{};

\draw (-0.2,-2.8) -- (0,-3) -- (0.2,-2.8);
              
%\draw[dashed,gray] (IIbotleft) -- (2/3,-4/3);
%\draw[dashed,gray] (IIbotright) -- (-2/3,-4/3);

%\draw[domain=-2:-0.4, smooth, variable=\x, line width=0.8mm] plot ({\x}, {sin(deg((\x/2-1)))-1});
%\draw[domain=-0.4:0.5, smooth, variable=\x] plot ({\x}, {sin(deg((\x/2-1)))-1});
%\draw[domain=0.5:2, smooth, variable=\x, line width=0.8mm] plot ({\x}, {sin(deg((\x/2-1)))-1});

%\draw[blue] (-0.4,-1.93) -- (-1.53,-3.06);
\node at (-0.4,-1.93) [circle, fill, inner sep=1.5 pt]{};
\node at (-0.35,-2) [label=below:$\S_\Le$]{};
\node at (-1.53,-3.06) [circle, fill, inner sep=1.5 pt]{};
\node at (-1.6,-3.1) [label=right:$\S_\Le'$]{};
%\node at (-1.2,-2.7) [label=right:$\color{blue} \Delta_\Le$]{};

%\draw[blue] (0.5,-1.69) -- (1.19,-2.38);
\node at (0.5,-1.69) [circle, fill, inner sep=1.5 pt]{};
\node at (0.5,-1.7) [label=below:$\S_\Ri$]{};
\node at (1.19,-2.38) [circle, fill, inner sep=1.5 pt]{};
\node at (1.1,-2.38) [label=right:$\S_\Ri'$]{};
%\node at (0.1,-2.2) [label=right:$\color{blue} \Delta_\Ri$]{};
\node at (0.17,-1.36) [circle, fill, inner sep=1.5 pt, red, label=above:$\color{red} M$]{};

%\node at (-1.4,-1.15) [label=below:$\Sigma_\Le$]{};
%\node at (1.3,-0.5) [label=below:$\Sigma_\Ri$]{};

\end{tikzpicture}
\caption{\footnotesize ``Parent'' screen configuration $\S_\Le'\cup \S_\Ri'$, located on the apparent horizons of the pode and the antipode. \label{entanglementwedgesa}}
\end{subfigure}\hfill
\begin{subfigure}[t]{0.3\linewidth}
\centering
\begin{tikzpicture}

\path
       +(2,4)  coordinate (IItopright)
       +(-2,4) coordinate (IItopleft)
       +(2,-4) coordinate (IIbotright)
       +(-2,-4) coordinate(IIbotleft)
      
       ;

\fill[fill=red!20] (-0.4,-1.93) -- (0.17,-1.36) -- (0.5,-1.69) -- (-0.07,-2.26);
\draw[dashed,red] (-1.53,-3.06) -- (0.17,-1.36) -- (1.19,-2.38) -- (-0.43,-4) -- (-0.59,-4) -- (-1.53,-3.06);

\fill[fill=blue!50] (-2,-0.33) -- (-0.4,-1.93) -- (-2,-3.53) --  cycle;

\fill[fill=blue!50] (-0.4,-1.93-0.05) -- (-0.4,-1.93+0.05) -- (0.17,-1.36+0.05) -- (0.17,-1.36-0.05) -- cycle;

%\begin{scope}[transparency group]
\begin{scope}[blend mode=multiply]
%\fill[fill=violet!20] (-2,-4) -- (2/3,-4/3) -- (0,0) -- cycle;
%\fill[fill=orange!20] (2,-4) -- (-2/3,-4/3) -- (0,0) -- cycle;
\end{scope}
%\end{scope}
       
\draw (IItopleft) --
          node[midway, above, sloped]    {}
      (IItopright) --
          node[midway, above, sloped] {}
      (IIbotright) -- 
          node[midway, below, sloped] {}
      (IIbotleft) --
          node[midway, above , sloped] {}
      (IItopleft) -- cycle;
      
\draw (IItopleft) -- (IIbotright)
              (IItopright) -- (IIbotleft) ;

%\draw (-2-0.5,-4+0.5) -- (-2,-4);
%\draw (-2+0.5,-4+0.5) -- (-2,-4);
%\draw (-2-0.5,-4+0.3) -- (-2-0.5,-4+0.5) -- node[midway, left, sloped] {$x^-$} (-2-0.3,-4+0.5) ;
%\draw (-2+0.5,-4+0.3) -- (-2+0.5,-4+0.5) -- (-2+0.3,-4+0.5) ;
%\node at (-1.6,-4+0.5) [label = right:$x^+$]{};

\draw (-0.2,-2.8) -- (0,-3) -- (0.2,-2.8);
              
%\draw[dashed,gray] (IIbotleft) -- (2/3,-4/3);
%\draw[dashed,gray] (IIbotright) -- (-2/3,-4/3);

%\draw[domain=-2:-0.4, smooth, variable=\x, line width=0.8mm] plot ({\x}, {sin(deg((\x/2-1)))-1});
%\draw[domain=-0.4:0.5, smooth, variable=\x] plot ({\x}, {sin(deg((\x/2-1)))-1});
%\draw[domain=0.5:2, smooth, variable=\x, line width=0.8mm] plot ({\x}, {sin(deg((\x/2-1)))-1});

%\draw[blue] (-0.4,-1.93) -- (-1.53,-3.06);
\node at (-0.4,-1.93) [circle, fill, inner sep=1.5 pt]{};
\node at (-0.35,-2) [label=below:$\S_\Le$]{};
\node at (-1.53,-3.06) [circle, fill, inner sep=1.5 pt]{};
\node at (-1.6,-3.1) [label=right:$\S_\Le'$]{};
%\node at (-1.2,-2.7) [label=right:$\color{blue} \Delta_\Le$]{};

%\draw[blue] (0.5,-1.69) -- (1.19,-2.38);
\node at (0.5,-1.69) [circle, fill, inner sep=1.5 pt]{};
\node at (0.5,-1.7) [label=below:$\S_\Ri$]{};
\node at (1.19,-2.38) [circle, fill, inner sep=1.5 pt]{};
\node at (1.1,-2.38) [label=right:$\S_\Ri'$]{};
%\node at (0.1,-2.2) [label=right:$\color{blue} \Delta_\Ri$]{};
\node at (0.17,-1.36) [circle, fill, inner sep=1.5 pt, red, label=above:$\color{red} M$]{};

%\node at (-0.4,-1) [label=below:$\color{blue!50} \Delta_{\S_{\Le}M}$]{};
%\node at (-1.4,-1.15) [label=below:$\Sigma_\Le$]{};
%\node at (1.3,-0.5) [label=below:$\Sigma_\Ri$]{};

\end{tikzpicture} 
\caption{\footnotesize A ``daughter'' screen configuration $\S_\Le\cup \S_\Ri$, located anywhere along the lightlike segments joining $\S_\Le'$ to $M$ and $\S_\Ri'$ to $M$.\label{entanglementwedgesb}}
\end{subfigure}\hfill
\begin{subfigure}[t]{0.3\linewidth}
\centering
\begin{tikzpicture}

\path
       +(2,4)  coordinate (IItopright)
       +(-2,4) coordinate (IItopleft)
       +(2,-4) coordinate (IIbotright)
       +(-2,-4) coordinate(IIbotleft)
      
       ;

%\fill[fill=red!20] (-0.4,-1.93) -- (0.17,-1.36) -- (0.5,-1.69) -- (-0.07,-2.26);
\draw[dashed,red] (-1.53,-3.06) -- (0.17,-1.36) -- (1.19,-2.38) -- (-0.43,-4) -- (-0.59,-4) -- (-1.53,-3.06);

\fill[fill=blue!50] (-2,0.81) -- (0.17,-1.36) -- (-2,-3.53) --  cycle;

%\begin{scope}[transparency group]
\begin{scope}[blend mode=multiply]
%\fill[fill=violet!20] (-2,-4) -- (2/3,-4/3) -- (0,0) -- cycle;
%\fill[fill=orange!20] (2,-4) -- (-2/3,-4/3) -- (0,0) -- cycle;
\end{scope}
%\end{scope}
       
\draw (IItopleft) --
          node[midway, above, sloped]    {}
      (IItopright) --
          node[midway, above, sloped] {}
      (IIbotright) -- 
          node[midway, below, sloped] {}
      (IIbotleft) --
          node[midway, above , sloped] {}
      (IItopleft) -- cycle;
      
\draw (IItopleft) -- (IIbotright)
              (IItopright) -- (IIbotleft) ;

%\draw (-2-0.5,-4+0.5) -- (-2,-4);
%\draw (-2+0.5,-4+0.5) -- (-2,-4);
%\draw (-2-0.5,-4+0.3) -- (-2-0.5,-4+0.5) -- node[midway, left, sloped] {$x^-$} (-2-0.3,-4+0.5) ;
%\draw (-2+0.5,-4+0.3) -- (-2+0.5,-4+0.5) -- (-2+0.3,-4+0.5) ;
%\node at (-1.6,-4+0.5) [label = right:$x^+$]{};

\draw (-0.2,-2.8) -- (0,-3) -- (0.2,-2.8);
              
%\draw[dashed,gray] (IIbotleft) -- (2/3,-4/3);
%\draw[dashed,gray] (IIbotright) -- (-2/3,-4/3);

%\draw[domain=-2:-0.4, smooth, variable=\x, line width=0.8mm] plot ({\x}, {sin(deg((\x/2-1)))-1});
%\draw[domain=-0.4:0.5, smooth, variable=\x] plot ({\x}, {sin(deg((\x/2-1)))-1});
%\draw[domain=0.5:2, smooth, variable=\x, line width=0.8mm] plot ({\x}, {sin(deg((\x/2-1)))-1});

%\draw[blue] (-0.4,-1.93) -- (-1.53,-3.06);
\node at (-0.4,-1.93) [circle, fill, inner sep=1.5 pt]{};
\node at (-0.35,-2) [label=below:$\S_\Le$]{};
\node at (-1.53,-3.06) [circle, fill, inner sep=1.5 pt]{};
\node at (-1.6,-3.1) [label=right:$\S_\Le'$]{};
%\node at (-1.2,-2.7) [label=right:$\color{blue} \Delta_\Le$]{};

%\draw[blue] (0.5,-1.69) -- (1.19,-2.38);
\node at (0.5,-1.69) [circle, fill, inner sep=1.5 pt]{};
\node at (0.5,-1.7) [label=below:$\S_\Ri$]{};
\node at (1.19,-2.38) [circle, fill, inner sep=1.5 pt]{};
\node at (1.1,-2.38) [label=right:$\S_\Ri'$]{};
%\node at (0.1,-2.2) [label=right:$\color{blue} \Delta_\Ri$]{};
\node at (0.17,-1.36) [circle, fill, inner sep=1.5 pt, red, label=above:$\color{red} M$]{};

%\node at (-1.4,-1.15) [label=below:$\Sigma_\Le$]{};
%\node at (1.3,-0.5) [label=below:$\Sigma_\Ri$]{};

\end{tikzpicture}
\caption{\footnotesize Minimal screen configuration, when both screens are located at the top vertex $M$ of the exterior causal diamond.\label{entanglementwedgesc}}
\end{subfigure}
\caption{\footnotesize Several screen configurations in a given equivalence class, defined by a pair of screens $\S_\Le'$ and $\S_\Ri'$ on the apparent horizons of the pode and the antipode. The light red shaded region is the causal diamond in the exterior region E, and the dark blue region is the entanglement wedge of the single left-screen system. The biggest entanglement wedge is obtained when both screens lie at $M$ (Figure (c)). It coincides with the full causal diamond of any slice in the entanglement wedge of any of the equivalent screen configurations.}
\end{figure}

Let us focus on $\S_{\Le}$ and the parent screen $\S_{\Le}'$. In the exterior region, the entanglement wedge of $\S_{\Le}$ consists of the lightlike segment between $\S_{\Le}$ and $M$, denoted by $\Delta_{\S_{\Le}M}$. See Figure~\ref{entanglementwedgesb}. The part of the entanglement wedge of $\S_{\Le}'$ in E is the union of $\Delta_{\rm L}$ and $\Delta_{\S_{\Le}M}$ -- see Figure~\ref{entanglementwedgesa}.
In fact, as shown in Figure~\ref{entanglementwedgesb}, the full entanglement wedge of $\S_{\Le}$ is bigger and contains the entanglement wedge of the parent screen $\S_{\Le}'$. Since there is unitary evolution among the Cauchy slices in the entanglement wedge of $\S_{\Le}$, the state on them can be determined from the state on a Cauchy slice in the entanglement wedge of the parent screen $\S_{\Le}'$ (since the latter lies in the entanglement wedge of $\S_{\Le}$). 

The contrary is also true. The state on slices in the entanglement wedge of the parent screen can be determined from the state on the Cauchy slices in the entanglement wedge of the ``daughter'' screen. Therefore, the extra degrees of freedom that are integrated out in order to obtain the effective holographic theory on the ``daughter'' screen are not really needed for the reconstruction of the bulk state on these slices.     

It may seem counterintuitive that the entanglement wedge of the ``parent'' screen is contained in the entanglement wedge of a ``daughter'' screen, as it seems to imply that a greater bulk region can be reconstructed from the ``daughter'' screen $\S_{\Le}$ with fewer degrees of freedom. As we have already emphasized above however, the ``parent'' screen encodes more information than the information needed to describe the state on Cauchy slices in its entanglement wedge. First, assuming entanglement wedge reconstruction, since these slices also lie in the entanglement wedge of the ``daughter'' screen, the state on them can be reconstructed from the smaller ``daughter'' holographic screen system. Also, from the structure of the Bousso wedges in the region of trapped spheres, one can see that the area of the minimal extremal sphere $M$ divided by $4G\hbar$ already provides an upper bound on the coarse-grained entropy on Cauchy slices in the entanglement wedge of the ``parent'' (and any of) the ``daughter'' screen(s). Therefore, not all degrees of freedom on the parent screen are necessary for reconstruction of the state on these bulk slices. And more importantly, the entanglement wedge of the ``parent'' screen does not delimit the bulk region that is fully reconstructible from the holographic degrees of freedom on it. The reconstructible region in the bulk is actually bigger. One has to determine the full set of Cauchy slices that are unitarily related to the Cauchy slices that span the entanglement wedge. As we explain below, this causal diamond is actually bigger and contains the entanglement wedges of the ``parent'' and all ``daughter'' screens along $\rm{\Delta}_{\Le}\cup \Delta_{\S_{\Le}M}$. It turns out that this causal diamond coincides with the entanglement wedge of an equivalent left single screen subsystem, when both screens are placed at $M$, as shown in Figure~\ref{entanglementwedgesc}, and delimits the bulk region that is reconstructible from any of these equivalent left single screen subsystems.     

Indeed, it is important to note that the entanglement wedge of a screen does not contain all Cauchy slices that are unitarily related (and hence reconstructible from each other), but only those passing through the screen.\footnote{This ensures that all points of the Cauchy slice are spacelike (or lightlike) separated from the screen, and the local bulk fields at these points are independent from the fields on the screen.} In particular, the causal diamond of a Cauchy slice in the entanglement wedge is in general bigger than the entanglement wedge itself. The entanglement wedge of $\S_{\Le}'$ is spanned by the slices $\hat{\Sigma}_{\Le}'\cup \Delta_{\rm L}\cup \Delta_{\S_{\Le}M}$ that pass through $\S_{\Le}'$. The causal diamond of these slices is actually bigger (containing slices that do not pass through $\S_{\Le}'$). In fact,  by examining Figure~\ref{entanglementwedgesb}, we can easily infer that all Cauchy slices in the entanglement wedge of the ``daughter'' screen $\S_{\Le}$ lie in the full causal diamond of the Cauchy slices $\hat{\Sigma}_{\Le}'\cup \Delta_{\rm L}\cup \Delta_{\S_{\Le}M}$. The full causal diamond of the slices $\hat{\Sigma}_{\Le}'\cup \Delta_{\rm L}\cup \Delta_{\S_{\Le}M}$ coincides with the entanglement wedge of an equivalent left single screen subsystem, when both screens are placed at the minimal extremal sphere $M$ -- see Figure~\ref{entanglementwedgesc}. So the state on Cauchy slices in the entanglement wedge of $\S_{\Le}$ can be reconstructed from the state on $\hat{\Sigma}_{\Le}'\cup \Delta_{\rm L}\cup \Delta_{\S_{\Le}M}$, and hence the ``parent'' screen system. The entanglement wedge of the left screen, when both screens are placed at the minimal extremal sphere $M$, is the maximum bulk region that can be reconstructed from any of the equivalent left single screen subsystems.

In particular, consider the bulk slice $\hat{\Sigma}_{\rm L}\cup \Delta_{\S_{\Le}M}$ (lying in the entanglement wedge of $\S_{\Le}$) and the bulk subsystem associated with this slice. The holographic theories on $\S_{\Le}\cup \S_{\Ri}$ and $\S_{\Le}'\cup \S_{\Ri}'$ lead to the same prediction concerning the fine-grained entropy of this slice to leading order. Since the first corrected entropy requires adding the semiclassical entropy on slices in the entanglement wedge of $\S_{\Le}$ and $\hat{\Sigma}_{\Le}'\cup \Delta_{\rm L}\cup \Delta_{\S_{\Le}M}$ lies in this entanglement wedge, we see that the equivalence holds at the level of quantum corrections.

Notice that among the equivalent choices is the choice where both screens are placed at $M$ (the top vertex of the exterior causal diamond of the parent screens). The entanglement wedge of the left screen at $M$ now coincides with the full causal diamond of its Cauchy slices. Furthermore, this is the minimal configuration with the smallest possible number of degrees of freedom that are necessary to reconstruct the dual bulk subsystems (associated with Cauchy slices in the left causal diamond of $M$). The density matrix in this minimal dual holographic theory is maximally entangled since the von Neumann entropy is of the same order as the area of the screen in Planck units -- the latter sets the number of degrees of freedom in the holographic theory. For the rest of the configurations, one uses a larger density matrix, but as we emphasized before, this should have many zero eigenvalues, with the non-trivial block of non-zero eigenvalues corresponding to the subspace associated with the minimal extremal area sphere $M$, reflecting the fact that the additional number of degrees of freedom do not participate in the entanglement pattern with the right screen.

The situation can be generalized in the case when the exterior causal diamond of the parent screens acquires a fifth edge, along past lightlike infinity for the bouncing cases ($\gamma <-1$), or the Big Crunch singularity for the Big Bang/Big Crunch cosmologies ($\gamma > 1$). For the bouncing cases, the generalization is straightforward since the relevant minimal extremal surface is the top vertex of the causal diamond. For the singular cosmologies $\gamma > 1$, note that the nesting property continues to hold. All subsequent causal diamonds share the same fifth edge along the Big Crunch singularity and are contained in the previous ones. Hence, the disentanglement phenomenon occurs for all equivalent configurations simultaneously, and moreover the subsequent entanglement wedge structures resulting from quantum corrections will continue to yield identical predictions for the entropy of bulk subsystems associated with the screens.

%{\magenta TO ADD IN SECTION 4?: We should say explicitly in section 4 that the fine-grained entropy on the cauchy slice to the left of $M$ is the entropy on $M$ (for the single-screen system)? The fine-grained entropy on the cauchy slice to the left of $\S'_\Le$ is not that on $\S_\Le$. So the next paragraphs is WRONG: 

%To go one step further, let us define for any $\Sigma'$ in $\F'$ the parts $\Sigma_\Le'$ and $\Sigma_\Ri'$ that lie respectively to the left of $\S_\Le'$ and to the right of $\S_\Ri'$, as shown in purple in \Fig{pushback}. The states on the bulk Cauchy slices $\Sigma_\Le$ and $\Sigma_\Le'\cup \Delta_\Le$ should be related by a unitary transformation. Hence, their respective fine-grained entropies should be the same, equal to the entanglement entropy of the screen on $\S_\Le$. From the previous paragraph, the latter is also equal to the entanglement entropy of the screen on $\S_\Le'$, which is the fine grained entropy of the slice $\Sigma_\Le'$ only. As a result, the state on $\Delta_\Le$ must be pure, and information on $\Sigma_\Le$ and $\Sigma_\Le'$ equivalent. (RV: the word ``information'' is vague...)  The same conclusions apply to $\Delta_\Ri$, $\Sigma_\Ri$ and $\Sigma_\Ri'$.}

\subsection{Discussion}
\label{FRW_discussion}

\begin{figure}[h!]
    \centering
\begin{tikzpicture}
\begin{axis}[
    axis lines = left,
    xlabel = {$\eta_\Le$},
    ylabel = {$\S_{\Le}$},
    no markers,
    xtick={0, 1},
    xticklabels={$|\gamma|\pi/2$, $|\gamma|\pi$},
    ytick={0,1},
    yticklabels={$0$,$\frac{\pi}{2G}$},
    legend style={at={(1,0.28)},anchor=east, font=\small},
    x post scale=1.5,
    axis line style={-{Latex[scale=1.3]}},
    %scale =1
]

%alpha=3
\addplot [
    domain=0:1, 
    samples=100, 
    color=cyan,
    ]
    {(\x < 2/3)*(sin(deg(3.14*(3*x+2)/4)))^(2) + (\x > 2/3)*(0)};
\addlegendentry{$w=1/2$}

%alpha=5
\addplot [
    domain=0:1, 
    samples=100, 
    color=lime,
    ]
    {(\x < 4/5)*(sin(deg(3.14*(5*x+4)/8)))^(4)+(\x >4/5)*(0)};
\addlegendentry{$w=1/4$}

%alpha=-1
\addplot [
    domain=0:1, 
    samples=100, 
    color=yellow,
]
{(sin(deg(3.14*(x+2)/4)))^(-2)};
\addlegendentry{$w=-1/2$}

%alpha=-1/2
\addplot [
    domain=0:1, 
    samples=100, 
    color=orange,
    ]
    {(sin(deg(3.14*(x/2+3/2)/3)))^(-3/2)};
\addlegendentry{$w=-2/3$}

%alpha=0
\addplot [
    domain=0:1, 
    samples=100, 
    color=red,
    ]
    {1};
\addlegendentry{$w=-1$}

\addplot[domain=0:1.2,samples=100,color=black,]{-0.2};

\end{axis}
\end{tikzpicture}
\caption{\footnotesize Entropy of a single screen subsystem as a function of the conformal time, for $n=2$ and various values of $w$. The two screens are placed on the apparent horizons and taken to lie at equal conformal times. The entropy is rescaled with respect to $|\gamma|\pi$.}
    \label{S_AH}
\end{figure}
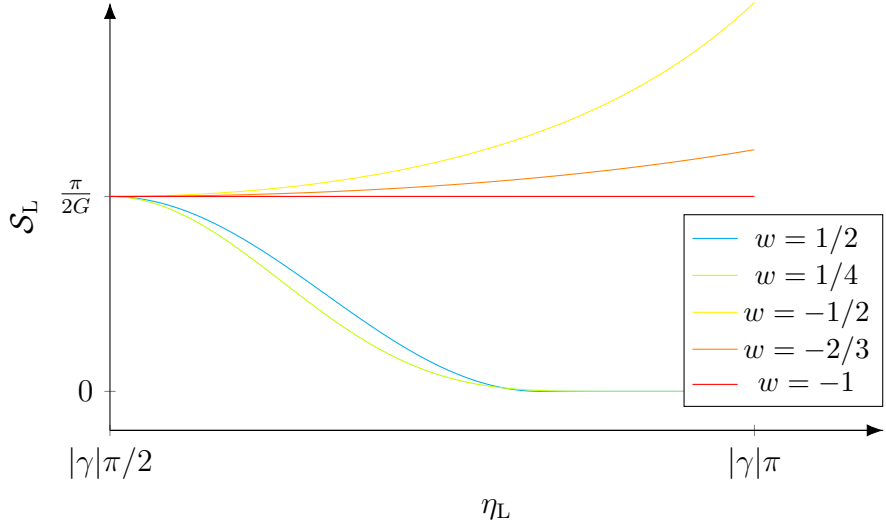

The entropies \eqref{sexpanding} and \eqref{entropy_FRW_alpha>0} are plotted in Figure~\ref{S_AH},
for three-dimensional cosmological spacetimes ($n=2$), and for several values of $w$ in the range $[-1,1]$. For simplicity, we consider the case where the screen trajectories coincide with the apparent horizons, and take the screen conformal times to be equal, $\eta_\Le=\eta_\Ri$.

\begin{figure}[h!]
\centering
\begin{subfigure}[t]{0.3\linewidth}
\centering
\includegraphics[height=37mm]{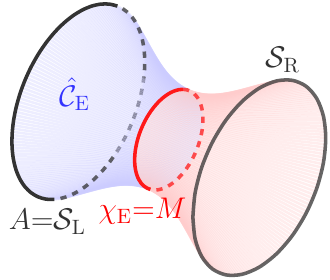}
\caption{\footnotesize For $(x^+_\Ri+x^-_\Le)/2 < \gamma\pi$, $M$ has a non-vanishing area.}
\end{subfigure}\hfill
\begin{subfigure}[t]{0.3\linewidth}
\centering
\includegraphics[height=32mm]{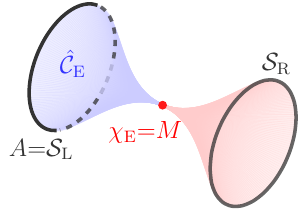}
\caption{\footnotesize When $(x^+_\Ri+x^-_\Le)/2 = \gamma\pi$, $M$ hits the Big Crunch singularity: its area vanishes and the bridge pinches and closes off.}
\end{subfigure}\hfill
\begin{subfigure}[t]{0.3\linewidth}
\centering
\includegraphics[height=27mm]{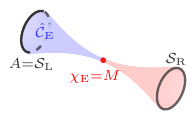}
\caption{\footnotesize The bridge remains disconnected for the rest of the cosmological evolution, for $(x^+_\Ri+x^-_\Le)/2>\gamma\pi$.}
\end{subfigure}\hfill
    \caption{\footnotesize Topology of the effective bridge in the exterior region connecting the screens $\S_\Le$ and $\S_\Ri$ via the minimal extremal surface $M$, depicted at different times in the contracting phase of a Big Bang/Big Crunch cosmology. The Cauchy slice $\hat\Sigma_\Ex$ is the union of the blue and red shaded regions. It corresponds to the limiting Cauchy slice consisting of the two upper edges of the exterior causal rectangle in the Penrose diagram.}
    \label{bridge}
\end{figure}

To summarize, the entanglement entropy between the two screens can be directly related to the geometry of an effective geometrical bridge in the exterior region. The effective geometrical bridge is the union of the two lightlike $\hat{\C}_{\rm E}$'s, which connect the screens with the minimal extremal homologous surface $M$, respectively. It corresponds to the limiting Cauchy slice on the boundary of the exterior causal diamond, which contains $M$. This is indeed the slice with the smallest bottleneck. In the Big Bang/Big Crunch cases, the bottleneck pinches and closes off, precisely when the two screen subsystems effectively disentangle. Indeed, when the leading classical entanglement entropy between the screens vanishes, the area of $M$ vanishes since this lies either at the Big Bang or the Big Crunch singularity. The presence of a complete geometrical bridge between the screen subsystems is a manifestation of the ER=EPR paradigm in a cosmological, time-dependent setting \cite{Maldacena:2013xja}. This is depicted in \Fig{bridge},
in the case where the two screens lie in the contracting phase of a Big Bang/Big Crunch cosmology.

Note that for the $\gamma <-1$ and $\gamma > 1$ cases, there is a certain period of time of finite duration, where the two screens are in causal contact and can exchange information. However, the screens are always spacelike separated, comprising two sets of independent degrees of freedom with possible interactions among them. 

The minimal extremal surface associated with a single screen subsystem in the exterior region E is lightlike separated with respect to the screens $\S_{\Le}$ and $\S_{\Ri}$. As a result, the entanglement wedge of $\S_{\Le}$ in the exterior has an infinitesimally small extent. This supports the fact that a single screen does not suffice to encode the full bulk spacetime. 

Let us conclude this section by commenting on the difference between maximal coarse-grained entropy and the fine grained entropy associated with a single screen. The first one provides a measure to the number of degrees of freedom on the screen. Since it is a maximal entropy, counting the underlying degrees of freedom, it is a property of the system itself rather than the state of the system at a given time. The second entropy quantifies the entanglement between the two screen subsystems. In general, these two entropies are not equal. The coarse-grained entropy is always greater or equal to the entanglement entropy. The latter, being a fine grained entropy is not in general additive and it does not obey a 2nd thermodynamical law. 

The maximal coarse-grained entropy of a single screen subsystem is given by the area of the screen in Planck units. Contrary to the case of de Sitter space, for which the number of degrees of freedom on the screens remains constant as the screens evolve along the cosmological horizons, for all other cases of closed FRW cosmologies, the number of degrees of freedom of the dual holographic theory on the screens is not constant. As we already remarked, this suggests that the evolution of the holographic dual to a generic closed FRW Universe is not unitary, a very interesting possibility that has been studied recently in \cite{Cotler:2022weg, Cotler:2023eza}.

The entanglement entropy between the two screens has been studied in this section. We find that the entanglement entropy to leading order is less than the maximal coarse-grained entropy (given by the area divided by $4G\hbar$), except for the cases where both screens coincide at the same point in the contracting phase of the cosmology. This highlights the particularity of the de Sitter case, for which, as in the examples of eternal AdS black holes, the entanglement entropy between the screens is given by the area of the horizon divided by $4G\hbar$. 
%The relation between these two entropies is not the same in the general singular and bouncing cases.
\begin{enumerate}
    \item In the Big Bang/Big Crunch cases, the area of the screens tends to zero as we approach the Big Bang or the Crunch singularity. The classical entanglement entropy vanishes well before, and the quantum corrections dominate. As the Universe contracts the number of fundamental holographic degrees of freedom decreases. The entanglement entropy between the screens should remain bounded by the smaller (quantum) area between the screens, as expected. Near the Big Bang/Big Crunch singularities, quantum corrections to the geometry as well as to the generalized entropy become important.
    \item In the bouncing cases, the number of holographic degrees of freedom goes to infinity as we approach the future or past null infinity, but the entanglement entropy between the screens saturates a finite bound. In other words, the entanglement entropy grows without ever becoming infinite, despite the fact that extra degrees of freedom keep being added on the two screens.
\end{enumerate}
The particularity of the de Sitter and the eternal black hole in AdS is due to the fact that the holographic dual to these spacetimes is expected to be in a thermofield double state. Indeed, taking a partial trace over the degrees of freedom of one of the two copies leads to a thermal or maximally entangled density matrix for the other copy. The corresponding von Neumann entropy is given by the area of the horizon in Planck units. In the de Sitter case, when the two screens coincide at the bifurcate horizon, their state is dual to the Bunch-Davies vacuum state, which can be expressed as a thermofield double state over degrees of freedom associated with the two static patches \cite{Bousso:2001mw}. However, we saw that in the general closed FRW spacetimes, the patches associated with the pode and antipode are not always causally disconnected. The states of the single screen systems are thus expected to be more complicated.

%%%%%%%%%%%%%%%%%%%%%%%%%%%%%%%%%%%%%%%%%

\section{Big Bang/Big Crunch cosmological evolutions with \bm $1/(n-1)\le \gamma < 1$}
\label{BBBC}
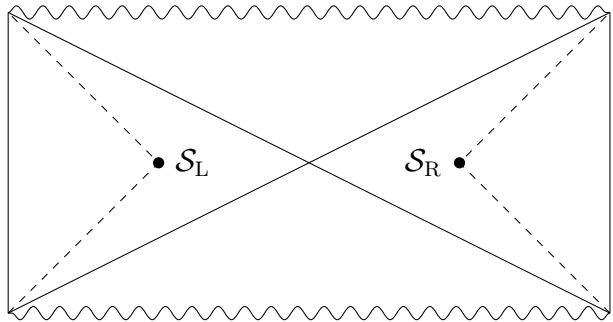
\begin{figure}[h!]
\centering
\begin{tikzpicture}
\path
       +(4,2)  coordinate (IItopright)
       +(-4,2) coordinate (IItopleft)
       +(4,-2) coordinate (IIbotright)
       +(-4,-2) coordinate(IIbotleft)
      
       ;
       
%\fill[fill=blue!50] (-4,2) -- (-2,0) -- (-4,-2) --  cycle;       
       
\draw[decorate,decoration=snake] (IItopleft) --
          node[midway, above, sloped]    {}
      (IItopright);
      
\draw (IItopright) --
          node[midway, above, sloped] {}
      (IIbotright);
      
\draw[decorate,decoration=snake]  (IIbotright) -- 
          node[midway, below, sloped] {}
      (IIbotleft);
      
\draw (IIbotleft) --
          node[midway, above , sloped] {}
      (IItopleft);
      
\draw (IItopleft) -- (IIbotright)
              (IItopright) -- (IIbotleft) ;

\draw[dashed] (IIbotleft)--(-2,0)--(IItopleft);
\draw[dashed] (IIbotright)--(2,0)--(IItopright);

\node at (-2,0) [circle,fill,inner sep=1.5pt, label = right:$\S_\Le$]{};
\node at (2,0) [circle,fill,inner sep=1.5pt, label = left:$\S_\Ri$]{};
\end{tikzpicture}
    \caption{\footnotesize FRW cosmologies for $1/(n-1)\leq \gamma<1$. The two screens $\S_\Le$ and $\S_\Ri$ are placed at the cosmological horizons (depicted by the dark dashed lines). The apparent horizons are the dark diagonal lines.}
    \label{Penrose_alpha<2}
\end{figure}

Let us discuss the holographic proposal for Big Bang/Big Crunch cosmologies with $1/(n-1)\leq \gamma<1$. This class includes the moduli dominated cases with $w=1$, $\gamma=1/(n-1))$, the matter dominated cases with $w=0$, $\gamma=2/(n-2)$, for $n>4$, as well as the radiation dominated cases $w=1/n$, $\gamma=2/(n-1)$,  for $n>3$. Recall from \Sect{basics_FRW} that the Penrose diagrams are wider than tall, and the causal patches associated with comoving observers at the pode and antipode, which are bounded by cosmological horizons, are non-overlapping and causally disconnected. No light signal from one patch will reach the other before the Big Crunch time at $\eta=\gamma \pi$. The apparent horizons are spacelike and lie in the region exterior of the causal patches. See Figure~\ref{Penrose_alpha<2}.
The new feature of these examples is that future directed left-moving light rays emanating from the pode apparent horizon, which constitute lightsheets of negative expansion, can terminate at the Big Crunch singularity without entering the causal patch of the pode. So we cannot in general apply Bousso's covariant entropy conjecture to bound the entropy on spacelike slices between the pode and the apparent horizon, without invoking new physics to describe the Big Crunch singularity. Furthermore, the apparent horizons lie outside the causal patches of the observers at the pode and antipode.  Instead, following our rules, we place the screens at the boundaries of the causal patches, namely the cosmological horizons. See Figure~\ref{Penrose_alpha<2}.
%Indeed the constant $\eta$ Cauchy slices will intersect both cosmological horizons and we can locate the two screens at the intersection points. 
Bousso's covariant entropy bound can be applied in order to encode holographically the interior regions. Notice that as the screens evolve, they remain out of causal contact and their trajectories never intersect, as shown in Figure~\ref{Penrose_alpha<2}. In particular, there is no instance at which the union of the Cauchy slices $\Sigma_{\Le}$ and $\Sigma_{\Ri}$, which Bousso's rule ensures to be covered holographically, amounts to a complete Cauchy slice. %We will argue in the following sections that the bilayer proposal is not applicable in this case, as two observers at the pode and antipode and the holographic systems on the two screens do not suffice to describe the full bulk cosmology. Additional observers may be needed in the exterior region, which is causally inaccessible by the two screens at the cosmological horizons.       

So, in contrast with the cases $|\gamma|\ge 1$, the screen trajectories never intersect. The distance between the screens starts small near the Big Bang singularity, grows to a maximum value of order $a_0$ and then decreases in the contracting phase. However, the screens remain out of causal contact for the entire cosmological evolution. In particular, no light signal from one screen will ever reach the other. The bulk picture suggests that the two screens cannot be in a pure state. Indeed, the region between the cosmological horizons remains inaccessible to the observers at the pode and antipode at all times. Particles created at the Big Bang singularity in the past may reach the Big Crunch singularity without entering the pode and antipode causal patches and without crossing either of the screen trajectories along the cosmological horizons.  In order to describe the exterior region, we would need additional observers following trajectories in the exterior region and the corresponding holographic screens. Therefore, we do not have evidence to believe that the degrees of freedom in the exterior region can be encoded holographically on a two-screen system at the cosmological horizons. Moreover, $\Sigma_{\Le}\cup \Sigma_{\Ri}$ never amounts to a complete bulk Cauchy slice. It seems plausible that the bilayer prescription, which requires both exterior and interior contributions, is not applicable in these cosmological cases, even though we do not have direct evidence for this from a replica bulk path integral computation. Indeed, applying this proposal blindly for these cases leads to paradoxical results from the point of view of bulk causality, as it would seem to imply that the entanglement wedge of the two screen system comprises  complete bulk Cauchy slices. Rather, the holographic entanglement entropy prescription could involve contributions from the two interior regions only.\footnote{It would be interesting to derive this result directly from a bulk replica path integral and contrast with the $|\gamma| \ge 1$ cases.} 

For the single screen subsystem, the dual homologous surface in region L is the empty surface, implying that the leading classical geometrical contributions to the entanglement entropy vanish. The entanglement entropy should arise from semiclassical entropy contributions associated with bulk field degrees of freedom on Cauchy slices connecting the pode and the screen. Indeed, we can imagine entangled spin pairs, which start in the early past and then separate further, entering the pode and antipode causal patches, respectively. As a result, there can be non-trivial quantum entanglement between the two causal patches. The spins could begin in the exterior region, outside of the two patches. Notice that this entanglement entropy is significantly smaller than the maximal possible entropy for a single screen subsystem, which is of order $(G\hbar)^{-1}$. 
%But unless there are exterior contributions, the semiclassical entropy is much smaller. \footnote{The singular cutoff dependent terms scale inversely proportional with a power of the short-distance cutoff, which we must take bigger than the Planck length to apply the semiclassical approximation.} 
Similar conclusions hold for the two-screen system. The leading geometrical entropy is zero but there are non-vanishing contributions at order $(G\hbar)^0$. Since the entanglement wedge does not extend in the exterior region, the semiclassical entropy associated with the two-screen system cannot be zero, showing that the screens are not in a pure state.

\section{Entanglement entropy at the semiclassical level}
\label{semicla}

To go beyond the classical, geometrical expression of the entanglement entropy of a holographic subsystem $A\subset(\S_\Le\cup\S_\Ri)$, we are led to consider the generalized entropy defined in \Eq{Sgen}. We already argued that the semiclassical contribution $S_{\rm semicl}$ is unambiguously defined, thanks to the unitary transformations relating the states, \ie the density matrices, on all possible choices of surfaces $\hat \C_\Le\cup\hat\C_\Ex\cup\hat\C_\Ri$. In this section, restricting to the case where $A$ and the Cauchy slices respect the SO$(n)$ symmetry of spacetime, we rederive this fact and find an explicit expression of $S_{\rm semicl}$, when the perfect fluid carries some coarse-grained  entropy generated at 1-loop. For instance, this applies when the fluid is made of radiation at finite temperature, for which $w=1/n$. However, the above assumption encompasses cases where the microscopic origin of the perfect fluid is a set of quantum fields at finite temperature, with different masses.\footnote{See~\Refs{Catelin-Jullien:2007ewh, Catelin-Jullien:2009duh} for examples. The FRW cosmological evolutions are induced by 1-loop free energies arising from string spectra, including the graviton, scalar fields, fermions, Kaluza-Klein modes and string oscillator modes.} In such instances, even if $w$ is not constant~\cite{Bourliot:2009na, Bourliot:2009ai, Estes:2010sh} and the time evolution of the scale factor $a$ is more involved than that described in \Sect{basics_FRW}, our final result for $S_{\rm semicl}$ applies.\footnote{Of course, the analysis of \Sect{holo_proposal} concerning the location of a pair of screens needs to be generalized.} 

We will focus on the first order corrected entropy, which can be obtained by adding to the leading geometrical contributions the semiclassical entropy on a Cauchy slice of the entanglement wedge, retaining for the extremal surfaces the classical ones \cite{Faulkner:2013ana}. In certain cases, we expect the quantum extremal surfaces, which minimize the generalized entropy, to be close to the classical ones. 

From now on, let us assume that the FRW cosmology is induced by a perfect fluid that carries some coarse-grained entropy of order $(G\hbar)^0$, generated at 1-loop by all bulk fields. The cosmological evolution being quasistatic (the pressure is well defined at every time) and adiabatic (no possible heat exchange), it is isentropic. In \Appendix{thermo}, we review how this fact as well as standard thermodynamical identities can be derived by applying the variational principle on a 1-loop effective action. Moreover, since the coarse-grained entropy is additive, we can define its local density $s_{\rm th}(\eta)$, which depends only on conformal time thanks to homogeneity and isotropy. As a result, the total coarse-grained entropy $S_{\rm th}$ of the universe satisfies  
\be
S_{\rm th}= s_{\rm th} \, \omega_n a^n=\mbox{constant},
\label{Scst}
\ee 
where $\omega_n$ is the volume of the unit sphere S$^n$. We can also define the entropy current 
\be
J^\mu= -s_{\rm th}\, n^\mu, \quad~~ \where~~ \quad n^\mu={\delta^{\mu 0}\over a}.
\ee
In this expression, $n^\mu$ is the future-directed unit vector normal to the constant $\eta$ Cauchy slice, which is a sphere~S$^n$. Since $n^\mu n_\mu=-1$, we have $s_{\rm th}=J^\mu n_\mu$. 

An arbitrary SO$(n)$-symmetric codimension-1 spacelike (possibly locally lightlike) connected surface $\hat \C$ can be described as a hypersurface
\be
    f(\theta,\eta)\equiv \eta-H(\theta)=0.
\ee
In this equation, $H$ is a function such that $|H'|\equiv |\d H/\d \theta|\le 1$, for the surface to be nowhere timelike. Connectedness implies that $\theta$ varies in a single interval $[\theta_{\rm i}, \theta_{\rm f}]$. Since $\left.\nabla_\mu f \right |_{f=0}$ is a covariant vector normal to $\hat \C$, the future-directed unit contravariant vector normal to $\hat \C$ is 
\be
\hat n^\mu={1\over a\sqrt{1-H'^2}}(1,H',0,\dots,0).
\ee
The fluid entropy passing through $\hat \C$ is then
\be
S_{\rm semicl}(\hat \C)=\int_{\theta_{\rm i}}^{\theta_{\rm f}} J^\mu \hat n_\mu\sqrt{(a\d \theta)^2-(aH'\d\theta)^2}\,\A,
\ee
where the scale factor $a$ and the S$^{n-1}$-sphere area $\A$ are respectively taken at $H(\theta)$ and $(\theta,H(\theta))$. In the integrand, the square root is the line element of the projection of $\hat \C$ on the Penrose diagram, as induced by the metric~(\ref{so met}). Taking into account the expressions of the current, normal vector and area, one obtains 
\be
S_{\rm semicl}(\hat \C)=\omega_{n-1}\int_{\theta_{\rm i}}^{\theta_{\rm f}}\d\theta\,s_{\rm th}\,a^n\, (\sin\theta)^{n-1}.
\ee
Thanks to the constancy of the coarse-grained entropy, \Eq{Scst}, we reach the final expression~
\be
S_{\rm semicl}(\hat \C)= S_{\rm th}\,{\omega_{n-1}\over \omega_n}\int_{\theta_{\rm i}}^{\theta_{\rm f}}\d\theta\, (\sin\theta)^{n-1},
\ee
which depends on the interval of definition  $[\theta_{\rm i}, \theta_{\rm f}]$ of $H$ but  not on its profile. 

The particular case where $\hat \C$ is an entire Cauchy slice $\hat \Sigma$ of spacetime provides a consistency check of this result. Indeed, setting $[\theta_{\rm i}, \theta_{\rm f}]=[0,\pi]$, one obtains
\be
S_{\rm semicl}(\hat \Sigma)= S_{\rm th}, 
\ee
which is the correct answer, since the entropy of the bulk fields through the whole Cauchy slice must be the entropy they induce in the entire spacetime. These considerations apply, for instance, to the case where the perfect fluid is made of pure radiation at finite temperature. The fluid index is $w=1/n$, which corresponds to a Big Bang/Big Crunch cosmological evolution characterized by the parameter $\gamma=2/(n-1)$. Restricting to the cases where $\gamma\ge 1$ imposes the dimension of space to take only two values, namely 
\begin{align}
n=2 ~\Longrightarrow~ \gamma=1 \qquad \mbox{or}\qquad n=3 ~\Longrightarrow~ \gamma=2.
\end{align}
The associated Penrose diagrams are respectively represented in \Figs{Pene} and~\ref{Peng}. For the two-screen system, we have seen in the previous section that  
\be\begin{aligned}
A&=\S_\Le\cup\S_\Ri, ~~~&A_\Le&=\S_\Le, ~~~&\chi_\Le&=\varnothing, ~~~&\hat \C_\Le&=\hat \Sigma_\Le,\\
&&A_\Ex&=\S_\Le\cup\S_\Ri, ~~~& \chi_\Ex&=\varnothing, ~~~& \hat \C_\Ex&=\hat \Sigma_\Ex, \\
&&A_\Ri&=\S_\Ri, ~~~& \chi_\Ri&=\varnothing, ~~~& \hat \C_\Ri&=\hat\Sigma_\Ri,
\end{aligned}\ee
where $\hat\Sigma_i$, $i\in\{\Le,\Ex,\Ri\}$,  is arbitrary. Hence, $\hat \C$ can indeed be identified with $\hat\C_\Le\cup\hat\C_\Ex\cup\hat\C_\Ri=\hat \Sigma$. Since the classical contribution to the fine-grained entropy vanishes, the generalized entropy~(\ref{Sgen}) for this configuration reduces to  
\be
S_{\rm gen}(\varnothing,\varnothing,\varnothing)=S_{\rm th}.
\ee
Notice that \apriori, it is not a good approximation of the fine-grained entropy at the semiclassical level, \Eq{Ssem}, which requires an extremization with respect to the three homologous surfaces. Indeed, the error can be $\O((G\hbar)^0)$, which is of the order of  $S_{\rm gen}(\varnothing,\varnothing,\varnothing)$. In any case, the semiclassical result must be strictly positive, since the classical area $\A$ as well as  $S_{\rm semicl}$ are non-negative functions that cannot vanish simultaneously. Positivity of the semiclassical von Neumann entropy follows from the fact that the state on the Cauchy slice $\hat \Sigma$ is a mixed thermal state rather than a pure state.  

The other SO$(n)$-symmetric holographic subsystem that can be considered is the single-screen system. The results of  \Sect{time_dep_EREPR} apply when the fluid is made of radiation and $n=2$ or 3, leading to
\be\begin{aligned}
A&=\S_\Le, ~~~&A_\Le&=\S_\Le, ~~~&\chi_\Le&=\varnothing, ~~~&\hat \C_\Le&=\hat \Sigma_\Le,\\
&&A_\Ex&=\S_\Le, ~~~& \chi_\Ex&=M, ~~~& \hat \C_\Ex&=\hat \Sigma_{\S_\Le\mbox{\scriptsize -} M}, \\
&&A_\Ri&=\varnothing, ~~~& \chi_\Ri&=\varnothing, ~~~& \hat \C_\Ri&=\varnothing,
\end{aligned}\ee
where $\hat\Sigma_\Le$ is arbitrary. As a result, we can take $\hat \C$ to be $\hat\C_\Le\cup\hat\C_\Ex\cup\hat\C_\Ri=\hat \Sigma_\Le\cup\hat \Sigma_{\S_\Le\mbox{\scriptsize -} M}$, so that $[\theta_{\rm i}, \theta_{\rm f}]=[0,\theta_M]$, where $\theta_M=(x^+_{\rm R}-x^-_{\rm L})/2$ is the $\theta$-angle of the minimal extremal homologous surface -- see Eq.~\ref{eta_M_position} -- and
\be
S_{\rm semicl}(\hat \Sigma_\Le\cup\hat \Sigma_{\S_\Le\mbox{\scriptsize -} M})= S_{\rm th}\times
 \left\{\!\begin{array}{ll}
\sin(\theta_M/2)  &~~\mbox{for}\quad n=2 , \\
\dis {1\over \pi}\Big(\theta_M-{\sin(2\theta_M)\over 2}\Big) &~~\mbox{for}\quad n=3 .\esp
\end{array}\right.
\ee 
The generalized entropy evaluated for the classical minimal extremal surfaces is
\be
S_{\rm gen}(\varnothing,M,\varnothing)={\A(M)\over 4G\hbar}+ S_{\rm semicl}(\hat \Sigma_\Le\cup\hat \Sigma_{\S_\Le\mbox{\scriptsize -} M}).
\ee
This is a good approximation of the semiclassical fine-grained entropy defined in \Eq{Ssem}. Indeed, the error can be $\O((G\hbar)^0)$, which is small compared to the leading, classical contribution. Notice the scaling of the classical geometrical and thermal entropy contributions. The first contribution obeys an area law, scaling with $a^{n-1}$, while the thermal contribution obeys a volume law scaling with $a^n$.

%%%%%%%%%%%%%%%%%%%%%%%%%%%%%%%%%%%%%%%%%

\section{Conclusion}
\label{conclu}
In this work we extend the arguments of \cite{Franken:2023pni,Susskind:2021esx,Shaghoulian:2021cef,Shaghoulian:2022fop} to classes of closed FRW spacetimes and conjecture that these spacetimes can be encoded by holographic degrees freedom located on two holographic screens associated with two antipodal observers. In the expanding phase of the FRW cosmologies, the holographic screens are uniquely defined to be located at the two apparent horizons. In the contracting phase however, there is an infinite number of choices for the trajectories of the holographic screens in the region of trapped surfaces. We show that these choices can be grouped in equivalence classes. The effective theories on the screens belonging to the same equivalence class are related by an RG flow while encoding the same regions of space. In each equivalence class, one can identify a ``parent'' two-screen configuration on the apparent horizons, containing redundant degrees of freedom that are integrated out in other elements of the equivalence class. The de Sitter case studied in \cite{Franken:2023pni} appears to be a particular case, where the trajectories of the holographic screens are uniquely defined to follow the cosmological horizons bounding two complementary static patches. 

In the cosmological classes in which the holographic prescription is defined, the covariant version of the bilayer proposal for holographic entanglement entropy computations \cite{Franken:2023pni} can be used. We proceed to apply this proposal in order to compute the entropy of a single screen subsystem. We find a generic behavior that manifests the ER=EPR correspondance of Maldacena and Susskind and strong connection between quantum entanglement and geometry, as described in \cite{Maldacena:2013xja, VanRaamsdonk:2010pw}. Indeed, quantum entanglement builds an effective bridge between the two screen subsystems via the trapped or anti-trapped sphere of smallest area on the boundary of the causal diamond in the region between the two screens. We argue that the area of this sphere divided by $4G\hbar$ determines the dominant geometrical contributions to the entanglement entropy of a single screen subsystem. It would be interesting to obtain direct evidence for this proposal via a replica bulk path integral computation. For the Big Bang/Big Crunch cosmologies, the effective bridge contracts and closes off when this minimal trapped (anti-trapped) sphere hits (touches) the Big Crunch singularity (Big Bang singularity). In the contracting phase, this moment occurs before the screens hit the Big Crunch with the geometrical contributions to the entanglement entropy vanishing.  Effectively when the two screen subsystems disentangle. For the bouncing cosmologies, the entanglement entropy between the two screens grows as the bridge expands, but saturates a bound, despite the fact that the area of the screens grows to infinity. This behavior is due to the fact that the minimal anti-trapped or trapped sphere asymptotes to a sphere of finite area.

As part of future work, it would be interesting to understand further semiclassical contributions to the entropy, as well as the effects of matter and/or radiation to the FRW cosmologies. In particular, our discussion on thermal radiation may be extended to ``radiation-like'' systems, where $\omega$ is time dependent. In the literature, there have been various proposals concerning the nature of the quantum theory dual to de Sitter space \cite{Banks:2000fe,Witten:2001kn,Banks:2006rx, Susskind:2021dfc, Susskind:2021esx, Susskind:2022dfz, Susskind:2022bia, Rahman:2022jsf, Narovlansky:2023lfz}. One should try to understand how the exterior region is encoded holographically in these models. Another interesting direction would be to investigate the nature of the holographic dual for the FRW cases. Indeed, the causal connection between the two screens indicates that there must be interactions between the two sets of degrees of freedom on the screens, which should turn off after a certain time. It would be nice to understand the relation of this transition time and the disentangling time further. It would be also interesting to extend the analysis to other subsystems of the two-screen system in order to probe further the non-local nature of the holographic dual theory and provide further tests of the classification into equivalence classes of the different constructions in the contracting phase of the cosmology. Finally, note that the evolution of the screen systems cannot be unitary in the generic FRW cases. This motivates us to investigate if this evolution can be isometric instead, as suggested in \cite{Cotler:2022weg, Cotler:2023eza}, and understand possible consequences.  

%%%%%%%%%%%%%%%%%%%%%%%%%%%%%%%%%%%%%%%%
\section*{Acknowledgements}
F.R. and N.T. would like to acknowledge hospitality by the Ecole Polytechnique, while H.P. and V.F. would like to thank the University of Cyprus for hospitality, where early stages of this work have been done. This work is partially supported by the Cyprus Research and Innovation Foundation grant EXCELLENCE/0421/0362.

%%%%%%%%%%%%%%%%%%%%%%%%%%%%%%%%%%%%%%%%
\begin{appendices} 
\numberwithin{equation}{section}

\section{Scale factor in closed FRW cosmology}
\label{FRW_cosmo}

In this appendix, we bring together  salient results concerning the FRW cosmological evolution of a closed universe filled with a perfect fluid. 

For a closed universe, the metric of an $(n+1)$-dimensional FRW cosmology takes the form 
\be
\d s^2=-N^2(x^0)(\d x^0)^2+a^2(x^0)\d\Omega_n^2,
\label{me}
\ee
where $N(x^0)$ is the lapse function, $a(x^0)$ is the scale factor and $\d\Omega_n^2$ is the metric of the sphere S$^n$ of radius 1. For $n\ge 2$, the metric of S$^n$ is related to that of S$^{n-1}$ as follows, 
\be
\d\Omega_n^2=\frac{\d r^2}{1-r^2}+r^2 \d\Omega_{n-1}^2.
\ee
Performing the change of coordinate
\be
\d\theta={\d r\over \sqrt{1-r^2}},
\ee
\Eq{me} takes   the alternative form
\begin{equation}
   \d s^2=-N^2(x^0)(\d x^0)^2+a^2(x^0)\big[\d\theta^2+\sin^2(\theta)\d\Omega_{n-1}^2\big].
\end{equation}

The Friedmann equations of motion for $N$ and $a$ are, in the gauge $N\equiv 1$,
\begin{align}
{n(n-1)\over 2} \left[\Big({\dot a\over a}\Big)^2+{1\over a^2}\right]&=\phantom{-}8\pi G\, \rho,\label{f1}\\
(n-1)\,{\ddot a\over a}+{(n-1)(n-2)\over 2} \left[\Big({\dot a\over a}\Big)^2+{1\over a^2}\right]&=-8\pi  G\,p, \label{f2}
\end{align}
where dots stand for derivatives with respect to cosmological time $t$, while $\rho$ and $p$ are respectively the energy density and pressure in the universe. The second equation can be replaced by
\be
(n-1)\,{\ddot a\over a}= -8\pi G\Big[{n-2\over n}\, \rho+p\Big].
\label{accel}
\ee
Alternatively, taking the time derivative of \Eq{f1}, one can show that \Eq{f2} can be replaced by 
\be
\dot \rho+n\, {\dot a\over a}\, (\rho+p)=0.
\label{conser}
\ee
This equation is nothing but the continuity equation for the stress-energy tensor, which in the present case expresses  the conservation of the energy in a universe in  adiabatic  evolution. 

Let us assume from now that the perfect fluid filling the entire universe satisfies the state equation~(\ref{seq}), where $w\in[-1,1]$ is a constant fluid index. In this case,  \Eq{conser} can be readily integrated, giving 
\be
\rho={C\over a^{n(1+w)}},
\ee
where $C> 0$ is a constant. Moreover, \Eq{accel} reduces to 
\be
(n-1)\, {\ddot a\over a}=8\pi G(w_{\rm c}-w)\rho,~~\quad \where~~\quad w_{\rm c}=-1+{2\over n}\in(-1,0].
\ee
This shows that as a function of cosmological time $t$, the evolution is accelerating when $-1\le w<w_{\rm c}$ and decelerating when $w_{\rm c}<w\le 1$. When the fluid index is critical, $w=w_{\rm c}$, the evolution is linear in $t$. 

In order to find explicitly the evolution of the scale factor, it turns out to be  relevant to work in conformal gauge, $N=a$. The Friedmann equation~(\ref{f1})  becomes 
\be
{n(n-1)\over 2} \left[\Big({a'\over a}\Big)^2+1\right]=8\pi G\, a^2\rho,
% (n-1)\left[{a''\over a}-\Big({a'\over a}\Big)^2\right]+{(n-1)(n-2)\over 2} \left[\Big({ a'\over a}\Big)^2+1\right]&=-8\pi  G\,a^2p,
\label{frieta}
\ee
where primes denote derivative with respect to conformal time $\eta$. Substituting for $\rho$, we obtain
\be
\mbox{for $w\neq w_{\rm c}:$} \quad \Big({a'\over a}\Big)^2=\Big({a\over a_0}\Big)^{n(w_{\rm c}-w)}-1,~~\quad \where ~~\quad a_0=\left({n(n-1)\over 16\pi C}\right)^{1\over n(w_{\rm c}-w)}.
\ee
Defining
\be
{a\over a_0}=A^{-{2\over n(w_{\rm c}-w)}},
\ee
the above equation simplifies to
\be
\Big({\d A\over \d u}\Big)^2+A^2=1, ~~\quad\where~~\quad u={n(w_{\rm c}-w)\over 2}\,\eta,
\ee
leading to the solution
\be
A(u)=|\:\!\!\sin u|, ~~\quad u\in[0,\pi].
\ee
As a result, the scale factor reads
\be
a(\eta)=a_0\!\left(\sin{\eta\over |\gamma|}\right)^\gamma, ~~\quad \eta\in\big[0,|\gamma|\pi\big], 
\ee
where we have defined
\be
\gamma={2\over n(w-w_{\rm c})}.
\ee

For completeness, let us mention that Friedmann equation~(\ref{frieta}) becomes
\be
\mbox{for $w= w_{\rm c}:$} ~~\quad\Big({a'\over a}\Big)^2={16\pi C\over n(n-1)}-1,
\ee
which admits solutions when the right hand side is non-negative. When this is the case, one obtains 
\be
a(\eta)=e^{\pm \eta \sqrt{{16\pi C\over n(n-1)}-1}}, ~~\quad \eta\in \R,
\label{acri}
\ee
for either choice of constant sign $\pm$. 

%%%%%%%%%%%%%%%%%%%%%%%%%%%%%%%%%%%%%%%%%

\section{Area extremization in a causal diamond}
\label{Lagrange_multipliers}

At fixed conformal coordinates $(\theta,\eta)$, the metric~(\ref{so met}) of the $(n+1)$-dimensional closed FRW spacetime reduces to the metric of S$^{n-1}$, whose area $\A$ is given in \Eq{AS}. In this appendix, we first find the extrema of this area, for $(\theta,\eta)$ in the Penrose diagram. We observe that for a cosmological evolution satisfying $|\gamma|>1$ and a generic Cauchy slice $\Sigma$, none of these solutions to the extremization condition $\d\A=0$ lies in the causal diamond of the Cauchy slice $\Sigma_\Ex$, except in the Big Bang/Big Crunch case, at the singularities. However, restricting the domain of definition of $\A$ to this causal diamond leads to the existence of maxima, minima and saddle points on its boundary, which nonetheless satisfy $\d\A\neq 0$. What we show is that Lagrange multipliers and ``auxiliary constants'' can be introduced to reconcile these points with the notion of extremization.

%%%%%%%

\subsection{Extrema of the area defined in the Penrose diagram}
\label{B1}

In this subsection, we define the area function $\A$ on the entire Penrose diagram. We will say that it has an extremum at $(\theta,\eta)$ if $\d \A=0$ at this point. Using \Eq{dAplusminus}, we find that: 

\noindent $\bullet$ For any $\gamma$, $\A$ is extremal at the bifurcate horizon $(\theta,\eta)=(\pi/2,|\gamma|\pi/2)$. When $n\ge 3$, other extrema are located at $(0,\eta)$ and  $(\pi,\eta)$, where $0<\eta<|\gamma|\pi$. 

\noindent $\bullet$ Moreover, when $\gamma>1/(n-1)$, additional extrema are found at the Big Bang and Big Crunch singularities, $\eta=0$ and $\eta=\gamma\pi$.

\noindent $\bullet$ Finally, in the particular case where $\gamma=1/(n-1)$, the only additional extremal points are the corners of the Penrose diagram: $(0,0)$, $(\pi,0)$, $(0,\gamma \pi)$, $(\pi,\gamma\pi)$.

\begin{figure}[h!]
   % \centering
\begin{subfigure}[t]{0.48\linewidth}
\centering
\scalebox{0.97}{
\begin{tikzpicture}

\path
       +(2,4)  coordinate (IItopright)
       +(-2,4) coordinate (IItopleft)
       +(2,-4) coordinate (IIbotright)
       +(-2,-4) coordinate(IIbotleft)

       +(-0.7,2) coordinate (SL)
       +(1.1,2.7) coordinate (SR)
      
       ;

\fill[fill=red!20] (SL) -- node[pos=0.4, above] {{\footnotesize $2$}}(0.55,3.25) -- node[pos=0.75, above] {{\footnotesize $4$}}(SR) -- node[pos=0.5, below] {{\footnotesize $1$}}(-0.15,1.45) -- node[pos=0.7, below] {{\footnotesize $3$}}(SL);
       
\draw[decorate,decoration=snake] (IItopleft) --
          node[midway, above, sloped]    {}
      (IItopright);
      
\draw (IItopright) --
          node[midway, above, sloped] {}
      (IIbotright);
      
\draw[decorate,decoration=snake]  (IIbotright) -- 
          node[midway, below, sloped] {}
      (IIbotleft);
      
\draw (IIbotleft) --
          node[midway, above , sloped] {}
      (IItopleft);
      
\draw (IItopleft) -- (IIbotright)
              (IItopright) -- (IIbotleft) ;

\draw (-2-0.5,-4+0.5) -- (-2,-4);
\draw (-2+0.5,-4+0.5) -- (-2,-4);
\draw (-2-0.5,-4+0.3) -- (-2-0.5,-4+0.5) -- node[midway, left, sloped] {$x^-$} (-2-0.3,-4+0.5) ;
\draw (-2+0.5,-4+0.3) -- (-2+0.5,-4+0.5) -- (-2+0.3,-4+0.5) ;
\node at (-1.6,-4+0.5) [label = right:$x^+$]{};

\draw (-0.2,3.2+0.25) -- (0,3+0.25) -- (0.2,3.2+0.25);

\draw (SL) to [bend right=10] (SR) ;
\node at (0.5,3.2) [label=below:$\Sigma_\Ex$]{};

\node at (SL) [circle,fill,inner sep=2pt, orange, label = above:$\S_\Le$]{};
\node at (SR) [circle,fill,inner sep=2pt, orange]{};
\node at (1.15,2.7) [label = right:$\S_\Ri$]{};
\node at (-0.15,1.45) [circle,fill,inner sep=2pt, violet]{};
\node at (0.55,3.25) [circle,fill,inner sep=2pt, red]{};

\end{tikzpicture}
}    
\caption{\footnotesize When $(x^+_\Ri+x^-_\Le)/2\le \gamma\pi$, the diamond has four edges $1, \dots , 4$.\label{extremaSL}}
\end{subfigure}
\quad \,
\begin{subfigure}[t]{0.48\linewidth}
\centering
\scalebox{0.97}{
\begin{tikzpicture}

\path
       +(2,4)  coordinate (IItopright)
       +(-2,4) coordinate (IItopleft)
       +(2,-4) coordinate (IIbotright)
       +(-2,-4) coordinate(IIbotleft)

       +(-1.1,2.8) coordinate (SL)
       +(1.5,3.4) coordinate (SR)
      
       ;

\fill[fill=red!20] (SL) -- node[pos=0.5, above] {{\footnotesize $2$}}(0.1,4) -- node[midway, above] {{\footnotesize $5$}}(0.9,4) -- node[pos=0.8, above] {{\footnotesize $4$}}(SR) -- node[pos=0.5, below] {{\footnotesize $1$}}(-0.1,1.8) -- node[pos=0.6, below] {{\footnotesize $3$}}(SL);

\draw[decorate,decoration=snake] (IItopleft) --
          node[midway, above, sloped]    {}
      (IItopright);
      
\draw (IItopright) --
          node[midway, above, sloped] {}
      (IIbotright);
      
\draw[decorate,decoration=snake]  (IIbotright) -- 
          node[midway, below, sloped] {}
      (IIbotleft);
      
\draw (IIbotleft) --
          node[midway, above , sloped] {}
      (IItopleft);
      
\draw (IItopleft) -- (IIbotright)
              (IItopright) -- (IIbotleft) ;     
\fill[fill=red] (0.1,4+0.05) -- (0.1,4-0.05) -- (0.9,4-0.05) -- (0.9,4+0.05) -- cycle;

\draw (-2-0.5,-4+0.5) -- (-2,-4);
\draw (-2+0.5,-4+0.5) -- (-2,-4);
\draw (-2-0.5,-4+0.3) -- (-2-0.5,-4+0.5) -- node[midway, left, sloped] {$x^-$} (-2-0.3,-4+0.5) ;
\draw (-2+0.5,-4+0.3) -- (-2+0.5,-4+0.5) -- (-2+0.3,-4+0.5) ;
\node at (-1.6,-4+0.5) [label = right:$x^+$]{};

\draw (-0.2,3.2+0.25) -- (0,3+0.25) -- (0.2,3.2+0.25);

\draw (SL) to [bend right=10] (SR) ;
\node at (0.2,3.05) [label=below:$\Sigma_\Ex$]{};
      
\node at (SL) [circle,fill,inner sep=2pt, orange, label = above:$\S_\Le$]{};
\node at (SR) [circle,fill,inner sep=2pt, orange]{};
\node at (1.75,3.4) [label = below:$\S_\Ri$]{};
\node at (-0.1,1.8) [circle,fill,inner sep=2pt, violet]{};
\node at (0.1,4) [circle,fill,inner sep=2pt, red]{};
\node at (0.9,4) [circle,fill,inner sep=2pt, red]{};

\end{tikzpicture}
}
\caption{\footnotesize When $(x^+_\Ri+x^-_\Le)/2> \gamma\pi$, the diamond has five edges $1, \dots , 5$.     \label{extremaSR}}
\end{subfigure}\\
\begin{subfigure}[t]{0.48\linewidth}
\centering
\scalebox{0.97}{
\begin{tikzpicture}

\path
       +(2,4)  coordinate (IItopright)
       +(-2,4) coordinate (IItopleft)
       +(2,-4) coordinate (IIbotright)
       +(-2,-4) coordinate(IIbotleft)
      
       ;

\fill[fill=red!20] (-3/4,3/2) -- node[pos=0.4, above] {{\footnotesize $2$}}(0.675,2.925) -- node[pos=0.7, above] {{\footnotesize $4$}}(1.2,2.4) -- node[pos=0.5, below] {{\footnotesize $1$}}(-0.225,0.975) -- node[pos=0.3, above] {{\footnotesize $3$}}(-3/4,3/2);
       
\draw (IItopleft) --
          node[midway, above, sloped]    {}
      (IItopright) --
          node[midway, above, sloped] {}
      (IIbotright) -- 
          node[midway, below, sloped] {}
      (IIbotleft) --
          node[midway, above , sloped] {}
      (IItopleft) -- cycle;
      
\draw (IItopleft) -- (IIbotright)
              (IItopright) -- (IIbotleft) ;

\draw (-2-0.5,-4+0.5) -- (-2,-4);
\draw (-2+0.5,-4+0.5) -- (-2,-4);
\draw (-2-0.5,-4+0.3) -- (-2-0.5,-4+0.5) -- node[midway, left, sloped] {$x^-$} (-2-0.3,-4+0.5) ;
\draw (-2+0.5,-4+0.3) -- (-2+0.5,-4+0.5) -- (-2+0.3,-4+0.5) ;
\node at (-1.6,-4+0.5) [label = right:$x^+$]{};

\draw (-0.2,2.8+0.4) -- (0,3+0.4) -- (0.2,2.8+0.4);

\draw (-3/4,3/2) to [bend right=10] (1.2,2.4) ;
\node at (0.45,2.8) [label=below:$\Sigma_\Ex$]{};

\node at (-3/4,3/2) [circle,fill,inner sep=2pt, orange, label = left:$\S_\Le$]{};
\node at (1.2,2.4) [circle,fill,inner sep=2pt, orange, label = right:$\S_\Ri$]{};
\node at (-0.225,0.975) [circle,fill,inner sep=2pt, red]{};
\node at (0.675,2.925) [circle,fill,inner sep=2pt, violet]{};

\end{tikzpicture} 
}
\caption{\footnotesize When $(x^+_\Ri+x^-_\Le)/2\le  |\gamma|\pi$, the diamond has four edges $1, \dots,  4$.\label{extremaBL}}
\end{subfigure}
\quad \,
\begin{subfigure}[t]{0.48\linewidth}
\centering
\scalebox{0.97}{
\begin{tikzpicture}

\path
      +(2,4)  coordinate (IItopright)
       +(-2,4) coordinate (IItopleft)
       +(2,-4) coordinate (IIbotright)
       +(-2,-4) coordinate(IIbotleft)

       +(-1.4,2.8) coordinate (SL)
       +(1.7,3.4) coordinate (SR)
       
       ;

\fill[fill=red!20] (SL) -- node[pos=0.4, above] {{\footnotesize $2$}}(-0.2,4) -- node[midway, above] {{\footnotesize $5$}}(1.1,4) -- node[pos=0.8, above] {{\footnotesize $4$}}(SR) -- node[pos=0.6, below] {{\footnotesize $1$}}(-0.15,1.55) -- node[pos=0.4, below] {{\footnotesize $3$}}(SL);
       
\draw (IItopleft) --
          node[midway, above, sloped]    {}
      (IItopright) --
          node[midway, above, sloped] {}
      (IIbotright) -- 
          node[midway, below, sloped] {}
      (IIbotleft) --
          node[midway, above , sloped] {}
      (IItopleft) -- cycle;

\fill[fill=violet] (1.1,4+0.05) -- (1.1,4-0.05) -- (-0.15,4-0.05) -- (-0.15,4+0.05) -- cycle; 
      
\draw (IItopleft) -- (IIbotright)
              (IItopright) -- (IIbotleft) ;      

\draw (-2-0.5,-4+0.5) -- (-2,-4);
\draw (-2+0.5,-4+0.5) -- (-2,-4);
\draw (-2-0.5,-4+0.3) -- (-2-0.5,-4+0.5) -- node[midway, left, sloped] {$x^-$} (-2-0.3,-4+0.5) ;
\draw (-2+0.5,-4+0.3) -- (-2+0.5,-4+0.5) -- (-2+0.3,-4+0.5) ;
\node at (-1.6,-4+0.5) [label = right:$x^+$]{};

\draw (-0.2,2.8+0.4) -- (0,3+0.4) -- (0.2,2.8+0.4);

\draw (SL) to [bend right=10] (SR) ;
\node at (0.3,3.05) [label=below:$\Sigma_\Ex$]{};

\node at (SL) [circle, fill, inner sep=2 pt,orange]{};
\node at (-1,2.4) [label=left:$\S_\Le$]{};
\node at (SR) [circle,fill,inner sep=2pt, orange]{};
\node at (1.75,3.4) [label = below:$\S_\Ri$]{};

\node at (-0.15,1.55) [circle,fill,inner sep=2pt, red]{};
\node at (-0.2,4) [circle,fill,inner sep=2pt, violet]{};
\node at (1.1,4) [circle,fill,inner sep=2pt, violet]{};

\end{tikzpicture}
}
\caption{\footnotesize When $(x^+_\Ri+x^-_\Le)/2> |\gamma|\pi$, the diamond has five edges $1, \dots , 5$.\label{extremaBR}}
\end{subfigure}

    \caption{\footnotesize Causal diamond of a Cauchy slice $\Sigma_\Ex$ located in the upper triangular region of the  Penrose diagram, when $|\gamma|>1$. In the Big Bang/Big Crunch cases~(a), (b), $\S_\Le$ and $\S_\Ri$ can be in the interior of the triangle. In the bouncing cases~(c), (d), they are on the apparent horizons. The S$^{n-1}$-area function defined in the causal diamond has minima (red), maxima (purple) and saddle points (orange) on the boundary of the diamond.  \label{extrema}}
\end{figure}
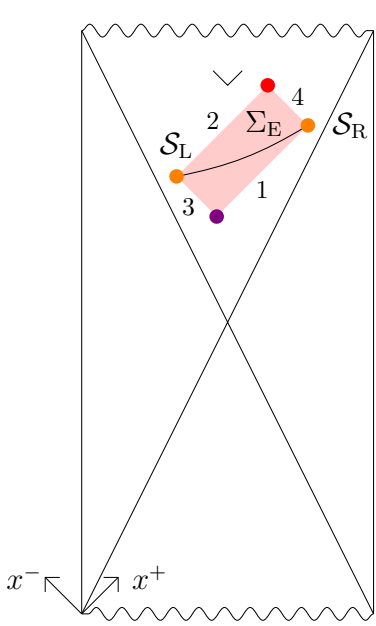
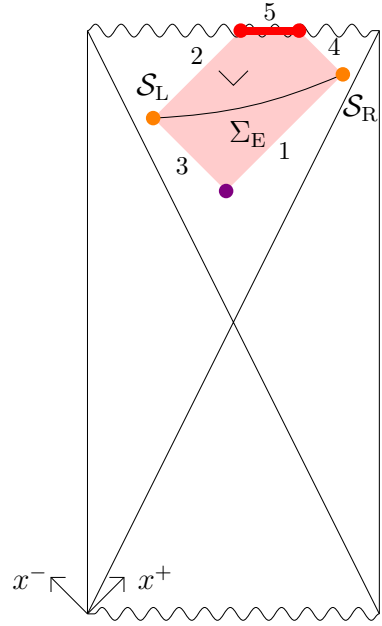
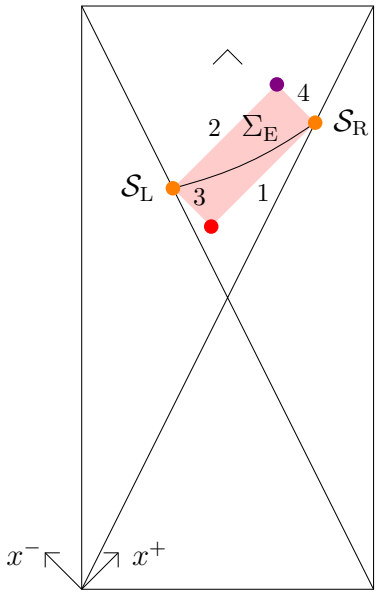
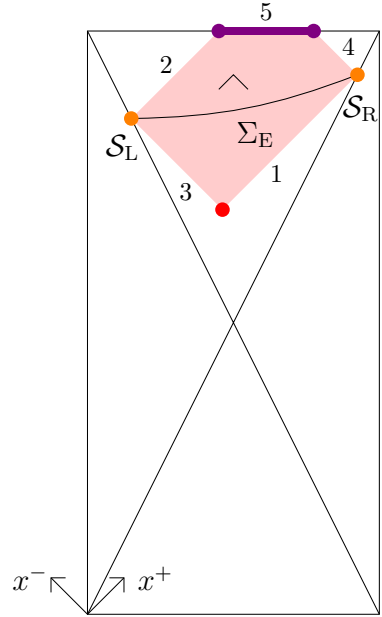

From now on, in this \Appendix{Lagrange_multipliers}, we limit ourselves  to the case where $|\gamma|>1$. We also denote $(x^+_\Le,x^-_\Le)$ and $(x^+_\Ri,x^-_\Ri)$ the null coordinates of $\S_\Le$ and $\S_\Ri$ defined in \Eq{x+-}.  Let us first consider  the case where the Cauchy slice $\Sigma_\Ex$ extends in the upper triangular region of the Penrose diagram in \Fig{extrema}. In the Big Bang/Big Crunch case $\gamma>1$, the boundaries $\S_\Le$ and $\S_\Ri$ of $\Sigma_\Ex$ can be located in the interior of this triangle, as shown in \Figs{extremaSL} and~\ref{extremaSR}. 
On the contrary, in the bouncing case, $\S_\Le$ and $\S_\Ri$ are restricted to lie on the apparent horizons of the pode and antipode, as shown in \Figs{extremaBL} and~\ref{extremaBR}.   When $(x^+_\Ri+x^-_\Le)/2\le |\gamma|\pi$, the causal diamond of $\Sigma_\Ex$ corresponds to the domain
 \be
\left\{\!\begin{array}{l}
x^+_\Le\le x^+\le x^+_\Ri \\
x^-_\Ri\le x^-\le x^+_\Le\esps
\end{array}\right. \!,
\label{rec}
\ee
as follows from the fact that $\S_\Le$ is to the left of $\S_\Ri$ and $\Sigma_\Ex$ is spacelike. It has the shape of a rectangle, whose edges are  
labelled $1,\dots,4$ in \Figs{extremaSL} and \ref{extremaBL}. However, when $(x^+_\Ri+x^-_\Le)/2>|\gamma|\pi$, the domain is further restricted to $\eta\le |\gamma|\pi$, \ie
\be
{x^+ + x^-\over 2}\le |\gamma|\pi .
\label{extra}
\ee
In this case, the boundary of the causal diamond has 5 edges, denoted $1,\dots,5$ in \Figs{extremaSR} and~\ref{extremaBR}. 

\begin{figure}[h!]
   % \centering
\begin{subfigure}[t]{0.48\linewidth}
\centering
\scalebox{0.97}{
\begin{tikzpicture}

\path
       +(2,4)  coordinate (IItopright)
       +(-2,4) coordinate (IItopleft)
       +(2,-4) coordinate (IIbotright)
       +(-2,-4) coordinate(IIbotleft)

       +(-1.3,-2.6) coordinate (SL)
       +(0.85,-1.7) coordinate (SR)
      
       ;

\fill[fill=red!20] (SL) -- node[pos=0.5, above] {{\footnotesize $3$}}(0.225,-1.075) -- node[pos=0.3, below] {{\footnotesize $1$}}(SR) -- node[pos=0.4, below] {{\footnotesize $4$}}(-0.675,-3.225) -- node[pos=0.7, below] {{\footnotesize $2$}}(SL);
       
\draw[decorate,decoration=snake] (IItopleft) --
          node[midway, above, sloped]    {}
      (IItopright);
      
\draw (IItopright) --
          node[midway, above, sloped] {}
      (IIbotright);
      
\draw[decorate,decoration=snake]  (IIbotright) -- 
          node[midway, below, sloped] {}
      (IIbotleft);
      
\draw (IIbotleft) --
          node[midway, above , sloped] {}
      (IItopleft);
      
\draw (IItopleft) -- (IIbotright)
              (IItopright) -- (IIbotleft) ;

\draw (-2-0.5,-4+0.5) -- (-2,-4);
\draw (-2+0.5,-4+0.5) -- (-2,-4);
\draw (-2-0.5,-4+0.3) -- (-2-0.5,-4+0.5) -- node[midway, left, sloped] {$x^-$} (-2-0.3,-4+0.5) ;
\draw (-2+0.5,-4+0.3) -- (-2+0.5,-4+0.5) -- (-2+0.3,-4+0.5) ;
\node at (-1.6,-4+0.5) [label = right:$x^+$]{};

\draw (-0.2,-3.2-0.3) -- (0,-3-0.3) -- (0.2,-3.2-0.3);

\draw (SL) to [bend right=10] (SR) ;
\node at (-0.1,-1.4) [label=below:$\Sigma_\Ex$]{};

\node at (SL) [circle,fill,inner sep=2pt, orange]{};
\node at (-1.15,-2.6) [label = left:$\S_\Le$]{};
\node at (SR) [circle,fill,inner sep=2pt, orange, label = right:$\S_\Ri$]{};
\node at (0.225,-1.075) [circle,fill,inner sep=2pt, violet]{};
\node at (-0.675,-3.225) [circle,fill,inner sep=2pt, red]{};

\end{tikzpicture}
}    
\caption{\footnotesize When $(x^+_\Le+x^-_\Ri)/2\ge 0$, the diamond has four edges $1, \dots , 4$.\label{extremaSL2}}
\end{subfigure}
\quad \,
\begin{subfigure}[t]{0.48\linewidth}
\centering
\scalebox{0.97}{
\begin{tikzpicture}

\path
       +(2,4)  coordinate (IItopright)
       +(-2,4) coordinate (IItopleft)
       +(2,-4) coordinate (IIbotright)
       +(-2,-4) coordinate(IIbotleft)

       +(-1.3,-2.6) coordinate (SL)
       +(1.6,-3.2) coordinate (SR)
      
       ;

\fill[fill=red!20] (SL) -- node[pos=0.6, below] {{\footnotesize $3$}}(-0.15,-1.45) -- node[midway, above] {{\footnotesize $1$}}(SR) -- node[pos=0.3, below] {{\footnotesize $4$}}(0.8,-4) -- node[midway, above] {{\footnotesize $5$}}(0.1,-4) -- node[pos=0.4, above] {{\footnotesize $2$}}(SL);

\draw[decorate,decoration=snake] (IItopleft) --
          node[midway, above, sloped]    {}
      (IItopright);
      
\draw (IItopright) --
          node[midway, above, sloped] {}
      (IIbotright);
      
\draw[decorate,decoration=snake]  (IIbotright) -- 
          node[midway, below, sloped] {}
      (IIbotleft);
      
\draw (IIbotleft) --
          node[midway, above , sloped] {}
      (IItopleft);
      
\draw (IItopleft) -- (IIbotright)
              (IItopright) -- (IIbotleft) ;     
\fill[fill=red] (0.1,-4+0.05) -- (0.1,-4-0.05) -- (0.8,-4-0.05) -- (0.8,-4+0.05) -- cycle;

\draw (-2-0.5,-4+0.5) -- (-2,-4);
\draw (-2+0.5,-4+0.5) -- (-2,-4);
\draw (-2-0.5,-4+0.3) -- (-2-0.5,-4+0.5) -- node[midway, left, sloped] {$x^-$} (-2-0.3,-4+0.5) ;
\draw (-2+0.5,-4+0.3) -- (-2+0.5,-4+0.5) -- (-2+0.3,-4+0.5) ;
\node at (-1.6,-4+0.5) [label = right:$x^+$]{};

\draw (-0.2,-3.2-0.3) -- (0,-3-0.3) -- (0.2,-3.2-0.3);

\draw (SL) to [bend right=10] (SR) ;
\node at (0,-2.2) [label=below:$\Sigma_\Ex$]{};
      
\node at (SL) [circle,fill,inner sep=2pt, orange]{};
\node at (-1.15,-2.6) [label = left:$\S_\Le$]{};
\node at (SR) [circle,fill,inner sep=2pt, orange]{};
\node at (1.7,-3.2) [label = above:$\S_\Ri$]{};
\node at (-0.15,-1.45) [circle,fill,inner sep=2pt, violet]{};
\node at (0.1,-4) [circle,fill,inner sep=2pt, red]{};
\node at (0.8,-4) [circle,fill,inner sep=2pt, red]{};

\end{tikzpicture}
}
\caption{\footnotesize When $(x^+_\Le+x^-_\Ri)/2<0$, the diamond has five edges $1, \dots , 5$.     \label{extremaSR2}}
\end{subfigure}\\
\begin{subfigure}[t]{0.48\linewidth}
\centering
\scalebox{0.97}{
\begin{tikzpicture}

\path
       +(2,4)  coordinate (IItopright)
       +(-2,4) coordinate (IItopleft)
       +(2,-4) coordinate (IIbotright)
       +(-2,-4) coordinate(IIbotleft)

       +(-1.1,-2.8) coordinate (SL)
       +(0.7,-2.1) coordinate (SR)
      
       ;

\fill[fill=red!20] (SL) -- node[pos=0.4, above] {{\footnotesize $3$}}(0.15,-1.55) -- node[pos=0.75, above] {{\footnotesize $1$}}(SR) -- node[pos=0.5, below] {{\footnotesize $4$}}(-0.55,-3.35) -- node[pos=0.25, above] {{\footnotesize $2$}}(SL);
       
\draw (IItopleft) --
          node[midway, above, sloped]    {}
      (IItopright) --
          node[midway, above, sloped] {}
      (IIbotright) -- 
          node[midway, below, sloped] {}
      (IIbotleft) --
          node[midway, above , sloped] {}
      (IItopleft) -- cycle;
      
\draw (IItopleft) -- (IIbotright)
              (IItopright) -- (IIbotleft) ;

\draw (-2-0.5,-4+0.5) -- (-2,-4);
\draw (-2+0.5,-4+0.5) -- (-2,-4);
\draw (-2-0.5,-4+0.3) -- (-2-0.5,-4+0.5) -- node[midway, left, sloped] {$x^-$} (-2-0.3,-4+0.5) ;
\draw (-2+0.5,-4+0.3) -- (-2+0.5,-4+0.5) -- (-2+0.3,-4+0.5) ;
\node at (-1.6,-4+0.5) [label = right:$x^+$]{};

\draw (-0.2,-2.8-0.5-0.2) -- (0,-3-0.5-0.2) -- (0.2,-2.8-0.5-0.2);

\draw (SL) to [bend right=10] (SR) ;
\node at (0,-1.65) [label=below:$\Sigma_\Ex$]{};

\node at (SL) [circle,fill,inner sep=2pt, orange]{};
\node at (-1.12,-2.6) [label = left:$\S_\Le$]{};
\node at (SR) [circle,fill,inner sep=2pt, orange, label = below:$\S_\Ri$]{};
\node at (0.15,-1.55) [circle,fill,inner sep=2pt, red]{};
\node at (-0.55,-3.35) [circle,fill,inner sep=2pt, violet]{};

\end{tikzpicture} 
}
\caption{\footnotesize When $(x^+_\Le+x^-_\Ri)/2\ge 0$, the diamond has four edges $1, \dots,  4$.\label{extremaBL2}}
\end{subfigure}
\quad \,
\begin{subfigure}[t]{0.48\linewidth}
\centering
\scalebox{0.97}{
\begin{tikzpicture}

\path
      +(2,4)  coordinate (IItopright)
       +(-2,4) coordinate (IItopleft)
       +(2,-4) coordinate (IIbotright)
       +(-2,-4) coordinate(IIbotleft)

       +(-1.1,-2.8) coordinate (SL)
       +(1.5,-3.4) coordinate (SR)
       
       ;

\fill[fill=red!20] (SL) -- node[pos=0.4, above] {{\footnotesize $3$}}(-0.1,-1.8) -- node[pos=0.5, above] {{\footnotesize $1$}}(SR) -- node[pos=0.2, below] {{\footnotesize $4$}}(0.9,-4) -- node[midway, above] {{\footnotesize $5$}}(0.1,-4) -- node[pos=0.5, below] {{\footnotesize $2$}}(SL);
       
\draw (IItopleft) --
          node[midway, above, sloped]    {}
      (IItopright) --
          node[midway, above, sloped] {}
      (IIbotright) -- 
          node[midway, below, sloped] {}
      (IIbotleft) --
          node[midway, above , sloped] {}
      (IItopleft) -- cycle;

\fill[fill=violet] (0.9,-4+0.05) -- (0.9,-4-0.05) -- (0.1,-4-0.05) -- (0.1,-4+0.05) -- cycle; 
      
\draw (IItopleft) -- (IIbotright)
              (IItopright) -- (IIbotleft) ;      

\draw (-2-0.5,-4+0.5) -- (-2,-4);
\draw (-2+0.5,-4+0.5) -- (-2,-4);
\draw (-2-0.5,-4+0.3) -- (-2-0.5,-4+0.5) -- node[midway, left, sloped] {$x^-$} (-2-0.3,-4+0.5) ;
\draw (-2+0.5,-4+0.3) -- (-2+0.5,-4+0.5) -- (-2+0.3,-4+0.5) ;
\node at (-1.6,-4+0.5) [label = right:$x^+$]{};

\draw (-0.2,-2.8-0.5-0.2) -- (0,-3-0.5-0.2) -- (0.2,-2.8-0.5-0.2);

\draw (SL) to [bend right=10] (SR) ;
\node at (0,-2.4) [label=below:$\Sigma_\Ex$]{};

\node at (SL) [circle, fill, inner sep=2 pt,orange]{};
\node at (-1.15,-2.6) [label=left:$\S_\Le$]{};
\node at (SR) [circle,fill,inner sep=2pt, orange]{};
\node at (1.75,-2.4) [label = below:$\S_\Ri$]{};

\node at (-0.1,-1.8) [circle,fill,inner sep=2pt, red]{};
\node at (0.9,-4) [circle,fill,inner sep=2pt, violet]{};
\node at (0.1,-4) [circle,fill,inner sep=2pt, violet]{};

\end{tikzpicture}
}
\caption{\footnotesize When $(x^+_\Le+x^-_\Ri)/2<0$, the diamond has five edges $1, \dots , 5$.\label{extremaBR2}}
\end{subfigure}

    \caption{\footnotesize Causal diamond of a Cauchy slice $\Sigma_\Ex$ located in the lower triangular region of the  Penrose diagram, when $|\gamma|>1$. In the Big Bang/Big Crunch cases~(a),~(b), $\S_\Le$ and $\S_\Ri$ are on the apparent horizons. In the bouncing cases~(c),~(d), they  can be in the interior of the triangle.  The S$^{n-1}$-area function defined in the causal diamond has minima (red), maxima (purple) and saddle points (orange) on the boundary of the diamond.  \label{extrema2}}
\end{figure}
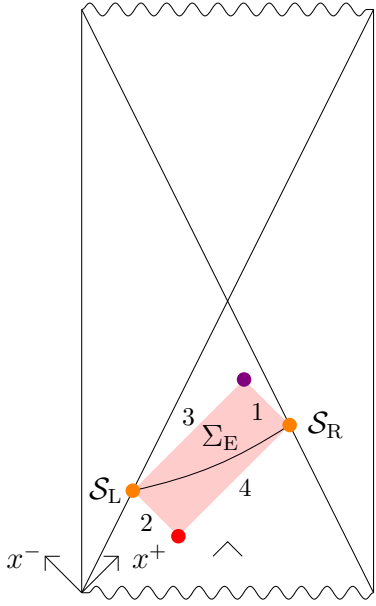
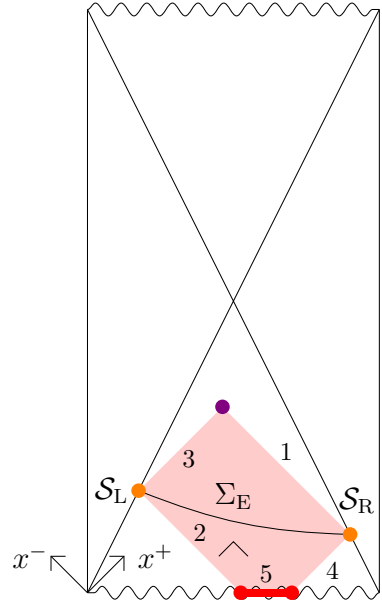
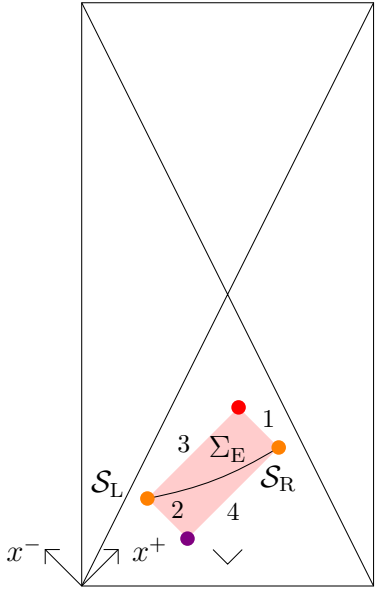
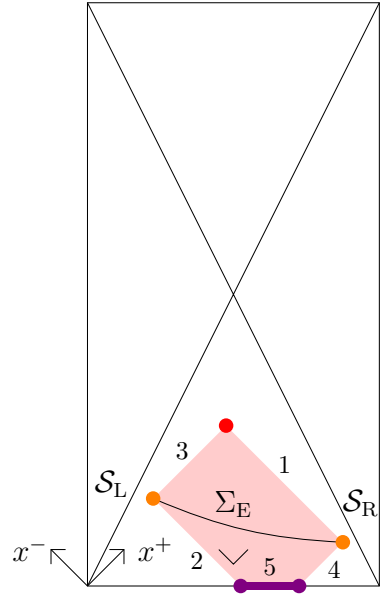

When $\Sigma_\Ex$ lies in the lower triangular region, in the Big Bang/Big Crunch case $\gamma>1$,  $\S_\Le$ and $\S_\Ri$ are located on the apparent horizons of the pode and antipode, as shown in \Figs{extremaSL2} and~\ref{extremaSR2}. 
In the bouncing case, the spheres $\S_\Le$ and $\S_\Ri$ can be placed in the interior of the lower triangle, as shown in \Figs{extremaBL2} and~\ref{extremaBR2}. When $(x^+_\Le+x^-_\Ri)/2\ge 0$, the causal diamond of $\Sigma_\Ex$ corresponds to the domain defined in \Eq{rec}, with  four edges, as indicated in \Figs{extremaSL2} and~\ref{extremaBL2}. When $(x^+_\Le+x^-_\Ri)/2< 0$, the diamond is further restricted~to~
\be
{x^+ + x^-\over 2}\ge 0
\label{extra2}
\ee
and has a fifth edge, as shown in \Figs{extremaSR2} and~\ref{extremaBR2}.

In all cases, assuming that the spacetime Cauchy slice $\Sigma$ is generic, \ie does not passes through the bifurcate horizon, we see that:

\noindent $\bullet$  In the bouncing case $\gamma<-1$, no extremum of the function $\A$ lies in the causal diamond of $\Sigma_\Ex$. 

\noindent $\bullet$ In the Big Bang/Big Crunch case $\gamma>1$, the extrema of $\A$ in the diamond are located along the edge 5, when the latter exists. These extrema are global minima since all S$^{n-1}$ (as well as the S$^n$) vanish at the singularity. 

%%%%%%%%%%%%

\subsection{Minima, maxima and saddle points of the area defined in a causal diamond}
\label{B2}

Another point of view is to choose the domain of definition of the area function $\A$ as the causal diamond of $\Sigma_\Ex$ only. Since this domain has a boundary, minima, maxima and saddle points can exist on this boundary, which however are not ``extrema,'' in the sense that $\d\A\neq 0$ at these points. When $\Sigma_\Ex$ is in a upper (lower) triangular region  of the Penrose diagrams  in \Figs{extrema} (\Fig{extrema2}), we can immediately conclude the following thanks to the orientation of the Bousso wedges.

\noindent $\bullet$ Bouncing case $\gamma<-1$:

-  The upper (lower) vertex of the diamond in \Fig{extremaBL} (\Fig{extremaBL2}) is a global maximum of $\A$. Moreover, since all S$^{n-1}$'s along the edge 5 in \Fig{extremaBR} (\Fig{extremaBR2}) have infinite areas, they correspond to global maxima.

-  The lower (upper) vertex is a global minimum.

- The left and right vertices are saddle points.\footnote{For instance, when $\Sigma_\Ex$ is in the upper triangular region, the function $\A$ initially at the left vertex increases when $x^+$ increases and decreases when $x^-$ decreases. }  

\noindent $\bullet$ Big Bang/Big Crunch case $\gamma>1$:

-  The upper (lower) vertex of the diamond in \Fig{extremaSL} (\Fig{extremaSL2}) is a global minimum of $\A$. Moreover, we already saw that all S$^{n-1}$'s along the edge 5 in \Fig{extremaSR} (\Fig{extremaSR2}) are vanishing and thus correspond to global minima.

-  The lower (upper) vertex is a global maximum.

- The left and right vertices are saddle points.

%%%%%%%%%

\subsection{Minima, maxima and saddle points as solutions of an extremization problem}
\label{B3}

By following the method already presented in \Refe{Franken:2023pni}, all minima, maxima and saddle points we  found for the area function $\A$ defined in the causal diamond of $\Sigma_\Ex$ can be retrieved from an extremization problem.
In the following, we show this explicitly, assuming that $\Sigma_\Ex$ lies in a upper triangular region of the Penrose diagrams in \Fig{extrema}. Indeed, the case where $\Sigma_\Ex$ is in a lower triangular region in \Fig{extrema2} can be analyzed in a similar way. 

\noindent $\bullet$ Let us  first consider the case where $(x^+_\Ri+x^-_\Le)/2\le  |\gamma|\pi$, for which the causal diamond is shown in \Figs{extremaSL} and~\ref{extremaBL}.  The constraints~(\ref{rec}) can be imposed by supplementing the area function with terms proportional to Lagrange multipliers~$\nu_I$, $I\in\{1,2,3,4\}$,
\be\begin{aligned}
\hA(x^+,x^-,\nu_I,a_I)=\A(x^+,x^-)&+\nu_1\!\left(x^--x^-_\Ri-a_1^2\right)+\nu_2\!\left(x^-_\Le -x^- -a_2^2\right)\\
&+\nu_3\!\left(x^+-x^+_\Le-a_3^2\right)+\nu_4\!\left(x^+_\Ri-x^+-a_4^2\right)\!.
\end{aligned}\label{hA}\ee
Beside the multipliers, the $a_I$'s are extra variables whose squares are the ``positive distances from $(x^+,x^-)$ to the edges $I$ in the Penrose diagrams'' in \Fig{extremaSL} and~\ref{extremaBL}. The introduction of the $a_I$'s is necessary because the constraints~(\ref{rec}) are inequalities rather than equalities. The function $\hA$ is defined in a domain without boundary, where $x^\pm$, $\nu_I$, $a_I$ are all spanning~$\R$. To find its extrema, \ie the points in $\R^{10}$ satisfying $\d\hA=0$, we first vary $\hA$ with respect to~$x^\pm$, which yields
\begin{subequations}
\label{x}
 \begin{align}
\nu_4-\nu_3&= \frac{\partial{\A}}{\partial x^+},\label{x+}\\
\nu_2-\nu_1&= \frac{\partial{\A}}{\partial x^-}.\label{x-}
\end{align}
\end{subequations}
Varying $\hA$ with respect to the Lagrange multipliers leads to the equations 
\begin{subequations}
\label{n}
 \begin{align}
a_1^2&= x^--x^-_\Ri\label{n1},\\
a_2^2&=x^-_\Le-x^-,\label{n2}\\
a_3^2&=x^+-x^+_\Le ,\label{n3}\\
a_4^2&=x^+_\Ri -x^+.\label{n4}
\end{align}
\end{subequations} 
When they are satisfied, the squares of the ``auxiliary constants'' $a_I$ are fixed and $\hA=\A$. Finally, the variations of $\hA$ with respect to the $a_I$'s give
\begin{subequations}
\label{a}
\begin{align}
\nu_1a_1&=0,\label{a1app}\\
\nu_2a_2&=0,\label{a2app}\\
\nu_3a_3&=0,\label{a3}\\
\nu_4a_4&=0.\label{a4}
\end{align}
\end{subequations}
To find all solutions to these equations, we organize our discussion from the location of $(x^+,x^-)$ in the entire $\R^2$-plane:

- When $(x^+,x^-)$ is not on the boundary of the diamond, \Eqs{n} impose it to lie in the interior of the diamond and determine the values of $a^2_I> 0$, $I\in\{1,2,3,4\}$. As a result, \Eqs{a} imply $\nu_I=0$. However, \Eqs{x} have no solution, as seen at the end of \Sect{B1}. 

- Take now $(x^+,x^-)$ in the interior of edge~1, \ie not at its endpoints. We have $x^-=x^-_\Ri$, while  \Eqs{n3} and~(\ref{n4}) determine $a^2_{3,4}> 0$. As a result, \Eqs{a3} and~(\ref{a4}) imply $\nu_{3,4}=0$. Hence,  \Eq{x+} reduces to $\partial\A/\partial x^+=0$, which cannot be satisfied, as the interior of edge 1 does not cross an apparent horizon. In a similar way, there is no solution to the equations when  $(x^+,x^-)$ lies in the interior of edges~2,~3,~4. 

- If $(x^+,x^-)$ is at the upper vertex of the diamond, \ie  at the intersection of edges~2 and 4, we have $(x^+,x^-)=(x^+_\Ri,x^-_\Le)$ and thus $a_{2,4}=0$ from \Eqs{n2}, (\ref{n4}). \Eqs{a2app}, (\ref{a4}) are thus satisfied. \Eqs{n1} and  (\ref{n3}) determine $a^2_{1,3}>0$, which implies $\nu_{1,3}=0$ from \Eqs{a1app}, (\ref{a3}). \Eqs{x+}, (\ref{x-}) then fix $\nu_{4,2}$.\footnote{In the bouncing case, we have $\nu_2\to +\infty$ and $\nu_4\to +\infty$ when the upper vertex approaches the line $\eta=|\gamma|\pi$. However, this is not an issue, as will be noticed below \Eq{ninf}.} We have thus found an extremum of $\hA$.\footnote{There are actually four degenerate extrema, since the fixed values of $a_1^2>0$, $a_3^2>0$ yield four solutions, $a_1=\pm\sqrt{a_1^2}$, $a_3=\pm\sqrt{a_3^2}$.} In a similar way, the other 3 vertices of the diamond correspond to extrema of $\hA$. All these solutions, which are represented by dots in \Figs{extremaSL} and~\ref{extremaBL},  are in agreement with the maxima, minima and saddle points of $\A$ described in \Sect{B2}, except that now the extremization condition $\d\hA=0$ is satisfied. 

\noindent $\bullet$ When  $(x^+_\Ri+x^-_\Le)/2> |\gamma|\pi$, the causal diamond is represented in \Figs{extremaSR} and~\ref{extremaBR}. We introduce extra variables $\nu_5$, $a_5$ in $\R$ and define
\be
\tA(x^+,x^-,\nu_I,\nu_5,a_I,a_5)=\hA(x^+,x^-,\nu_I,a_I)+\nu_5\!\left(2|\gamma|\pi-x^+-x^--a_5^2 \right)
\ee
to impose the inequality constraint~(\ref{extra}). 
Solving the equation $\d\tA=0$ amounts to satisfying~
\begin{subequations}
\label{x'}
 \begin{align}
\nu_5+\nu_4-\nu_3&= \frac{\partial{\A}}{\partial x^+},\label{x+'}\\
\nu_5+\nu_2-\nu_1&= \frac{\partial{\A}}{\partial x^-},\label{x-'}
\end{align}
\end{subequations} 
\Eqs{n} and 
\be
a_5^2=2|\gamma|\pi-x^+-x^-,
\label{n5}
\ee
along with \Eqs{a} and   
\be
\nu_5\,a_5=0.
\label{a5}
\ee
When \Eqs{n} and~(\ref{n5}) are fulfilled, we have $\tA=\A$. As before, we organize our discussion from the location of $(x^+,x^-)$ in the $\R^2$-plane:

- When $(x^+,x^-)$ is not on the line $\eta=|\gamma|\pi$, \Eq{n5} imposes it to be below this line and determines $a_5^2> 0$. Therefore, $\nu_5=0$ from \Eq{a5} and we have $\tA=\hA$. We can then apply all steps of the analysis presented below \Eq{a4} to the full rectangular diamond defined by \Eq{rec}, and simply omit the solution located above the line $\eta=|\gamma|\pi$, which is the upper vertex of the rectangular diamond. We have thus found extrema of $\tA$ located at the left, right and bottom vertices of the diamonds depicted in \Fig{extremaSR} and \ref{extremaBR}.

- Take now $(x^+,x^-)$ along the line $\eta=|\gamma|\pi$, but not at the boundary endpoints of edge~5. \Eqs{n} impose it to lie in the interior of edge 5 and determine $a_I^2>0$, $I\in\{1,2,3,4\}$. Hence, $\nu_I=0$ from \Eq{a}. Since $x^++x^-=2|\gamma|\pi$, we also have $a_5=0$ from \Eq{n5}, while \Eq{a5} is satisfied. Finally, \Eqs{x+'} and~(\ref{x-'}) are equivalent, giving
\be\begin{aligned}
\nu_5&=-\sign(\gamma)\,{n-1\over 2}\,\omega_{n-1}\,a_0^{n-1}\, (\sin\theta)^{n-1} \left(\sin{\eta\over |\gamma|}\right)^{\gamma(n-1)-1}\\
&=\left\{\!\begin{array}{ll}
0&~~\when\quad \gamma>1, \\
+\infty&~~\when\quad \gamma<-1.
\end{array}\right.
\end{aligned}\label{ninf}\ee
The interior of edge~5 thus corresponds to degenerate extrema of $\tA$. Notice that in the bouncing case $\gamma<-1$, the fact that $\nu_5$ is infinite should not be an issue. Indeed, since homologous surfaces at infinite distances (and thus infinite sizes) should be allowed in limiting cases, other variables such as Lagrange multipliers should also be allowed to be infinite. 

- If $(x^+,x^-)$ is at the intersection of edges~2 and~5, we have $x^-=x^-_\Le$ and $x^+=-x^-_\Le+2|\gamma|\pi$, and thus $a_{2,5}=0$ according to  \Eqs{n2},~(\ref{n5}). \Eqs{a2app},~(\ref{a5}) are thus satisfied. 
\Eqs{n1}, (\ref{n3}), (\ref{n4}) also fix $a^2_{1,3,4}> 0$, which implies $\nu_{1,3,4}=0$ from \Eqs{a1app}, (\ref{a3}), (\ref{a4}). \Eq{x+'}, which reduces to \Eq{ninf}, determines $\nu_5$, while (\ref{x-'}) yields $\nu_2=0$. Hence, the left endpoint of edge~5 corresponds to an extremum of $\tA$. In a similar way, the right endpoint of edge~5 corresponds to an extremum of $\tA$. 

\noindent To summarize, we have recovered all the maxima, minima and saddle points of $\A$ found in \Sect{B2}, except that now the extremization condition $\d\tA=0$ is satisfied. Indeed, the Lagrange multipliers are tuned in order to compensate the derivatives of $\A$ and thus obtain  extrema of $\hA$ or $\tA$.

%%%%%%%%%%%%%%%%%%%%%%%%%%%%%%%%%%%%%%%%%%%%%%%%%%

\section{Thermodynamics and FRW cosmology}
\label{thermo}

When the perfect fluid responsible for the FRW cosmological evolution arises from quantum fields at finite temperature, its energy density and pressure satisfy thermodynamical identities. In this appendix, we review how these relations can be fully derived from the variational principle~\cite{Catelin-Jullien:2009duh}. However, if scalar fields exist, we restrict to the case where they are stabilized at the bottom of their potential. This is only for the sake of simplicity, since the analysis can be generalized to take into account coherent motions of scalar fields, such as run away behaviors, oscillations at bottom wells, freezing along plateaus, etc~\cite{Bourliot:2009ai, Estes:2010sh}.   

Let us consider the general relativity action divided by $\hbar$,
\be
{I\over \hbar} =\int \d x^{1+n}\sqrt{-g}\left({R\over 16\pi G\hbar}-\F\right)\!,
\ee
where $R$ is the Ricci curvature and $\F$ is the Helmholtz free energy density of quantum fields at finite temperature $T$. The source term $\F$ can be computed at 1-loop \via a path integral on the background $\mbox{S}^1\times\mbox{S}^n$, where time is Euclidean and compactified on the circle S$^1$ of circumference $\beta=\hbar/T$ and S$^n$ is space.\footnote{Higher order quantum corrections can also be taken into account if one wishes to go beyond the semiclassical level considered in the present work.} By analytic continuation back to real time, the FRW metric for a closed universe takes the form~(\ref{me}), in the gauge $N(x^0)\equiv \beta(x^0)$. In general, the free energy density can depend on the (inverse) temperature and the scale factor, $\F(\beta,a)$. Varying $I$ with respect to $\beta$ and $a$ then leads to the equations of motion
 \begin{align}
{n(n-1)\over 2} \left[\Big({\o a\over a}\Big)^2+{\beta^2\over a^2}\right]&=\phantom{-}8\pi G \hbar\, \beta^2\rho,\label{f1t}\\
(n-1)\,{\oo a\over a}+{(n-1)(n-2)\over 2} \left[\Big({\o a\over a}\Big)^2+{\beta^2\over a^2}\right]-(n-1){\o a\over a}{\o\beta\over \beta}&=-8\pi  G \hbar\,\beta^2p, \label{f2t}
\end{align}
where $*$-derivatives are with respect to time $x^0$, while the energy density and pressure are given by\footnote{When $\F$ depends only on $\beta$, which is often the case, one obtains $\rho+p=T\partial p/\partial T$ and $p=-\F$.}
\be
\rho=\F+\beta\,{\partial \F\over \partial \beta}, \qquad p=-\F-{a\over n}\,{\partial \F\over \partial a}.
\label{rp}
\ee
These relations are equivalent to the usual thermodynamical identities valid at thermal equilibrium 
\be
U=\left({\partial (\beta F)\over \partial \beta}\right)_{\!\!V}\! , \qquad p=-\left({\partial F\over \partial V}\right)_{\!\!\beta}\! .
\ee
In these expressions, $U=\rho V$ and $F(\beta,V)=\F V$ are the internal energy and Helmholtz free energy of the entire closed universe of volume $V=\omega_n a^n$, where $\omega_n$ is the volume of the unit~S$^n$. 

In the gauge $N\equiv 1$, the equations of motion reduce to \Eqs{f1} and~(\ref{f2}), where the expressions~(\ref{rp}) of the source terms $\rho$, $p$ hold. Since $\F$ is a function of $\beta$ and $a$ only, we have
\be
\dot \F={\partial \F\over \partial \beta}\, \dot \beta+{\partial \F\over \partial a}\, \dot a.
\ee
Subtracting term by term the above equation from \Eq{conser}, we  obtain 
\begin{align}
(\dot \rho-\dot\F)+n\, {\dot a\over a}\, (\rho-\F)&=-{\partial\F\over \partial \beta}\, \dot\beta\\
&=-(\rho-\F)\, {\dot \beta\over \beta},
\end{align}
where \Eq{rp} is used in the second line. By integrating, we have 
\be
\ln\!\big[(\rho-\F)a^n\beta\big]\!=\mbox{constant}
\ee
or, equivalently,
\be
{U-F\over T}\equiv S_{\rm th}=\mbox{constant}.
\label{isentro}
\ee
In this result, we recognize the definition of the thermal entropy $S_{\rm th}$ of the system in terms of energy, free energy and temperature. The FRW cosmological evolution is thus isentropic, which has to be the case since it is quasi static ($p$ is well defined at every time) and adiabatic.

\end{appendices} 
 
%%%%%%%%%%%%%%%%%%%%%%%%%%%%%%%%%%%%%%%%%%%%%%%%%%%%%%%%%%%%%%%%%%%%%%%%%%%%%%%%%%%%

\normalem
\bibliographystyle{jhep}
\bibliography{FRW}

\end{document}